\documentclass[11pt,a4paper]{article}                           
\pdfoutput=1

\usepackage{hyperref}
\usepackage{bookmark}
\usepackage{tikz}
\usepackage{tikz-3dplot}
\usetikzlibrary{calc}
\usetikzlibrary{decorations} %
\usepackage[UKenglish]{babel}
\usepackage[toc,page]{appendix}
\usepackage{amsmath}
\usepackage{amssymb}
\usepackage{graphicx}
\usepackage{hhline}
\usepackage[bf]{caption}
\usepackage{cite}
\usepackage[vcentermath]{youngtab}
\usepackage{geometry}
\DeclareSymbolFont{bbold}{U}{bbold}{m}{n}
\DeclareSymbolFontAlphabet{\mathbbold}{bbold}

 \def\be{\begin{equation}}
\def\ee{\end{equation}}
\def\bseq{\begin{subequations}}
\def\eseq{\end{subequations}}

\def\bea{\begin{eqnarray}}
\def\eea{\end{eqnarray}}

\def\bseq{\begin{subequations}}
\def\eseq{\end{subequations}}

\usepackage{todonotes}

\numberwithin{equation}{section}
\numberwithin{table}{section}

\newcommand{\tw}{{\rm w}}

\newcommand{\executeiffilenewer}[3]{%
 \ifnum\pdfstrcmp{\pdffilemoddate{#1}}%
 {\pdffilemoddate{#2}}>0%
 {\immediate\write18{#3}}\fi%
}

\def\X {X}
\def\Y {Y}
\def\B  {B}

\def\d {{\rm d}}

\def\calc         {{\cal C}}
\def\cald         {{\cal D}}

\def\cali         {{\cal I}}

\def\calo         {{\cal O}}

\def\cals         {{\cal S}}

\def\calw         {{\cal W}}

\def\tw   {{\rm w}}

\def\del          {\partial}

\def\ii           {{\rm i}}

\def\Im           {{\rm Im\hskip0.1em}}

\def\sqr#1#2{{\vcenter{\vbox{\hrule height.#2pt
 \hbox{\vrule width.#2pt height#1pt \kern#1pt \vrule width.#2pt}\hrule
 height.#2pt}}}}

\def\aDo{\overline{\rm D1}}

\def\d{\text{d}}

\def\slashchar#1{\setbox0=\hbox{$#1$}           
\dimen0=\wd0                                 
\setbox1=\hbox{/} \dimen1=\wd1               
\ifdim\dimen0>\dimen1                        
\rlap{\hbox to \dimen0{\hfil/\hfil}}      
#1                                        
\else                                        
\rlap{\hbox to \dimen1{\hfil$#1$\hfil}}   
/                                         
\fi}

\begin{document}

\baselineskip=14pt
\parskip 5pt plus 1pt

\vspace*{-1.5cm}
\begin{flushright}    
  {\tiny DFPD-2015/TH/13

  }
\end{flushright}

\vspace{2cm}
\begin{center}        
  {\LARGE Non-perturbative selection rules in F-theory}
\end{center}

\vspace{0.75cm}
\begin{center}        
Luca Martucci$^1$ and  Timo Weigand$^2$
\end{center}

\vspace{0.15cm}
\begin{center}        
\emph{ 
$^1$ Dipartimento di Fisica e Astronomia `Galileo Galilei', Universit\`a di Padova, \\
\& I.N.F.N.\ Sezione di Padova, via Marzolo 8, I-35131 Padova, Italy\\ \vspace{2mm}
$^2$Institut f\"ur Theoretische Physik, Ruprecht-Karls-Universit\"at, \\
             Philosophenweg 19, 69120
             Heidelberg, Germany
             }
\end{center}

\vspace{2cm}


\begin{abstract}
\noindent  
We discuss the structure of charged matter couplings in 4-dimensional F-theory compactifications. 
Charged matter is known to arise from M2-branes wrapping fibral curves on an elliptic or genus-one fibration $Y$.
If a set of fibral curves satisfies a homological relation in the fibre homology, a coupling involving the
states can arise without exponential volume suppression due to a splitting and joining of the M2-branes. 
If the fibral curves only sum to zero in the integral homology of the full fibration, no such coupling is possible. In this case an M2-instanton wrapping a 3-chain bounded by the fibral matter curves can induce a D-term which is volume suppressed. 
We elucidate the consequences of this pattern for the appearance of massive $U(1)$ symmetries in F-theory and 
analyse the structure of discrete selection rules in the coupling sector. The weakly coupled analogue of said M2-instantons is worked out to be given by D1-F1 instantons.
The generation of an exponentially suppressed F-term requires the formation of half-BPS bound states of M2 and M5-instantons. This effect and its description in terms of fluxed M5-instantons is discussed in a companion paper.

\end{abstract}

\thispagestyle{empty}
\clearpage
\setcounter{page}{1}


\newpage

\tableofcontents
\section{Introduction}

Among the most intriguing aspects of F-theory \cite{Vafa:1996xn,Morrison:1996na,Morrison:1996pp} is the ingenuity with which the symmetries of the effective action are encoded in the topological and geometrical properties of the compactification space.
For instance, non-abelian gauge symmetries are related to the fibre structure of the F-theory genus-one fibration in complex codimension-one, while the massless abelian gauge symmetries 
are in one-to-one correspondence with the Mordell-Weil group of rational sections. This endows abstract structures in algebraic and arithmetic geometry with a direct physics interpretation.

From the perspective of 4-dimensional effective actions an equally rich and important concept is the notion of a massive $U(1)$ symmetry.
By this one means a global abelian symmetry of the effective action which becomes gauged at scales above a certain mass scale of the theory. Indeed it is conjectured that in any consistent quantum theory of gravity, global symmetries in the infrared must arise from a gauged symmetry in the ultra-violet (see e.g. \cite{Banks:2010zn} and references therein), and string theory is no exception. For instance, in perturbative string compactifications, abelian gauge symmetries can become massive via a St\"uckelberg coupling with an axionic field. At energies below the St\"uckelberg mass scale, the $U(1)$ symmetry persists as a perturbative selection rule of the effective action. It is broken only by non-perturbative effects  involving the exponential of the St\"uckelberg axion \cite{Blumenhagen:2006xt,Ibanez:2006da,Florea:2006si,Haack:2006cy,Blumenhagen:2009qh} to a discrete subgroup $\mathbb Z_k$ \cite{BerasaluceGonzalez:2011wy,Camara:2011jg,BerasaluceGonzalez:2012zn,BerasaluceGonzalez:2012vb,Berasaluce-Gonzalez:2013bba,Berasaluce-Gonzalez:2013sna}. 
This includes the case $k=1$, in which the $U(1)$ symmetry is completely broken even though the couplings breaking it in the effective action are exponentially suppressed. 
The non-perturbative breakdown of the massive $U(1)$ to a discrete subgroup leads to a potentially interesting hierarchy of couplings in the effective action with many applications in string model building including the more recent \cite{Ibanez:2012wg,Anastasopoulos:2012zu,Honecker:2013hda,Honecker:2015ela} and references therein.

Given this rich structure, a natural question is to what extent this intriguing pattern and a possible generalisation thereof is realised also in non-perturbative string compactifications. F-theory is our key laboratory to study such effects. In this article we address this and other related questions by carefully re-examining the conditions for a coupling involving charged matter fields to appear in the F-theory effective action. Since charged matter arises from the excitations of M2-branes along fibral curves (see, e.g., \cite{Witten:1996qb,Intriligator:1997pq,LieF}), the structure of interactions is rooted in the possible joinings and splittings of such fibral curves \cite{Marsano:2011hv}.
We will argue that it is key to distinguish between a perturbative and a non-perturbative homological relation for fibral curves wrapped by M2-branes: 
A set of fibral curves satisfies a \emph{perturbative} homological relation if, loosely speaking, their sum is trivial in the fibre homology. In the \emph{non-perturbative} case, by contrast, triviality holds only in the homology ring of the full genus-one fibration. 
This implies that the states associated with M2-branes wrapping such fibral curves can only interact in conjunction with an interpolating M2-brane instanton wrapping a 3-chain $\Gamma$ bounded by the respective curves. The 3-chain $\Gamma$ in question has the important property that its volume does not vanish in the F-theory limit. Such  couplings are therefore non-perturbatively  suppressed  by a factor $e^{-2\pi {\rm vol}_3(\Gamma)}$ (in natural units). 
This allows us to make a clear distinction between the operators associated with the two types of homological relations. 
This behaviour is indeed expected from the perspective of M-theory on $G_2$-manifolds \cite{BerasaluceGonzalez:2011wy,Camara:2011jg,Berasaluce-Gonzalez:2013bba,Berasaluce-thesis}. However, since in F-theory an M2-instanton along a 3-chain on a Calabi-Yau 4-fold is non-BPS - unlike the situation for $G_2$-manifolds -  M2-instanton induced couplings can arise  at the level of D-terms in the 4-dimensional F-theory effective action, but {\em not} as F-term couplings. Consequently, the superpotential and other F-terms are in fact  {\em protected} against such non-perturbative corrections generated by M2-instantons.  

On the other hand, the instantonic M2-branes along the 3-chains $\Gamma$ can form supersymmetric bound states with M5-instantons, described by an  M5-instanton supporting a non-trivial world-volume flux.  Such fluxed M5-instanton {\em can} contribute to the F-term sector of the effective action, producing couplings exponentially suppressed by $e^{-2\pi {\rm vol}_6({\rm M5})}$. We will discuss such corrections, the associated selection rules as well as their weakly coupled description in terms of fluxed D3-brane instantons in a companion paper \cite{paper2}.

The viewpoint  developed in the present work allows us, in particular, to investigate the selection rules associated with massive $U(1)$s directly in F/M-theory.
As in perturbative compactifications, one obvious way to obtain an effectively massive $U(1)$ symmetry is by means of a flux-induced St\"uckelberg mechanism. More challenging from a conceptual point of view is the realisation of a massive $U(1)$ symmetry even in absence of gauge fluxes. The perturbative analogue of this mechanism is sometimes called `geometric St\"uckelberg mechanism' and involves axionic fields from reduction of the orientifold odd 2-forms $C_2$.
This can be mimicked in F/M-theory by means of a set of non-harmonic 2- and 3-forms in terms of which the M-theory 3-form $C_3$ is expanded \cite{Grimm:2010ez,Grimm:2011tb}.
More precisely a $k$-torsional  \cite{Camara:2011jg,BerasaluceGonzalez:2012vb} 3-form $\alpha_3$ with $k \, \alpha_3 = \d \tw_2$ gives rise to a vector field $A$ and  a St\"uckelberg axion $c$ by expanding $C_3 = A \wedge \tw_2 + c \, \alpha_3$. The relation between $\alpha_3$ and $\tw_2$ ensures that $A$ is massive by `eating' the axionic field. 
This structure has been explicitly confirmed in \cite{Mayrhofer:2014laa} for the simplest case of a $\mathbb Z_2$ symmetry in a six-dimensional F-theory compactification.\footnote{The geometry associated with a $\mathbb Z_k$ symmetry is much richer, though, and involves a set of $k-1$ additional genus-one fibrations classified by the Tate-Shafarevich group \cite{Braun:2014oya} of the original elliptic fibration, whose associated effective action becomes equivalent only in the F-theory limit \cite{Morrison:2014era,Mayrhofer:2014haa,Mayrhofer:2014laa}.} 
Another, in fact earlier, example in this spirit is the realisation of a `$\mathbb Z_1$' symmetry presented in \cite{Braun:2014nva}.
It is worth noting that in all these setups the massive $U(1)$ symmetry can be interpreted as a Kaluza-Klein mode of the 11-dimensional field $C_3$. Indeed the geometric mass scale  for the modes $A$ and $c$ is generically the compactification scale. In this sense the quantum gravity folk theorem about the inevitable gauging of a global symmetry in the ultra-violet is explicitly respected. 

In the explicit 4-dimensional examples with gauge group $G \times \mathbb Z_k$ for various non-abelian gauge groups $G$ in \cite{Klevers:2014bqa,Garcia-Etxebarria:2014qua,Mayrhofer:2014haa} all $\mathbb Z_k$ compatible couplings arise already at the perturbative level. More generally, whenever the $\mathbb Z_k$ symmetry can be unhiggsed to a massless $U(1)$ via a localised Higgs field of charge $k$ as in \cite{Morrison:2014era}, all $\mathbb Z_k$ compatible couplings are realised perturbatively provided in the unhiggsed model all Yukawa couplings involving the Higgs field allowed by gauge invariance appear as perturbative couplings.  Indeed, for generic choice of base spaces the fibrations \cite{Klevers:2014bqa,Garcia-Etxebarria:2014qua,Mayrhofer:2014haa} are of this type \cite{Morrison:2014era,Anderson:2014yva,Klevers:2014bqa,Garcia-Etxebarria:2014qua,Mayrhofer:2014haa,Mayrhofer:2014laa,Cvetic:2015moa}. 
Nonetheless for specific patterns of intersection numbers in the base, it can happen that the full set of operators allowed by the $\mathbb Z_k$ symmetry arises only once non-perturbative couplings are taken into account. This then describes a perturbative enhancement of the $\mathbb Z_k$ symmetry, possibly to a full massive $U(1)$.
Such instances are of the form familiar from 
weakly coupled Type IIB orientifolds, where massive $U(1)$ symmetries at the perturbative level are abundant and broken by a St\"uckelberg mechanism at the non-perturbative level \cite{BerasaluceGonzalez:2011wy}. In particular all Type IIB orientifolds, once uplifted to F-theory, continue to exhibit said characteristic property. These possibilities admit a natural description within our framework. We will show that whether or not the couplings allowed by a $\mathbb Z_k$ symmetry are realised at the perturbative level is directly related to the specific type of homological relations obeyed by the fibral curves.

The remainder of this article is organised as follows:
In section \ref{sec:general} we introduce the notion of perturbative versus non-perturbative homological  relations crucial for our analysis and outline the relevance of these concepts for the appearance of non-perturbatively suppressed couplings in F-theory compactifications with massive $U(1)$ symmetries. 
Two local examples in section \ref{sec:local} illustrate these ideas in a simple setting. A slightly more involved example of perturbatively versus non-perturbatively realised triple couplings is presented in the context of an $SU(5)$ F-theory model in section \ref{sec:SU(5)ex}.  This example also shows the necessary condition for the appearance of an effective perturbative selection rule in terms of the base geometry.
An application to discrete symmetries is discussed in section \ref{sec:Z2} involving the two different \cite{Morrison:2014era,Mayrhofer:2014haa,Mayrhofer:2014laa} realisations of a $\mathbb Z_2$ symmetry in F-theory: In particular we will again work out the condition for  the appearance of a perturbative massive $U(1)$  symmetry in F-theory which is broken only by said M2-instanton effects.  
A challenge for the explicit description of these M2-instantons is the global nature of the 3-chains wrapped by the instanton. In section \ref{sec:Weakcoupling} we describe the weak coupling limit of such instantons.
We begin by discussing their counterpart in IIB language given by specific ${\rm D1}$-instantons with charged fermionic zero modes at the pointlike intersection with spacetime filling 7-branes.
The requirement of saturating these zero modes translates into an obstruction of lifting the ${\rm D1}$-instantons to M/F-theory which is overcome only for a specific intersection structure with the 7-branes. This condition is indeed shown to be equivalent to the statement that the M2-instanton must wrap a 3-chain $\Gamma$ with 2 legs in the base bounded by a combination of matter fibral curves.  
We then embed these insights into the model of \cite{Braun:2014nva} of a massive $U(1)$ (`$\mathbb Z_1$ symmetry'). Analyzing the fate of the M2-instanton in the stable degeneration introduced in \cite{Clingher:2012rg} (see also \cite{Braun:2014pva,Braun:2014nva})
provides further important information on the details of the 3-chain $\Gamma$ corroborating our general picture. Along the way we use the Mayer-Vietoris exact sequence in the stable degeneration limit to obtain a better understanding of the uplift of orientifold-odd cycles to F-theory. Its dual, cohomological formulation is found to precisely encode the geometric St\"uckelberg mechanism.
Some of the technical details of our analysis are relegated to the appendices. Our conclusions are presented in section \ref{sec:Conclusions}.

\section{Resolution homology, (massive) $U(1)$s and effective field theory}
\label{sec:general}

In the following section we introduce the main concepts studied in this work. 
In section \ref{subsec_masslessU1} we assemble the required terminology and briefly review the key geometrical structures underlying 4-dimensional F-theory compactifications.
The notion of perturbative and non-perturbative homological relations is introduced in a general language in section \ref{sec:geomsr}. 
Section \ref{sec_Zksymmetry} establishes the role of these relations in the context of F-theory compactifications with discrete symmetries.

\subsection{Massless $U(1)$ gauge symmetries} \label{subsec_masslessU1}

The definition of an F-theory compactification requires the specification of a genus-one fibration over a base space $\B_n$. We can pass to the associated Jacobian fibration, which is an elliptic fibration over the same base $B_n$ with the same discriminant. The Jacobian fibration can be written as an - in general singular \cite{Braun:2014oya} - Weierstrass model, which may or may not allow for a Calabi-Yau resolution. 
With this in mind we can thus phrase our discussion in terms of an 
elliptically fibred $(n+1)$-fold 
\bea
\pi:Y_{n+1}\rightarrow B_n
\eea
with zero-section 
$\sigma_0: B_{n} \rightarrow Y_{n+1}$.\footnote{While in most of the paper we will therefore assume the existence of such a zero-section, our results generalise immediately to fibrations with multi-sections as discussed in section \ref{sec:Z2}.}
To avoid unwieldy notation we assume that the non-abelian gauge algebra ${\mathfrak g}$ of the F-theory model stems from a stack of 7-branes wrapping a single divisor $W$ in the base. For definiteness we will have the physically most relevant situation of a Calabi-Yau 4-fold fibred over a three-complex-dimensional base in mind; with this understanding we will drop the subscripts in $Y_{n+1}$ and $B_n$ in the sequel.

Let us assume for a moment that $\Y$ admits a flat Calabi-Yau resolution $\hat \Y$.
In particular resolution of the singularity in the fibre  of $\Y$ over $W$ requires the introduction of independent resolution divisors $E_i$, $i=1, \ldots, {\rm rk}({\mathfrak g})$, which are rationally fibred over $W$ within $\hat \Y$.\footnote{
One also defines a divisor $E_0$ such that $\pi^{-1} W =  E_0  + \sum_{i=1}^{{\rm rk}{\mathfrak g}} a_i \, E_i$ where $a_i$  are the Dynkin labels of ${\mathfrak g}$. The fibre of $E_0$  then constitutes the affine node of the Dynkin diagram of ${\mathfrak g}$.}
By the Shioda-Tate-Wazir theorem, the group of divisors on $H^{1,1}(\hat \Y,\mathbb R)$ is spanned by the zero-section divisor $Z$, the vertical divisors $B_\alpha = \pi^{-1}(b_\alpha)$ with $b_\alpha$ a basis of  divisors on $\B$, the divisor classes associated with the generators of the free part of the Mordell-Weil group of rational sections of $\hat \Y$
as well as the resolution divisors $E_i$.
Dimensional reduction of the M-theory 3-form $C_3$ in terms of the respective Poincar\'e dual harmonic 2-forms gives rise to massless vector fields in the M-theory effective action. 
Associated with the zero-section $Z$ is a $U(1)$ gauge field in the M-theory effective action which is to be interpreted as the Kaluza-Klein $U(1)$ in the reduction from F-theory to M-theory. The vector multiplets associated with $B_\alpha$ dualise to chiral multiplets associated with the K\"ahler moduli. More information on the 4-dimensional ${\cal N}=1$ supergravity effective action can be found in \cite{Grimm:2010ks}.

Finally, each of the
\be \label{def-r}
r=h^{1,1}(\hat \Y)-h^{1,1}(\B)-1
\ee
remaining independent divisor classes $D_A$ 
 gives rise  to a massless 1-form in the M-theory effective action that uplifts to a $U(1)$ gauge group factor in F-theory \cite{Morrison:1996na,Morrison:1996pp}. This follows by writing 
\be
C_3 = A \wedge \tw_A + \ldots
\ee
in terms of the dual 2-form $\tw_A$.
The divisors $D_A$ with this property are orthogonal to $Z$ and $B_\alpha$ with respect to the height pairing given by forming the intersection product on $\hat \Y$ and projecting to the base.
The set of $U(1)$ symmetries obtained in this way includes both the $U(1)$s in the Cartan of the non-abelian gauge algebra ${\mathfrak g}$  and extra, non-Cartan $U(1)$ gauge groups. The Cartan $U(1)$s are associated with the resolution divisors $E_i$.
Thus the massless non-Cartan $U(1)$ gauge group factors in the F-theory effective action are in one-to-one correspondence with the independent elements of the Mordell-Weil group of rational sections of $ \hat \Y$ \cite{Morrison:1996na,Morrison:1996pp}. 
Indeed, the Shioda map \cite{Shioda:1989} (see also \cite{Park:2011ji}) associates to each rational section $\sigma: \B \rightarrow \hat \Y$ a divisor $\varphi(\sigma)$ on $\hat \Y$ with vanishing height pairing with the divisors $Z$, $B_\alpha$ and $E_i$, which is the correct condition for a non-Cartan $U(1)$ to arise. The explicit realisation of $U(1)$ gauge symmetries in 4-dimensional F-theory compactifications has been investigated systematically recently \cite{Grimm:2010ez,Grimm:2011tb,Braun:2011zm,Krause:2011xj,Grimm:2011fx,Krause:2012yh,Morrison:2012ei,Cvetic:2012xn,   Mayrhofer:2012zy,Braun:2013yti,Borchmann:2013jwa,Cvetic:2013nia,Braun:2013nqa,Cvetic:2013uta,Borchmann:2013hta,Cvetic:2013jta,Cvetic:2013qsa,Mayrhofer:2013ara,Krippendorf:2014xba,Braun:2014nva,Morrison:2014era,Martini:2014iza,Bizet:2014uua,Kuntzler:2014ila,Klevers:2014bqa,Lawrie:2014uya,Lawrie:2015hia}. Early studies, mostly in six dimensions, include \cite{Klemm:1996hh,Aldazabal:1996du}.

The presence of massless Cartan and of  non-Cartan $U(1)$s affects the structure of fibral curves.
Indeed the existence of independent resolution divisors and extra rational sections implies the appearance of dual 2-cycles. Thus a massless $U(1)$ gauge group can equivalently be characterised in terms of the structure of 2-cycles.
To be more precise, recall that each resolution divisor $E_i$ is rationally fibred over the base divisor $W$ supporting the gauge algebra ${\mathfrak g}$. The rational fibres of the $E_i$ constitute independent fibral curves $\mathbb P^1_i$, $i=1, \dots {\rm rk}({\mathfrak g})$ arising in codimension-one. These are well-known to span the root lattice of the non-abelian gauge algebra ${\mathfrak g}$.\footnote{The global structure of the gauge group $G$, as opposed to the gauge algebra ${\mathfrak g}$, is determined by the torsional part of the Mordell-Weil group \cite{Aspinwall:1998xj,Mayrhofer:2014opa}.} 
This means that over the codimension-one locus $W$ in the base, the universal fibre ${\mathfrak f}$ splits as  ${\mathfrak f} \rightarrow \sum_{i=0}^{\rm rk(\mathfrak g)} a_i \, \mathbb P^1_i$ with $a_i$ the Dynkin labels of ${\mathfrak g}$.

In the fibre over certain codimension-two loci on the base, some of these fibral curves $\mathbb P^1_i$, $i=0, \ldots, {\rm rk(\mathfrak g)}$, split further.
In the presence of non-Cartan abelian gauge group factors, the fibre can split in addition over codimension-two loci  away from the divisor $W$ supporting the non-abelian part of the gauge group. The set of all codimension-two fibral curves with zero intersection with $Z$ spans the weight lattice of the full gauge group including abelian and non-abelian factors. The restriction to fibral curves with zero intersection number with $Z$ ensures that M2-branes wrapping these curves have vanishing Kaluza-Klein charge upon circle compactification to M-theory. Given an arbitrary fibral curve, this can always be achieved by adding or subtracting suitable multiples of the universal fibre class.
Over various points on the base, i.e. in codimension-three, further splittings of the fibral curves can occur. We will discuss the structure of these fibres in more detail in the next subsection.

Consider the set of all such fibral curves obtained by direct splitting of the universal fibre in codimension-one, -two and -three and let us denote by
 $C_a$, $a=1,\ldots,p$ with $p\geq r$, the subset thereof with zero intersection with $Z$ and, by construction, with  the vertical 
divisors $B_\alpha$.
Hence, although the divisors $D_A$ introduced below (\ref{def-r}) are not uniquely defined and could be modified by a linear combination of $Z$ and $B_\alpha$, their intersection numbers
\be \label{int-numbers}
\cali_{aA}\equiv C_a\cdot D_A
\ee
 with the two-cycles $C_a$ are non-ambiguous. Clearly, the $p\times r$ matrix $\cali_{aA}$ has rank $r$ and then there must be $p-r$ homological relations 
 \be \label{relations1}
 m^a  \, [C_a] =0\ (\text{mod torsion})\ \in H_{2}(\hat \Y,\mathbb{Z}).
 \ee
In particular, as noted already, the existence of non-Cartan $U(1)$s implies the presence of additional fibral curves beyond the ones descending from the rational fibres of the $E_i$. Wrapped M2-branes along these curves give rise to massless matter charged under the $U(1)$ gauge symmetry.  The charge is given precisely by the intersection numbers (\ref{int-numbers}). Conversely if $p> r$, i.e.\ if there are more fibral curves than massless $U(1)$s, then $(p-r)$ combinations of these curves are either trivial or pure torsion in $H_{2}(\hat \Y,\mathbb{Z})$.

\subsection{Geometrical selection rules and massive $U(1)$s}   
\label{sec:geomsr}  
  
Let us now analyse the homological relations (\ref{relations1}) between the fibral curves in more detail.
An important point for our discussion is that one  can distinguish between two kinds of homological relations (\ref{relations1}) which we call  {\em perturbative} and { {\em non-perturbative}, respectively, for reasons that will be presently clear.

Consider a point $p$ in the base. The fibre $\mathfrak f_p = \pi^{-1} p$ takes the form of a normal crossing variety with components
$\mathbb{P}^1_i$ with certain multiplicities $\alpha^i$.
Thus $H_2(\mathfrak f_p, \mathbb Z) = \bigoplus_i \mathbb Z^{\oplus {\alpha_p^i}}$. 
Suppose that a set of fibral  curves $C_a = \sum_i k^i_a\, \mathbb{P}^1_i$  in the fibre $\mathfrak f_p$ obey a homological relation $\sum_a n^a C_a= 0 \in H_2(\hat \Y, \mathbb Z)$. 
This implies that there exists a 3-chain $\Gamma$ in $\hat \Y$ with $\partial \Gamma =-\sum_a n^a C_a$. 
A special case is that the homological relation holds not only within $H_2(\hat \Y, \mathbb Z)$, but also as the relation
$\sum_a n^a [C_a] = 0 \in H_2({\mathfrak f}_p, \mathbb Z)$. In this case the 2-cycles are even locally homologous in the fibre ${\mathfrak f}_p$.

More generally we can make the following distinction:
Consider a set of fibral curves $C_a$ with  zero intersection with $Z$ such that
\bea \label{homrelb}
\sum_a n^a [C_a] = 0 \in H_2(\hat \Y, \mathbb Z).
\eea
If one can continuously move the $C_a$ such as to lie in the fibre over the same point $p$ while preserving holomorphicity
and 
\bea
\sum_a n^a [C_a]= 0 \in H_2(\mathfrak f_p, {\mathbb Z}),
\eea
 then we  call the homological relation (\ref{homrelb}) {\em perturbative}. Equivalently, we can say that $\sum_a n^a C_a$ is perturbatively trivial if it is the boundary of  a 3-chain that can be obtained by fibering fibral resolution curves over one-dimensional paths in the base. This 3-chain then has vanishing volume in the F-theory limit.\footnote{ Indeed in the F-theory limit the typical radius of the base and the fibre, $R_B$ and $R_{\mathfrak f}$, scale as $R_B \rightarrow x^{-1/2} R_B$,    $R_{\mathfrak f} \rightarrow x R_{\mathfrak f}$ with $x\rightarrow 0$ such that in particular vertical divisors $\pi^{-1} D$ with $D$ a base divisor have finite volume \cite{Witten:1996bn}.} Our conventions are to denote by $\Gamma_{\rm p}$ such 3-chain with 
 $\del\Gamma_{\rm p}=  - \sum_a n^a C_a$, and by $-\Gamma_{\rm p}$ the orientation reversed object. 
 
Otherwise we call the relation (\ref{homrelb})  {\em non-perturbative}. Hence, $\sum_a n^a C_a$ is non-perturbatively trivial if there exists a 3-chain $\Gamma_{\rm np}$ with $\del\Gamma_{\rm np}=-\sum_a n^a C_a$ which has non-vanishing volume in the F-theory limit, i.e.\ it has (a component with) 
`two legs' in the base.  
We will occasionally abbreviate these two types of relations as 
\be
[\sum_a n^a \, C_a]_{\rm p} = 0 \qquad {\rm or} \qquad [\sum_a n^a \, C_a]_{\rm np} = 0,
\ee
respectively.
Notice that two fibral curves which are non-perturbatively equivalent could contain some components which are actually equivalent already at the perturbative level. 
In this sense, a definition of non-perturbative homological equivalence could be introduced only mod the perturbative one.

The physical relevance of these definitions is the following:
Let us schematically denote by $\Phi_a$ any of the physical states associated with an M2-brane wrapping the fibral curve  $C_a = \sum_i k^i_a\, \mathbb{P}^1_i$.\footnote{In order for the wrapped M2-branes to describe physical particles, the curve classes $C_a$ must be effective ($k^i_a > 0$) or anti-effective  ($k^i_a < 0$) classes in the fibre. In the latter case,  the curves are wrapped by anti-M2 branes, which become BPS in the limit of vanishing fibre volume as required in F-theory.}  If we have a set of two-cycles $C_a$ obeying a {perturbative} relation $[\sum_a n^a C_a]_{\rm p} =0$ in the above sense, then different M2-brane states can interact locally by joining and merging, producing a {perturbative} interaction of the form 
\be\label{pint}
\calo_{\rm p}=A\prod_a(\Phi_a)^{n^a}
\ee
 in the effective theory, with the coefficient $A$ expected to be of order one. An example of such perturbative interactions are the holomorphic Yukawa couplings produced via splitting and joining of M2-branes in the fibre as in \cite{Marsano:2011hv}.
 
On the other hand, suppose that the two-cycles $C_a$ 
obey a {non-perturbative} relation $[\sum_a n^a C_a]_{\rm np}=0$.
Now, the states $\Phi_a$ can still interact, but not at the perturbative level. Rather, their interaction can be mediated by an instantonic M2-brane inserted at spacetime point $x_0$.
Associated with the incoming physical states are M2-branes with worldvolumes
$\gamma_{a,k}^{\rm in} \times C_a$, $k=1,\ldots,n^a$, 
where   $\gamma_{a,k}^{\rm in}$ are semi-infinite worldlines in the uncompactified spacetime dimensions ending on $x_0$, i.e.\ $\del\gamma_{a,k}^{\rm in}=x_0$,  representing the incoming effective point particles. 
These incoming M2-branes can couple to an instantonic M2-brane wrapping the 3-chain $\Gamma_{\rm np}$ at $x_0$. Indeed the boundaries of $\sum_{a,k} \gamma^{\rm in}_{a,k} \times C^a$ and $\Gamma_{\rm np}$ at $x_0$ just cancel because
$\partial ( \sum_{a,k} \gamma^{\rm in}_{a,k} \times C_a )= \sum_an^a C_a = - \partial \Gamma_{\rm np}$. 
The resulting term in the effective theory will have the schematic form\footnote{See the last paragraph of this section.}
\be\label{npint}
\calo_{\rm np}=A\prod_a(\Phi_a)^{n^a}e^{-2\pi{\rm vol}(\Gamma_{\rm np})+2\pi\ii\int_{\Gamma_{\rm np}}C_3}
\ee
in units where $\ell_{\rm P}=1$, with $A$ of order one. 
Likewise the conjugate operator is induced by an M2-instanton on $-\Gamma_{\rm np}$ (i.e. an anti-instanton on $\Gamma_{\rm np}$) coupling to a set of outgoing M2-branes with worldvolumes $\gamma_{a,k}^{\rm out} \times C_a$, $k=1,\ldots,n^a$, for $\del\gamma_{a,k}^{\rm out} = -x_0$.

By going to the F-theory limit, 
the M2-brane wrapping $\Gamma_{\rm np}$ dualises  to a tensionful  bound-state of D1-F1 instantonic strings  on the base $\B$.
This will be discussed in more detail in the following sections. Hence, the exponential prefactor $e^{-2\pi{\rm vol}(\Gamma_{\rm np})}$ in (\ref{npint})  provides an exponential suppression of the coupling  
in the supergravity regime, i.e. small-$\alpha'$ limit. On the other hand, 
in the case of a perturbative relation $[\sum_a n^a C_a]_{\rm p} =0$, the corresponding   chain $\Gamma_{\rm p}$ 
has vanishing volume  in the F-theory limit, and sometimes, in fact, already in M-theory. At any rate no exponential suppression follows in the 4-dimensional effective action.

We stress that the term \emph{non-perturbative} (versus \emph{perturbative}) refers to the exponential volume suppression in (\ref{npint}). Note that in this sense, a `perturbative' coupling involving multi-pronged strings may not be present at the perturbative level in Type IIB orientifolds even though  in F-theory it is not exponentially volume suppressed. An example of such a coupling  is the famous triple Yukawa coupling ${\bf 10 \, 10 \, 5\, }$ in F-theory GUT models \cite{Donagi:2008ca,Beasley:2008dc,Beasley:2008kw,Donagi:2008kj} (see also the reviews \cite{Heckman:2010bq,Weigand:2010wm,Maharana:2012tu}), which can be localised at points of $E_6$ singularity enhancement and is not exponentially volume suppressed. In our terminology such a coupling is called `perturbative' even though the coupling cannot be realised in Type IIB orientifolds.

We then arrive at the conclusion that, for a non-perturbative relation $[\sum_a n^a C_a]_{\rm np}=0$, 
even though there is no massless $U(1)$ prohibiting an interaction containing $\prod_a(\Phi_a)^{n^a}$ a perturbative interaction of the form (\ref{pint}) would be forbidden or, more precisely, it should be dressed by an exponentially suppressed factor as in  (\ref{npint}).

From the effective low-energy viewpoint, such a selection rule is typically associated with a massive $U(1)$. This interpretation is possible in the F-theory context as well \cite{Grimm:2010ez,Grimm:2011tb}.
Indeed, one could in principle KK-reduce the three-form potential $C_3$ by keeping not only the $r$ massless $U(1)$s and the $b^3(\hat \Y)-b^3(\B)$ massless axions,
but also the full tower of massive KK-excitations of $C_3$. The latter can be regarded as an infinite number of massive KK $U(1)$s which get their mass by eating KK axions.  
The full tower of KK $U(1)$ gauge transformations is then encoded in the gauge transformation $C_3\rightarrow C_3+\d\Lambda_2$. The exponential factor in (\ref{npint})
is charged under these gauge transformations  in the sense that
\be
e^{2\pi\ii\int_{\Gamma_{\rm np}}C_3}\rightarrow e^{ - 2\pi\ii  \sum_a n^a \int_{C_a}\Lambda_2}\,e^{2\pi\ii\int_{\Gamma_{\rm np}}C_3},
\ee
where we have used that $\partial \Gamma_{\rm np} = - \sum_an^a C_a$.
Analogously, the combination $\prod_a(\Phi^a)^{n_a}$ will be charged as follows
\be
\prod_a(\Phi_a)^{n^a}\rightarrow e^{  2\pi\ii  \sum_a n^a\int_{ C_a}\Lambda_2}\prod_a(\Phi_a)^{n^a} .
\ee
Only the combination (\ref{npint}) is gauge invariant and then the above selection rule which forbids the `naked' perturbative interaction (\ref{pint}) 
can be seen as the result of some massive $U(1)$ gauge symmetry.  

Our framework in fact gives a convenient way to determine the massive $U(1)$ symmetries:
The relevant massive $U(1)$s  are in one-to-one correspondence with the independent non-perturbative homological relations  $[ \sum_a n^a C_a]_{\rm np}=0$. In particular, the associated $U(1)$ symmetry phase $e^{\ii\lambda}$ is identified by
\be
\lambda=\sum_a n^a \int_{C_a}\Lambda_2 .
\ee 
We will test this proposal by explicitly exhibiting this correspondence in the weak coupling limit.  
In fact, the proposal is an extension of the observation elaborated on in \cite{Braun:2014oya} that F-theory models with massive $U(1)$ gauge groups do not possess a Calabi-Yau resolution - provided the model contains localised matter charged only under massive $U(1)$s and no massless gauge group. The underlying reason is that in such a case no supersymmetric flat direction in the Coulomb branch of M-theory is available to make the massless matter massive. Such a direction would correspond to a Calabi-Yau resolution. Conversely, in the alternative non-K\"ahler resolution available in the massive $U(1)$ model of \cite{Braun:2014oya} the fibral curve associated with the matter in question is homologically trivial in the full compactification space, though non-trivial in the fibre. 
This is then the purest realisation of a non-perturbative homological relation, and indeed associated to a massive $U(1)$ symmetry.

Finally, let us  incorporate  the so far ignored supersymmetry in the discussion.
The interaction  terms $\calo_{\rm p}$ and $\calo_{\rm np}$ discussed above should be organised into D-terms or F-terms of the 4-dimensional effective action.
It is known that the perturbative terms   $\calo_{\rm p}$ can contribute to  F-terms as $\int\d^2\theta \, W$, see for instance \cite{Donagi:2008ca,Beasley:2008dc,Beasley:2008kw,Donagi:2008kj,Hayashi:2009ge,Hayashi:2010zp,Marsano:2011hv}.
On the other hand, the instantonic M2-branes generating $\calo_{\rm np}$  always break the background supersymmetry. 
This implies that terms of the form (\ref{npint})  are generically part of the D-terms in the effective theory, while they can never  be associated with F-terms.  
Then, if  $[n^a C_a]_{\rm np}=0$, the associated perturbative selection rule for F-terms {\em cannot} 
be violated by intantonic M2-branes alone. On the other hand, non-trivial non-perturbative F-terms that violate the perturbative selection rules {\em can} be generated by BPS bound states of instantonic M2 and M5-branes, or equivalently by M5-instantons with certain 3-form field strength.
We will discuss such fluxed M5-brane instantons in detail in the companion paper \cite{paper2}.

\subsection{$\mathbb Z_k$ gauge symmetries and torsion in (non-)perturbative homology}
\label{sec_Zksymmetry}

We now consider in more detail the implications of a non-vanishing torsional subgroup  
\be\label{tordec}
{\rm Tor} H_{2}(\hat\Y,\mathbb Z)\simeq \mathbb{Z}_{p_1}\times\ldots\times\mathbb{Z}_{p_n}\neq 0\,.
\ee 
The relation between torsional (co)homology and discrete gauge symmetries has first been discussed in \cite{Camara:2011jg,BerasaluceGonzalez:2012vb,BerasaluceGonzalez:2012zn,Berasaluce-Gonzalez:2013bba,Berasaluce-Gonzalez:2013sna}, originally in the context of discrete symmetries in the Ramond-Ramond sector of Type IIB compactifications and then generalizing to many dual settings including M-theory on $G_2$-manifolds.
The appearance of torsion in F-theory models with discrete symmetry has been confirmed in \cite{Mayrhofer:2014laa} for a six-dimensional compactification with $\mathbb Z_2$ symmetry.\footnote{Note that if the Weierstrass model has $\mathbb Z_p$ torsion cohomology of this type, there exist another $p-1$ different genus-one fibrations with the same discriminant \cite{Braun:2014oya}. Together these form $p$ elements of the Tate-Shafarevich group, which is $\mathbb Z_p$. These genus-one fibrations yield \emph{different} compactifications of M-theory, which become equivalent only in the F-theory limit \cite{Morrison:2014era,Mayrhofer:2014laa}. In this sense $\mathbb Z_p$ discrete symmetry in F-theory can also be described in terms of these genus-one fibrations with multi-sections \cite{Morrison:2014era,Anderson:2014yva,Klevers:2014bqa,Garcia-Etxebarria:2014qua,Mayrhofer:2014haa,Mayrhofer:2014laa,Cvetic:2015moa} even though the associated M-theory compactifications do not possess any $\mathbb Z_p$ symmetry. The precise relation between these descriptions has been given in \cite{Mayrhofer:2014laa}. In particular, in these genus-one fibrations torsion cohomology arises in a more subtle way modulo the universal fibre class.} In this work also M2/M5-branes wrapping torsional cycles and their bounding chains in F/M-theory have been discussed. 
In \cite{Grimm:2015ona} the role of torsion, including that of the base (co)homology, has been investigated in the context of non-abelian discrete symmetries in F-theory.

Here we propose a refined notion of torsion within the formalism introduced in the previous section and describe its imprint on the couplings in 4-dimensional F-theory compactifications.
If we focus on torsion which is not simply induced by torsion in the base $B$,
then there may be a combination of fibral curves $\sum_an^a C_a $ which is purely torsional so that $[\sum_an^a C_a]\neq 0 $ but $[k\sum_an^a C_a]= 0$ in $H_{2}(\hat\Y,\mathbb Z)$  for some (minimal) integer  $k>0$.  We can again distinguish between perturbative and non-perturbative homology. Namely we may have $[k\sum_an^a C_a]_{\rm p}= 0$ or just $[k\sum_an^a C_a]_{\rm np}= 0$. 

 In the first case, $[k\sum_in_i\Sigma^i]_{\rm p}= 0$,  from the above general discussion we can have a (possibly F-term) {\em perturbative} coupling of the form
\be\label{pintk}
\calo^k\quad~~~~\text{with}\quad \calo=A\prod_a(\Phi_a)^{n^a}.
\ee
By contrast, terms of the form $\calo^m$, with powers  $m=1,\ldots, k-1$, cannot be generated. 
Clearly such selection rule can be interpreted as the result of an effective discrete symmetry under which the operator $\calo$ is charged.  

In the second case, $[k\sum_an^a C_a]_{\rm np}= 0$, we have $k\sum_an^a C_a=-\del\Gamma_{\rm np}$ and an instantonic M2-brane wrapping $\Gamma_{\rm np}$ can generate a non-perturbatively suppressed (D-term) coupling of the form
\be\label{npintk}
\calo^k \,e^{-2\pi{\rm vol}(\Gamma_{\rm np})+2\pi\ii\int_{\Gamma_{\rm np}}C_3}.
\ee
In such a case the discrete symmetry manifests itself only through non-pertubatively generated couplings and is in fact enhanced to a full massive $U(1)$ symmetry at the perturbative level.

The link between such effective discrete symmetries  and some underlying massive $U(1)$ gauge symmetry is provided by
the universal coefficient theorem, which states that ${\rm Tor} H^{3}(\hat\Y,\mathbb Z)\simeq {\rm Tor} H_{2}(\hat\Y,\mathbb Z)$. 
For simplicity let us consider only a single $\mathbb Z_p$ factor in (\ref{tordec}) and take a torsional 3-form $\alpha_3$  generating  $\mathbb{Z}_p$ so that
\be \label{k-torsion1}
\d \tw_2 = p \, \alpha_3
\ee
for some non-closed 2-form $\tw_2$.  For later purposes note  that the integer $k$ introduced above must satisfy $\frac{p}{k} \in \mathbb Z$.
A naive dimensional reduction of the kinetic term for the M-theory 3-form by expanding
\bea \label{k-torsion2}
C_3 = A \wedge \tw_2 + c \, \alpha_3 + \ldots
\eea
then reproduces a  kinetic term of the schematic form $\int (\d c - p A)^2$ in the F-theory effective action \cite{Grimm:2010ez,Camara:2011jg,Grimm:2011tb}. The pair $(A, c)$ can thus be identified with a massive $U(1)$ gauge field $A$ that eats, via a St\"uckelberg mechanism, an axionic field $c$ of charge $p$. In particular, the associated gauge transformation is provided by $C_3\rightarrow C_3+\d\Lambda_2$, with $\Lambda_2=\lambda\tw_2$, which indeed corresponds to $A\rightarrow A+\d\lambda $ and $c\rightarrow c+p\lambda$. This is precisely the field theoretic hallmark of a $\mathbb Z_p$ gauge symmetry \cite{Banks:2010zn}. 

One can make the link between discrete selection rules and St\"uckelberg massive gauge symmetries more concrete by adapting  the approach of \cite{Camara:2011jg} to the present context.
See the analysis \cite{BerasaluceGonzalez:2011wy,BerasaluceGonzalez:2012vb,BerasaluceGonzalez:2012zn,Berasaluce-Gonzalez:2013bba,Berasaluce-Gonzalez:2013sna} for related considerations in other types of string compactifications.
 The universal coefficient theorem states that an element $\alpha_3\in {\rm Tor} H^{3}(\hat\Y,\mathbb Z)$ is unambiguously identified by a map that associates to each $C\in {\rm Tor} H_{2}(\hat\Y,\mathbb Z)$ a phase $e^{2\pi\ii\langle\alpha_3,C\rangle}$. More precisely, one can choose a representative of $\alpha_3$ of the form   $\alpha_3=\delta_3(R)$, where $R\in {\rm Tor} H_{5}(\hat\Y,\mathbb Z)$ is a torsional 5-cycle Poincar\'e dual to $\alpha_3$ and $\delta_3(R)$ is a delta-like 3-form localised on it, and write 
\be\label{deltaR}
p\,\delta_3(R)=\d \tw_2
\ee
 for some 2-form $\tw_2$. Hence, the phase of the universal coefficient theorem is provided by
\be \label{deftorsionlink}
e^{2\pi\ii\langle\alpha_3,C\rangle}=e^{\frac{2\pi\ii}{p}\int_C\tw_2}=e^{-\frac{2\pi\ii}{k}\int_\Gamma\alpha_3}
\ee
where we have used $k C=-\del\Gamma$.
Clearly, $\langle\alpha_3,C\rangle$ is defined up to an integer.
Notice that $\tw_2$ is defined up to a closed 2-form, which does not affect such definition. Furthermore, by (\ref{deltaR}) we see that such definition does not depend on the choice of homology representatives $C$ and $R$ within their corresponding homology classes.  

By using such $\tw_2$ and $\alpha_3$ in the decomposition (\ref{k-torsion2}), the corresponding gauge transformation uplifts to $C_3\rightarrow C_3+\d\Lambda_2$, with $\Lambda_2=\lambda \tw_2$. According to the general discussion of section \ref{sec:geomsr}, an operator $\calo$ associated with an M2 wrapping  $C$ transforms by a phase
\be
e^{2\pi\ii\int_C\Lambda_2}=e^{2\pi\ii\lambda\int_C\tw_2}=e^{2\pi\ii p\, \langle\alpha_3,C\rangle\lambda}
\ee
under such gauge transformation. The pairing $\langle\alpha_3,C\rangle$ is only defined modulo $\mathbb{Z}$ and (\ref{deltaR}) and (\ref{deftorsionlink}) imply $p\,\langle\alpha_3,C\rangle\in\mathbb{Z}$ (because $\frac{p}{k} \in \mathbb Z$).  We then arrive at the conclusion that $\calo$ carries a charge 
\be\label{torcharge}
q=p\,\langle\alpha_3,C\rangle\in \mathbb{Z} \quad(\text{mod} \, \, p)
\ee
under such $U(1)$ gauge transformation. This gauge transformation is spontaneously broken to a $\mathbb{Z}_p$ subgroup generated by $e^{\frac{2\pi\ii}{p}}$, under which $\calo$ transforms by a phase $e^{\frac{2\pi\ii q}{p}}$.  Notice that (\ref{deltaR}) implies that $\frac{k q}{p}=k\langle\alpha_3,C\rangle$ is integer and then  the operator $\calo^k$  is invariant under  the surviving $\mathbb{Z}_p$ discrete gauge symmetry, consistently with  the $\mathbb{Z}_k$ selection rules discussed above for the operators (\ref{pintk}) and  (\ref{npintk}). Hence the universal coefficient theorem provides an interpretation of such selection rules  in terms of a distinguished finite set of massive $U(1)$s, associated with each element $\alpha_3\in{\rm Tor} H^{3}(\hat\Y,\mathbb Z)$, under which an M2-brane wrapping a torsional 2-cycles $C$ have charge (\ref{torcharge}).\footnote{Notice that, even though in our discussion we have started from delta-like representatives of  $\alpha_3\in {\rm Tor} H^{3}(\hat\Y,\mathbb Z)$, $\langle\alpha_3,C\rangle$ can be interpreted as the torsional linking number of $C$ and a torsional 5-cycle $R$ that is Poincar\'e dual to $\alpha_3$, which is a well defined topological number. Hence the charge (\ref{torcharge}) is a genuinely topological quantity and  is independent of the spacific way the massive $U(1)$ gauge symmetry has been identified.}

We would like to stress that the above identification of the massive $U(1)$s responsible for the perturbative and non-perturbative selection rules is only schematic and would require a consistent KK reduction including non-harmonic forms, see for instance   \cite{Camara:2011jg} for a discussion. Furthermore,  the interpretation of the massive gauge fields and axions from a IIB perspective is not always obvious (although we will have more to say about it in the following). However, our formalism allows us to completely characterise the discrete selection rules, relying directly on their higher-dimensional origin and without any need of interpreting them as the result of 4-dimensional St\"uckelberg massive gauge fields. In fact,  generically, such interpretation appears in principle problematic within an effective 4-dimensional approach, since the masses of the massive states are generically expected to be around the KK scale.

We therefore arrive at the following intriguing pattern concerning the structure of couplings with respect to discrete gauge symmetries in F-theory:
If torsion in $H_2(\hat Y_4, \mathbb Z)$ is fully generated by relations in perturbative homology, then the associated operators allowed by the $\mathbb Z_p$ symmetry arise already at the perturbative level, i.e.\ without exponential volume suppression. 
This is indeed the case in the F-theory models with discrete symmetry analysed in \cite{Morrison:2014era,Anderson:2014yva,Klevers:2014bqa,Garcia-Etxebarria:2014qua,Mayrhofer:2014haa,Mayrhofer:2014laa} for a \emph{generic} choice of base space $B$. These models have the property that the $\mathbb Z_p$ symmetry can be enhanced to a massless $U(1)$ symmetry by unhiggsing  a localised Higgs field of charge $- p$, so that 
the perturbative term  ${\cal O}$  of charge $p$ in the $\mathbb{Z}_p$ model is  inherited from a coupling of the form $\varphi \, {\cal O}$, where $\varphi$ is a $-p$-charged Higgs field. 

However, it can happen that  the perturbative interactions per se do not exhaust the full set of terms allowed by all massless gauge symmetries and the additional discrete symmetry; rather the full $\mathbb Z_k$ symmetry may be manifest in the couplings sector only once all types of interactions are taken into account. In this case the symmetry group enhances at the perturbative level, for example to a full massive $U(1)$ symmetry.
We will exemplify this in more detail in section \ref{sec:Z2}.
This phenomenon is familiar from experience with weakly coupled Type IIB compactifications, which form a proper subset of the landscape of F-theory vacua; however, it need not be restricted to F-theory compactifications with a well-defined Type IIB limit. As will be reviewed in section \ref{axgauging}, in Type IIB orientifolds
 a massive $U(1)$ gauge group can arise via the geometric St\"uckelberg mechanism, with the axion $c$ appearing in (\ref{k-torsion2}) being associated with the R-R 2-form $C_2$. The term massive $U(1)$ is justified to the extent that the perturbative couplings respect a full global $U(1)$ symmetry which is broken to a discrete $\mathbb Z_k$-subgroup only in the non-perturbative sector.  
 Therefore, 
 the integral $\int_{\Gamma_{\rm np}} C_3$ is to be identified with the axion arising from the F-theory analogue of the R-R $C_2$ in weakly-coupled IIB.

\section{Simple local examples}
\label{sec:local}

To exemplify the previous general discussion, we start by presenting the probably simplest realisations of perturbative and non-perturbative homological relations. These examples involve just intersecting (mutually local) D7-branes and then admit a simple weakly coupled IIB interpretation.

\subsection{Local examples of perturbative relations}

As already mentioned, in this paper we focus on F-theory models in four dimensions, corresponding to an elliptic fibration over a 3-dimensional base $\B$. A trivial realisation of a perturbative homological relation is obtained by considering two single D7-branes intersecting on a base curve  $\calc$. This supports a family of conifold singularities, which can be locally resolved by a small resolution introducing a fibral $\mathbb{P}^1$. Then we obviously have the tautological identity  $[\mathbb{P}^1]-[\mathbb{P}^1]=0 \in H_2({\mathfrak f}_p,\mathbb Z)$ at any point $p\in C$ which, in our language, is a  {\em perturbative} relation. The corresponding operator, which can be generated with a coefficient of order 1, is just a quadratic term $\sim \Phi \, \tilde\Phi$, where $\Phi$ and $\tilde\Phi$ are associated with an M2 and an $\overline{\rm M2}$, respectively, wrapping $\mathbb{P}^1$. The presence of such a mass term can be read off by computing the spectrum of massless states $\Phi$ and $\tilde\Phi$. Methods to compute in particular the vectorlike spectrum in F-theory have been proposed in \cite{Bies:2014sra,Collinucci:2014taa}.

Then the simplest realisation of a non-trivial perturbative homological relation can be obtained by considering three D7-branes intersecting at a point $p\in \B$. This will lead to a prototypical example of a standard Yukawa coupling in codimension-three, which has been studied in detail in the F-theory literature.
An F-theory local description of this configuration starts from an elliptic fibration in Weierstrass form
\bea \label{Weier1}
y^2 = x^3 + f \, x \, z^4 + g \, z^6
\eea
parametrised, as usual, by
\bea  \label{Weier2}
f =   - \frac{1}{3} b_2^2 + 2 b_4, \qquad  g = \frac{2}{27} b_2^3 - \frac{2}{3} b_2 b_4 + b_6.
\eea
Here $b_i$ are sections of $\bar { K}^i$ with ${ K}$ the canonical bundle of the base. Setting $z=1$, we can bring this into the form
\be\label{ys0}
y^2=s^3+b_2 \, s^2+2 \, b_4 \, s+b_6
\ee
by defining $s=x-\frac13 b_2$  (see for instance \cite{Braun:2014nva}).
We can just focus on a local coordinate system $(u_1,u_2,u_3)$ of $\B$ such that the point $p$ corresponds to $u_1=u_2=u_3=0$, while the three D7-branes wrap the there divisors $D_i=\{u_i=0\}$. 
A local F-theory realisation of such configuration is obtained by choosing
\be
b_2=1+u_1 \, u_2 \, u_3\,,\quad b_4=u_1\, u_2 \, u_3\,,\quad b_6=u_1 \, u_2 \, u_3
\ee
so that locally
\be\label{ys}
y^2=s^3+s^2+u_1\, u_2 \, u_3 \, (1+s)^2.
\ee
Similar local models have been discussed in \cite{Braun:2014nva}. There are three curves of conifold singularities $\calc_{A}=\{u_1=u_2=0\}$, $\calc_{B}=\{u_2=u_3=0\}$ and $\calc_{C}=\{u_1=u_3=0\}$ intersecting at an
 $I_3$  singularity at $u_1=u_2=u_3=y=s=0$. As is well-known, this singular space $\Y$ can be resolved, for instance  by performing a double small resolution. The $I_2$-fibres of the resulting smooth model $\hat \Y$  contain the rational curves $\mathbb{P}^1_A$,  $\mathbb{P}^1_B$ and $\mathbb{P}^1_C$ fibred over the base curves 
 $\calc_A$, $\calc_B$ and $\calc_C$ respectively. However, $\mathbb{P}^1_A$,  $\mathbb{P}^1_B$ and $\mathbb{P}^1_C$ are not homologically independent. Indeed they can be localised at the common base point $p=\{u_1=u_2=u_3=0\}$ where they satisfy a homological relation. The specific form of this relation depends on which of the three possible double small resolutions we are considering.
 For instance, in one such resolution $\mathbb P^1_C$ splits into $\mathbb P^1_A$ and $\mathbb P^1_B$ over  $p$, consistent with the {\em perturbative} relation
  \be \label{homrelex1}
 [\mathbb{P}^1_A]+[\mathbb{P}^1_B] - [\mathbb{P}^1_C]=0 \in H_2(\mathfrak f_p,\mathbb Z).
 \ee
 If we denote by $\Phi_A$, $\Phi_B$ and $\Phi_C$ the states associated with M2-branes along $\mathbb P^1_A$ and $\mathbb P^1_B$ and an $\overline{\rm M2}$ on $\mathbb P^1_C$, this relation is associated with a possible low-energy interaction term of the form $\sim\Phi_{A}\Phi_{B}\Phi_{C} + c.c. $, as for instance a Yukawa term. One can think of this term as being induced by an interpolating M2-brane along a perturbative 3-chain\footnote{Here and in the following, for notational simplicity,  we often avoid to explicitly write the subscripts ${}_{\rm p}$ and ${}_{\rm np}$ introduced in the previous section to distinguish perturbative and non-perturbative chains. In such cases the distinction should be clear from the context.} $\Gamma$ with $\partial \Gamma = -  (\mathbb{P}^1_A + \mathbb{P}^1_B - \mathbb{P}^1_C)$ which is collapsed at the triple intersection point $p$; since this collapsed chain is of zero volume, the M2-brane can be BPS and therefore induce an F-term. 
The other two resolutions related via a flop transition change the position of the minus sign in (\ref{homrelex1}). Clearly this leads to the same interaction in F-theory after a  field redefinition.

\subsection{Local examples of non-perturbative relations}
\label{sec:localnp}

We now describe simple non-perturbative homological relations which can be regarded as the non-perturbative analogue of the two types of perturbative relations of the previous subsection.

We first consider an example with two D7-branes. Suppose that they intersect at two disjoint curves, $\calc_A$ and $\calc_B$, which  do not intersect other 7-branes. Then  $\calc_A$ and $\calc_B$ support two fibral curves, $\mathbb{P}^1_A$ and $\mathbb{P}^1_B$.
Our general arguments imply that there is at least a homological relation between them.   Indeed, all matter sitting at these curves is charged at most under one single $U(1)$, the relative one between the two D7-branes.  On the other hand, since $\calc_A$ and $\calc_B$ are disconnected, $\mathbb{P}^1_A$ and $\mathbb{P}^1_B$ cannot be continuously moved onto the same base point.  Then we expect a {\em non-perturbative} relation between $\mathbb{P}^1_A$ and $\mathbb{P}^1_B$. Notice that this conclusion  should not depend on the global structure of $\hat \Y$. These arguments, as well as the following discussion,  can be generalised in an obvious manner to the case of two D7-branes intersection at $N>2$ mutually disconnected curves, in which case one expects $N-1$ non-perturbative relations.

One can construct an explicit local model that realises this idea as follows. We start again with the elliptic fibration in the form (\ref{ys0}) and consider a local coordinate patch $(w,v,u)$. We then choose 
\be
b_2=1+\Delta_{\rm D7},\,\quad b_4=\Delta_{\rm D7},\,\quad b_6=\Delta_{\rm D7}
\ee
with 
\be
\Delta_{\rm D7}=w^2 - (v-a_1)^2 (v-a_2)^2.
\ee
By computing the discriminant of the elliptic fibration one obtains $\Delta\propto \Delta_{\rm D7}$, with $ \Delta_{\rm D7}=0$ representing the location of two D7-branes
\be
\begin{aligned}
{\rm D7}_1&=\{w=(v-a_1)(v-a_2)\},\\
{\rm D7}_2&=\{w=-(v-a_1)(v-a_2)\}.
\end{aligned}
\ee
These two D7-branes intersect at the two curves $\calc_A=\{w=0\} \cap \{v=a_1\}$ and $\calc_B=\{w=0\} \cap \{v=a_2\}$, see fig.~\ref{fig:np1}.
 \begin{figure}[t!]
  \centering
    \includegraphics[width=0.55\textwidth]{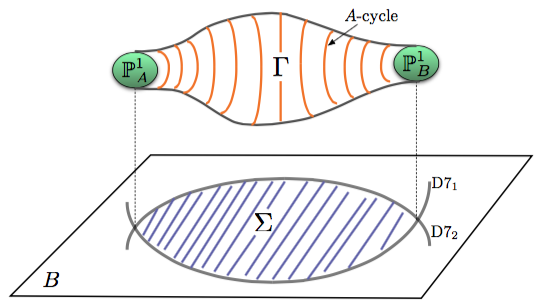}
      \caption{\footnotesize 3-chain $\Gamma$ connecting $\mathbb P^1_A$ and $\mathbb P^1_B$.}\label{fig:np1}
\end{figure}
The elliptic fibration becomes
\be
y^2=s^2+\Delta_{\rm D7}(1+s)^2+s^3,
\ee
which clearly exhibits the presence of two families of conifold singularties fibred over $\calc_A$ and $\calc_B$. 
These can be locally resolved by a single small resolution, which pops-up two fibral curves $\mathbb{P}^1_A$ and $\mathbb{P}^1_B$ over 
$\calc_A$ and $\calc_B$ respectively. 

Now, the above argument implies that $\mathbb{P}^1_A$ and $\mathbb{P}^1_B$ are homologous. Indeed, one can explicitly construct a 3-chain $\Gamma\subset \hat \Y$ interpolating between them by first choosing a 2-chain $\Sigma$ in the base, stretching it between the two D7-branes as in fig.~\ref{fig:np1}, and then fibering over it the $A$-type 1-cycle of the elliptic fibre. (For instance, if $a_1$ and $a_2$ are real, $\Sigma$ can be identified by restricting to the real slice $\Im w=\Im v=0$.) 
 Since the $A$-cycle of the elliptic fibre shrinks to zero at the $I_1$ singularities, $\Gamma$ is perfectly smooth over the D7-branes, excluding the intersection curves  ${\cal C}_A$ and ${\cal C}_B$, over which it acquires  $\mathbb{P}^1_A$ and $\mathbb{P}^1_B$ as boundaries, see fig.~\ref{fig:np1}. Hence $\del\Gamma= - (\mathbb{P}^1_A-\mathbb{P}^1_B)$ and there exists the {\em non-perturbative} relation
$$
[\mathbb{P}^1_A-\mathbb{P}^1_B]_{\rm np}=0.
$$
The instantonic  M2-branes wrapping the chain $\Gamma$ can generate an exponentially suppressed quadratic term $\sim  \Phi_A \, \tilde\Phi_B e^{-2\pi{\rm vol}(\Gamma)}$. 
Being non-BPS, the M2-branes cannot contribute to the F-terms, but merely to the D-terms of the 4-dimensional effective action. Therefore they do not generate a mass term for the component fields inside the chiral superfields $\Phi_A$ and $ \tilde\Phi_B$. Such M-theory instantons  correspond, in Type IIB, to worldsheet instantons  given by open Euclidean  F1-strings wrapping the 2-chain $\Sigma$. Such IIB interpretation of these instantons, as opposed to D1-instantons,  is due to the fact that the fibred 1-cycle is the $A$-cycle collapsing at the singularity. In later sections we will instead focus on M2-instantons that will be interpreted as particular D1-instantons  from the IIB perspective. The nature of the 3-chain will be different in these cases. 

Analogously, one can construct a local model corresponding to three D7-branes intersecting at three disjoint curves, by choosing for instance
\be
\Delta_{\rm D7}=w \, v \, (w+v-a),
\ee
The three D7-branes are located at the divisors $\{w=0\}$, $\{v=0\}$ and $\{w+v=a\}$. They intersect at the disjoint curves $\calc_A=\{u=v=0\}$, $\calc_B=\{v=w-a=0\}$ and $\calc_C=\{u=v-a=0\}$, over which $\Y$ has conifold singularities.  By resolving $Y$ into $\hat \Y$,  one finds three fibral curves $\mathbb{P}^1_A,\mathbb{P}^1_B,\mathbb{P}^1_C$  over the corresponding base curves. The corresponding matter states $\Phi_A,\Phi_B,\Phi_C$ have equal charges
under the three relatives $U(1)$s supported by the D7-branes, only two of which are independent. Correspondingly, only two of these fibral curves can be independent. This means that, the curves $\calc_A,\calc_B,\calc_C$ being disconnected, there must exists a {\em non-perturbative} homological relation
\be
[\mathbb{P}^1_A+\mathbb{P}^1_B-\mathbb{P}^1_C]_{\rm np}=0,
\ee
where again the position of the minus sign depends on the choice of resolution as in the perturbative example. 
Indeed, one can construct a non-perturbative 2-chain $\Gamma$ such that $\del\Gamma=-(\mathbb{P}^1_A+\mathbb{P}^1_B -\mathbb{P}^1_C)$ by fibering the $A$-cycle of the elliptic fibre over the  2-chain $\Sigma$ in the base that stretches between the three D7-branes. As above, an open M2-brane instanton wrapping $\Gamma$ can generate a term $\sim \Phi_A\Phi_B\Phi_C \, e^{-2\pi{\rm vol}(\Gamma)}$ (where $\Phi_A$ and  $\Phi_B$ correspond to M2-branes and $\Phi_C$ to an $\overline{\rm M2}$ brane on the respective fibral curves)
in the effective theory, which corresponds to an open world-sheet instanton wrapping $\Sigma$. Such open world-sheet instantons are analogous to those considered for instance in \cite{Cremades:2003qj,Cvetic:2003ch} in the context of intersecting D6-branes. However, differently from that case, our world-sheet instantons cannot be supersymmetric and then can generate only D-terms.

\section{Application to a global $SU(5)$ example} \label{sec:SU(5)ex}

Let us now study perturbative and non-perturbative relations in a phenomenologically flavored example. This setup illustrates the relevance of the global structure in the determination of the non-perturbative relations. Furthermore, differently from the examples of the previous section, the non-perturbative relations have a less obvious Type IIB interpretation. We will come back to this point in section \ref{sec:Weakcoupling}. 

We start from a Weierstrass model  in Tate form describing an F-theory compactification with gauge group $SU(5)$. Such models have been analysed in detail in the context of F-theory GUT model building \cite{Andreas:2009uf,Donagi:2009ra,Blumenhagen:2009yv,Grimm:2009yu,Esole:2011sm,Marsano:2011hv,Krause:2011xj,Grimm:2011fx,Hayashi:2013lra}.
The fibre of the Tate model is modeled as the hypersurface given by the vanishing locus of
\bea \label{PTdef}
P_T: = x^3 - y^2 - a_1 \,  x \, y \, z + a_2 \, x^2 \, z^2 - a_3 \, y \, z^3  + a_4 \, x \, z^4 + a_6 \, z^6,
\eea
where $[x : y : z]$ are projective coordinates in  the fibre ambient space $ \mathbb P_{2,3,1}$ and $a_i$ are sections of $\bar {K}^i$.
As in the previous section, we focus on F-theory compactification to four dimensions. 
The specification
\be
a_2 = a_{21} \, w, \qquad a_3 = a_{32} \, w^2, \qquad a_4 = a_{43} \, w^3, \qquad a_6 = a_{65} \, w^5,
\ee
where $w$ is the section of a certain line bundle over the base $\B$, 
induces an $A_4$ singularity in the fibre over the divisor $W: \{w=0\}$ on $\B$ \cite{Bershadsky:1996nh}. Over the curves
\be\label{C10C5}
\begin{aligned}
\calc_{\bf 10}&: \{w = 0\} \cap \{a_1=0\}\,,\\
 \calc_{\bf 5}&: \{w = 0\} \cap \{\cald=0\} \,,  \quad \text{with}\quad \cald = a_{32} (a_{21} a_{32} - a_1 a_{43}    ) + a_1^2 a_{65} \,,
\end{aligned}
\ee
the singularity type enhances to $D_5$ and $A_5$, respectively. This indicates the presence of massless matter in representations ${\bf 10}$ and ${\bf 5}$.
These curves intersect at two types of points on $W$ given by
\be\label{p1p2}
 p_1: \{w=0\} \cap \{a_1 = 0\} \cap \{a_{21}=0\}, \qquad  \quad  p_2: \{w=0\} \cap \{a_1 =0 \} \cap \{a_{32}=0\},
 \ee
 where the singularity type enhances to $E_6$ and $D_6$, respectively. It has become common knowledge \cite{Donagi:2008ca,Beasley:2008kw,Hayashi:2009ge,Hayashi:2009bt,Hayashi:2010zp,Marsano:2011hv} that these points support Yukawa couplings of the form ${\bf 10 \, 10 \, 5}$ and ${\bf 10 \, \bar 5 \, \bar 5}$, respectively.
 In the sequel we re-examine the appearance of these couplings as an example for the role played by perturbative and non-perturbative homological relations.

The fibre structure of the resolution of this model is well-known from the analysis of \cite{Esole:2011sm,Marsano:2011hv,Krause:2011xj}. Here we merely collect the information needed to exemplify the homological fibre relations.  In the notation of \cite{Krause:2011xj}, to which we refer for further details, the proper transform of the resolved hypersurface takes the form
\be\label{PTeq}
\begin{aligned}
\hat P_T =& e_1 \, e_2^2 \, e_3 \, x^3 - e_3 \, e_4 \, y^2 - a_1 \, x \, y \, z + a_{21} \, e_0 \, e_1 \, e_2 \, x^2 \, z^2 - 
   a_{32} \, e_0^2 \, e_1 \, e_4 \, y \, z^3 + a_{43} \, e_0^3 \, e_1^2 \, e_2 \, e_4 \, x \, z^4  \\
   &  +a_{65} \, e_0^5 \, e_1^3 \, e_2 \, e_4^2 \, z^6.
\end{aligned}
\ee
The divisors $E_i: \{e_i = 0 \}$ are obtained by fibering rational curves $\mathbb P^1_i$ over the divisor $W$ in $\B$ and correspond to the resolution divisors.
We are particularly interested in the fibre over $\calc_{\bf 5}$ and its further modification as we approach the $D_6$ Yukawa point.

As discussed in appendix \ref{app:split}, over $\calc_{\bf 5}$ the fibre $\mathbb P^1_3$ of the resolution divisor $E_3$ splits into two rational curves,
\be\label{P1split}
\mathbb P^1_3 \rightarrow \mathbb P^1_{3D}+\mathbb P^1_{3C}\, .
\ee
By wrapping M2-branes on the two new fibral curves $\mathbb P^1_{3D}$  and  $\mathbb P^1_{3C}$ one gets specific states in the ${\bf  5}$ and ${\bf  \bar 5}$ representations of $SU(5)$ respectively, identified by the following weights\footnote{\label{footnoteweights}The specific weight vectors associated with the fibre curves depend on which of the six inequivalent  triangulations of the toric resolution is chosen. Here we are referring to the triangulation called $T_{11}$ in \cite{Krause:2011xj}. Our convention for the weight vectors is $\mu_{10} = [0,1,0,0]$,  $\mu_{5} = [1,0,0,0]$ and for the simple roots  $ - \alpha_1 = [-2,1,0,0]$, $ - \alpha_2 = [1,-2,1,0]$, $ - \alpha_3= [0,1,-2,1]$, $- \alpha_4 = [0,0,1,-2]$. The entries denote the Cartan charges and can be computed as the intersection number between the fibral curve associated with the weight and the resolution divisors $E_i$. In particular, an M2-brane wrapping the fibre $\mathbb P^1_i$ of $E_i$ is associated with $-\alpha_i$.  }
\be
 \mathbb P^1_{3D}:    \mu_5 - \alpha_1 - \alpha_2 - \alpha_3\,,   \qquad \qquad \mathbb P^1_{3C}:   - \mu_5 + \alpha_1 + \alpha_2\,.
\ee
On the other hand, on $\calc_{\bf 10}$ there is a fibral curve $\mathbb P^1_{32}$ that represents the  state of the ${\bf 10}$ representation associated with the weight
\be
\mathbb P^1_{32}: \mu_{10} - \alpha_1 - 2 \alpha_2 - \alpha_3 .
\ee
By comparing the weights of the above curves, we can conclude that there must exist a homological relation\footnote{For generic models ${\rm Tor}H_2(\hat \Y, \mathbb Z)=0$ and so the homological relations between fibral 2-cycles are unambiguously specified by their Cartan charges.}
\be
[\mathbb P^1_{32} +  \mathbb P^1_{3C}-\mathbb P^1_{3D}]=0\in H_2(\hat \Y, \mathbb Z).
\ee

We would like to show that this homological relation is actually valid at the {\em perturbative} level. In order to see this, let us move the curves ${\mathbb P}^1_{32} $ (along $\calc_{\bf 10}$) as well as ${\mathbb P}^1_{3D}$ and ${\mathbb P}^1_{3C}$ (along ${\cal C}_{\bf 5}$) to  the $D_6$ Yukawa point $p_2$.
Notice that, due to the complicated structure of the curve ${\cal C}_{\bf 5}$, there are actually two different types of paths $\gamma_\pm\subset{\cal C}_{\bf 5}$  that one can follow while moving ${\mathbb P}^1_{3D}$ and ${\mathbb P}^1_{3C}$. Along the first type $\gamma_+$ we encounter the splitting
\be
\begin{aligned}
\mathbb P^1_{3D} &\rightarrow \mathbb P^1_{32} + \mathbb P^1_{3+}\,, \\
\mathbb P^1_{3C} &\rightarrow \mathbb P^1_{3-}\,.
\end{aligned}
\ee
On the other hand, along the other type $\gamma_-$ the splitting takes the form
\be
\begin{aligned}
\mathbb P^1_{3D} &\rightarrow \mathbb P^1_{32} + \mathbb P^1_{3-}, \\
\mathbb P^1_{3C} &\rightarrow \mathbb P^1_{3+}.
\end{aligned}
\ee
We prove this behaviour and further specify the two types of paths $\gamma_+$ and $\gamma_-$ in appendix \ref{app:split}. 
This complicated splitting behaviour is a consequence of the fact the curve ${\cal C}_{\bf 5}$ self-intersects at $p_2$ and explains the presence of two copies of the {\bf 5}-state in the fibre over $p_2$.
The self-intersection of ${\cal C}_{\bf 5}$ at the $D_6$ Yukawa point has been studied before, e.g. in  \cite{Hayashi:2009bt}. Finally,  note that also some of the other resolution $\mathbb P^1$s split.
The final result for the fibre structure over $p_2$ is depicted in figure \ref{figureYukpoint}.

\begin{figure}[t]
    \centering
  \begin{tikzpicture}[scale=3.9,main node/.style={circle,
  draw,minimum size=4mm}]
  \node[main node] (1) {$3-$};
  \node[main node] (2) [above left of=1] {$32$};
  \node[main node] (7) [above right of=2] {$3+$};
  \node[main node] (3) [left of=2] {$24$};
  \node[main node] (4) [left of=3] {$14$};
  \node[main node] (5) [above left of=4] {$0A$};
  \node[main node] (6) [below left of=4] {$4D$};
    \draw[ultra thick] (1) -- (2);
      \draw[ultra thick] (2) -- (7);
    \draw[ultra thick] (2) -- (3);
    \draw[ultra thick] (3) -- (4);
    \draw[ultra thick] (4) -- (5);
    \draw[ultra thick] (4) -- (6);
   \end{tikzpicture}
      \caption{Fibre   topology at the Yukawa  point $p_2 = \{w=0\} \cap \{a_1 =0 \} \cap \{a_{32}=0\}$ in the $SU(5)$ Tate model.}  \label{figureYukpoint}
\end{figure}
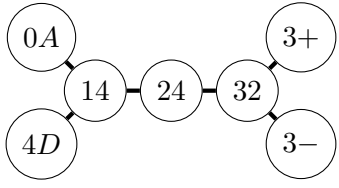

We are now in a position to exemplify the perturbative and non-perturbative homological relations.
As is clear from the discussion, the fibral curves
$\mathbb P^1_{3D}$, $\mathbb P^1_{3C}$, $\mathbb P^1_{32}$ and $\mathbb P^1_{3\pm}$ obey  the following perturbative homological relations
\be
 [ \mathbb P^1_{3D} - \mathbb P^1_{32} -  \mathbb P^1_{3+}]_{\rm p} = 0\,, \qquad [\mathbb P^1_{3C} - \mathbb P^1_{3-} ]_{\rm p} =0\,
 , \qquad [\mathbb P^1_{3+} - \mathbb P^1_{3-} ]_{\rm p} =0 .
\ee
In our language, this translates into the existence of a {\em perturbative} 3-chain $\Gamma_{\rm p}$ such that 
\be\label{pert1055}
\del\Gamma_{\rm p}=-( \mathbb P^1_{32} +  \mathbb P^1_{3C}-\mathbb P^1_{3D}).
\ee
 Notice that $\Gamma_{\rm p}$ has a component that connects 
$\mathbb P^1_{3+}$ and  $\mathbb P^1_{3-}$, which is obtained by fibering $\mathbb P^1_{3C}$ over a closed non-trivial path on ${\cal C}_{\bf 5}$.
According to the general discussion of section \ref{sec:general}, the perturbative relation (\ref{pert1055}) corresponds to  a perturbative coupling $\sim \Phi_{\bf 10}\Phi_{\bf \bar 5}\Phi_{\bf \bar 5}$ (after appropriate covariantisation), typically written as ${\bf 10 \, \bar 5 \,  \bar 5}$.
Physically, at $p_2$ an incoming M2-brane on  $\mathbb P^1_{3D} $ can split \cite{Marsano:2011hv} in, as our analysis shows, a path-dependent manner into an M2-brane on $\mathbb P^1_{32}$ and $\mathbb P^1_{3+}$ or $\mathbb P^1_{3-}$. The M2-brane on $\mathbb P^1_{32}$ can continue to propagate on the curve ${\cal C}_{\bf 10}$ as a well-defined state in the ${\bf 10}$ representation. If the splitting of $\mathbb P^1_{3D}$ occurred via $\gamma_\pm$, then the outgoing M2-brane on $\mathbb P^1_{3\pm}$ must leave the Yukawa point along a path $-\gamma_\mp$ and propagate as a state along $\mathbb P^1_{3C}$ away from $p_2$.

The other triple interaction consistent with $SU(5)$ gauge invariance is the coupling ${\bf 10 \, 10 \, 5}$. As we now show, couplings with such a structure can be generated either at the perturbative or non-perturbative level, depending on the level of genericity of the base $\B$.
We start by noticing that the fibre over $\calc_{\bf 10}$ contains the curves $ \mathbb P^1_{4D}$ and $\mathbb P^1_{24}$ (see again \cite{Krause:2011xj} for more details on the present notation or alternatively \cite{Esole:2011sm,Marsano:2011hv}), which correspond to the following weights of  the ${\bf 10}$ and ${\bf \overline{10}}$ representations, respectively,
\be
 \mathbb P^1_{4D}: \mu_{10} - \alpha_2 - \alpha_3 - \alpha_4, \qquad \mathbb P^1_{24}: -\mu_{10} + \alpha_1 + \alpha_2 + \alpha_3 .
\ee
By comparing these weights with the weights of $\mathbb P^1_{3C}$, we see that there must exist a homological relation 
\be \label{10105np}
[-  \mathbb P^1_{4D} +  \mathbb P^1_{24} + \mathbb P^1_{3C}]   = 0 \in H_2(\hat \Y, \mathbb Z).
\ee

In order to understand its nature in terms of our refined perturbative homology, we must again study the behaviour of the fibral curves as we approach the common base points $p_1$ and $p_2$. Let us first move the fibral curves along base paths ending on $p_2$. From the structure of the elliptic fibre over $p_2$ described in figure \ref{figureYukpoint} it is evident that for any choice of paths we end up with $-  \mathbb P^1_{4D} +  \mathbb P^1_{24} + \mathbb P^1_{3C}\neq 0$ within $H_2(\mathfrak f_{p_2}, \mathbb Z)$. Hence, no perturbative homological relation can arise via the point $p_2$.

On the other hand, by approaching an  $E_6$ Yukawa point $p_1$ one can see that a relation analogous to (\ref{10105np}) is in fact realised at the perturbative level. Hence, if the   triple intersection (\ref{p1p2}) defining the points $p_1$ has a non-trivial solution (as expected for generic base $\B$), then our general arguments imply the possible existence of a perturbative F-term coupling of the form $\Phi_{\bf 10}\Phi_{\bf 10}\Phi_{\bf 5}$. This is just the well known statement that $p_1$, and not $p_2$, supports the perturbative Yukawa coupling ${\bf 10 \, 10\,  5}$  \cite{Donagi:2008ca,Beasley:2008kw,Hayashi:2009ge,Hayashi:2010zp,Marsano:2011hv}.
Recent progress in the quantitative evaluation of these couplings in local approaches has been made in \cite{Font:2012wq,Font:2013ida,Marchesano:2015dfa} and references therein.

However, there do exist examples where the $E_6$-point $p_1$ is absent as a result of the specific structure of the base $\B$. In particular this is guaranteed to be the case if $\B$ is the orientifold quotient of a smooth Calabi-Yau 3-fold $\X$ such that the orientifolded  Type IIB on $\X$ describes the Sen limit of the F-theory model. An example of such a base space has been constructed in \cite{Blumenhagen:2009up} and will be discussed in more detail in section \ref{subsec:explicitD1}.
In such a situation the homological relation (\ref{10105np}) is realised only at the {\em non-perturbative} level and  there exists a non-perturbative 3-chain $\Gamma_{\rm np}\in \hat \Y$ such that 
\be
\del\Gamma_{\rm np}=- (-  \mathbb P^1_{4D} +  \mathbb P^1_{24} + \mathbb P^1_{3C} ) .
\ee
It is clear from the above discussion that $\Gamma_{\rm np}$ is not just spanned by fibral curves moving  along the matter curves $\calc_{\bf 5}$ and $\calc_{\bf 10}$. Hence, in the F-theory limit the volume of such a chain remains finite and  an M2-brane instanton wrapping $\Gamma_{\rm np}$ may generate an exponentially suppressed D-term coupling of the form  $\Phi_{\bf 10}\Phi_{\bf 10}\Phi_{\bf 5}e^{-2\pi{\rm vol}(\Gamma_{\rm np})}$. On the other hand, as we will discuss in more detail in \cite{paper2}, a  ${\bf 10 \, 10\,  5}$ F-term coupling can still be generated by fluxed M5-brane instantons,  similarly to what happens in Type II $SU(5)$ models  \cite{Blumenhagen:2007zk}.

\section{$\mathbb Z_2$-symmetry and instantons} \label{sec:Z2}

Another example where our distinction between perturbative and non-perturbative  homological relations become important is in the context of F-theory compactifications with discrete gauge symmetries. 
In the sequel we will exemplify the general structure outlined already in section \ref{sec_Zksymmetry} in the context of F-theory compactifications with $\mathbb Z_2$ discrete symmetries. 
In particular we will demonstrate how in sufficiently \emph{generic} situations in four dimensions all couplings consistent with the discrete symmetry are allowed at the perturbative level; for specific backgrounds, however, it is only due to non-perturbative homological relations that the full set of operators allowed by the discrete symmetry is realised. We will trace this phenomenon back  to an enhancement of the discrete symmetry to a \emph{perturbative} $U(1)$ symmetry, which can only be broken non-perturbatively.

 Our conclusions build specifically on the analysis of $\mathbb Z_2$ discrete symmetry in F-theory as presented in \cite{Mayrhofer:2014haa,Mayrhofer:2014laa}, to which we refer for further details. See also \cite{Anderson:2014yva,Klevers:2014bqa,Garcia-Etxebarria:2014qua} for discussions of this model.
Starting from the F-theory compactification with gauge group $U(1)$ considered in \cite{Morrison:2012ei}, two distinct Higgsings can be performed \cite{Morrison:2014era,Mayrhofer:2014haa,Mayrhofer:2014laa} to arrive at either a genus-one fibration with a bisection \cite{Braun:2014oya} or to a Weierstrass model with $\mathbb Z_2$ torsional elements in $H_2(\hat \Y, \mathbb Z)$ \cite{Mayrhofer:2014laa}.
M-theory compactified on the genus-one and the Weierstrass model  to three dimensions gives rise to gauge group $U(1)$ and, respectively, $U(1) \times \mathbb Z_2$. In the F-theory limit both 
theories reduce to a 4-dimensional model with $\mathbb Z_2$ symmetry \cite{Mayrhofer:2014laa}.

\subsection{Fibration with two rational sections}

We will first analyse the $U(1)$ model prior to Higgsing from the perspective of the homological relations and then discuss the two Higgsed models in turn.
The $U(1)$ model is defined in terms of a ${\rm Bl}^1\mathbb P_{1,1,2}[4]$-fibration first introduced and analysed in \cite{Morrison:2012ei}. The Calabi-Yau 4-fold $\hat \Y$ is described as the vanishing locus of the hypersurface constraint
\be \label{1124model}
P_{U(1)}:=s  \, w^2   + b_{0}  \, s^2 \, u^2 \,  w  + b_1 \, s \, u \, v \, w  + b_2  \, v^2 \, w + c_0 \, s^3 \, u^4   + c_1 \, s^2  \, u^3 \, v  + c_2 \, s \, u^2 \, v^2   + c_3 \, u \, v^3
\ee
with $b_i$ and $c_j$ suitable sections on a base space $\B$. The fibre coordinates $[u: v : w]$ are homogeneous coordinates on $\mathbb P_{1,1,2}$ and $S: \{s=0\}$ represents a rational section independent of the holomorphic zero-section $U: \{u=0\}$. The zero-section $U$ acts, as always, as the generator of the Kaluza-Klein $U(1)$ symmetry of the 3-dimensional effective action obtained by compactification of M-theory on $\hat \Y$. The KK symmetry is associated with the $S^1$ reduction from the 4-dimensional effective F-theory compactification to three dimensions.
The combination $S - U - \bar K - [b_2]$ generates a massless $U(1)$ gauge symmetry, which survives as such in the F-theory limit.

The fibre degenerates to an $I_2$-fibre over two matter curves $\calc_{I}$ and $\calc_{II}$ on $\B$  \cite{Morrison:2012ei}.
We will stick to  the notation $A_{I}$, $B_{I}$ as well as $A_{II}$, $B_{II}$ for the two fibral curves over generic points of these matter curves as in figure 1 of \cite{Mayrhofer:2014laa}.
In the 3-dimensional effective action, M2-branes wrapping these fibral curves give rise to chiral fields with the following charges under $U(1)_U \times U(1)_{S-U}$:
\be \label{Mtheory-fields}
\Phi_{A_{I}}: (1,-2), \qquad \Phi_{B_{I}}: (0,2),\qquad \Phi_{A_{II}}: (1,-1),\qquad \Phi_{B_{II}}: (0,1).
\ee

In the 4-dimensional effective action there exist two ${\cal N}=1$ chiral superfields $\Phi_I$ and $\Phi_{II}$ whose KK zero modes in three dimensions are given by $\Phi_{B_{I}}$ and $\Phi_{B_{II}}$. The fields $\Phi_{A_{I}}$ and $\Phi_{A_{II}}$ form the first level of the KK tower of these fields. The remaining fields in the KK tower arise from M2-branes with arbitrary wrappings of the full fibre. The importance of the KK charge of the states in this context has been stressed, from different perspectives, in \cite{Anderson:2014yva,Garcia-Etxebarria:2014qua,Mayrhofer:2014haa,Mayrhofer:2014laa}.

For a generic base $\B$, the curves $\calc_I$ and $\calc_{II}$ intersect at the point set $p = \calc_I \cap \calc_{II}$ given by 
\be \label{tripleintU1model}
p = \{b_2 = 0\} \cap \{c_3 = 0\} \cap \{b_1^2 \,  c_0 - b_ 0 \, b_1 \, c_1 + b_0^2 \, c_2 + c_1^2 - 4 \, c_0 \, c_2 =0\}.
\ee
Here the fibre enhances to an $I_3$-fibre with three rational curves $F_a$, $F_b$, $F_c$, see figure \ref{U(1)Z2fibers}.
 \begin{figure}[t!]
  \centering
    \includegraphics[width=0.60\textwidth]{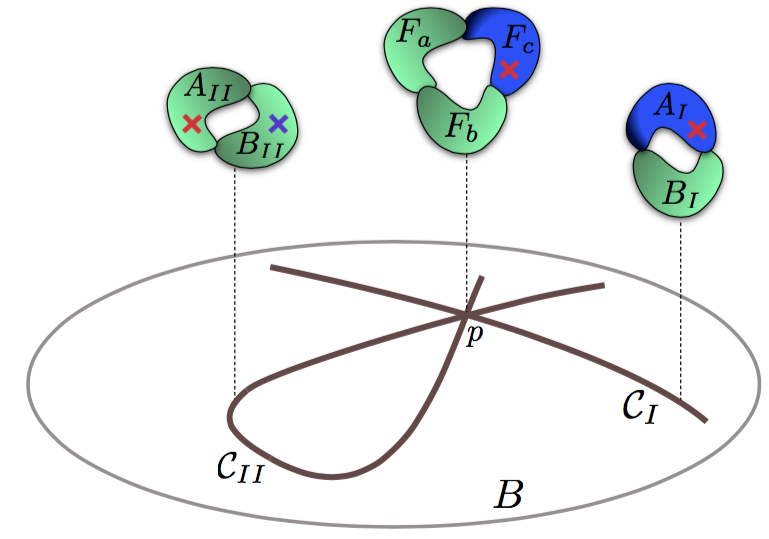}
      \caption{\footnotesize Fibre enhancement over the Yukawa points in the $U(1)$ model introduced in  \cite{Morrison:2012ei}. The { red} and { blue } crosses denote intersections with the sections ${ U}$ and, respectively, $S$, which also wraps $F_c$ and $A_I$ as indicated. The self-intersection of ${\cal C}_{II}$ at $p$ is responsible for the existence of the two splitting paths (\ref{pertCIIa}) and (\ref{pertCIIb}) for the fibral curves. }\label{U(1)Z2fibers}
\end{figure}
This is due to a factorisation of the hypersurface equation as \cite{Klevers:2014bqa,Mayrhofer:2014haa} 
\be\label{triplepointhypersurf}
P_{U(1)}|_{p} =   s \,  \calw_+\calw_-|_{p},
\ee
where
\be
\calw_\pm = w + \frac{1}{2} \big(b_0 s u^2 + b_1 u v\big)\pm \frac{u}{\sqrt{b_1^2-4c_{2}}}\Big[(c_{1} - \frac{1}{2}b_{0}b_1)su + 2(c_{2}-\frac{1}{4}b_1^2)v\Big].
\ee
These fibral curves are related to the fibral curves over generic points of the matter curves as follows: Approaching the point $p$ along $\calc_I$ only $B_I$ splits,
\bea \label{pertCI}
A_I \rightarrow F_c, \qquad 
B_I \rightarrow F_a + F_b.
\eea
This is reflected in the perturbative homological relations
\bea \label{homrelCI}
[F_{c}-A_{I} ]_{\rm p} =0, \qquad   \qquad [ F_a + F_b-B_{I} ]_{\rm p}=0.
\eea
The splitting pattern coming on $\calc_{II}$ is more complicated and has not been analysed explicitly so far. By analogy to the model discussed in section \ref{sec:SU(5)ex}, there must again exist two equivalence classes of paths approaching the Yukawa point $p$.
Along the first class of trajectories the splitting is 
\bea \label{pertCIIa}
A_{II} \rightarrow F_c + F_a, \qquad B_{II} \rightarrow F_b,
\eea
while along a second type of trajectories
\bea \label{pertCIIb}
A_{II} \rightarrow F_c + F_b, \qquad B_{II} \rightarrow F_a.
\eea
As for the model in section \ref{sec:SU(5)ex} the reason for the appearance of two different paths is the fact that ${\cal C}_{II}$ self-intersects at $p$, see figure 3 in \cite{Mayrhofer:2014haa} and the above figure \ref{U(1)Z2fibers} here.
The various splittings induce again corresponding perturbative  homological relations,
\bea \label{pertFaFb}
 [F_c + F_a-A_{II} ]_{\rm p} = 0, \qquad      [F_a -  B_{II}]_{\rm p} =0, \qquad    [F_a - F_b]_{\rm p} = 0.
\eea

These perturbative relations yield a standard perturbative Yukawa coupling in the F-theory effective action of the form ${\tilde \Phi_I} \, \Phi_{II} \, \Phi_{II}$.
Indeed in terms of the M-theory fields (\ref{Mtheory-fields}) combining (\ref{homrelCI}) and (\ref{pertFaFb}) implies a coupling of the schematic form
\bea \label{YukawaI-II-II}
\tilde \Phi_{B_I} \Phi_{B_{II}} \Phi_{B_{II}}.
\eea
The existence of the coupling (\ref{YukawaI-II-II}) has already been predicted, based on the $I_3$ fibre (\ref{triplepointhypersurf}), in \cite{Klevers:2014bqa,Mayrhofer:2014haa}. 
As we see here, at the microscopic level an incoming M2-brane on $B_{II}$ splits, locally at $p$, into two M2-branes on $F_a$ and $F_b$. Both describe a copy of the state $\Phi_{B_{II}}$ on $\calc_{II}$ due to the existence of the two splitting processes (\ref{pertCIIa}) and (\ref{pertCIIb}). The interpretation of this process in terms of perturbative 3-chains is as in section \ref{sec:SU(5)ex}.

 Likewise in M-theory the perturbative couplings involving the full KK tower of states with higher KK-charges summing up to zero follows. An example is the coupling
\bea \label{Yuk-coupling2}
\tilde \Phi_{A_{II}} \Phi_{A_{I}} \Phi_{B_{II}}.
\eea
By combining (\ref{pertCI}) with, say, (\ref{pertCIIa}) we see that an incoming M2-brane on $A_{II}$ splits into a copy of the state $\Phi_{A_{I}}$ along $F_c$  and an M2-brane wrapping $F_a$.

\subsection{Bisection versus torsional Weierstrass models}

Let us now follow these couplings through the two possible Higgsings.
The first Higgsing corresponds to a conifold transition in which first $A_I$  shrinks to zero volume and is then deformed into an $S^3$ \cite{Morrison:2014era,Anderson:2014yva,Klevers:2014bqa,Garcia-Etxebarria:2014qua,Mayrhofer:2014haa}. Field-theoretically this  corresponds to giving a VEV to the massless pair $\Phi_{A_I} \tilde \Phi_{A_I}$. The deformation turns the fibration into a genus-one fibration with a bisection \cite{Braun:2014oya} by first setting $s\equiv 1$ in (\ref{1124model})  and then adding the term $c_4 \, v^4$. The bisection is given by $U_{\rm bi} = \{u=0\}$ in the deformed hypersurface. It can be thought of as the sum $S+U$ of the $U(1)$ model and intersects each fibre in the two points $\{u=0\} \cap \{ w + \frac{1}{2} v^2 \, (b_2 \pm \sqrt{b_2^2 - 4 c_4}) = 0 \}$.
These two points cannot be distinguished globally due to a monodromy responsible for the gluing of both points in a bisection \cite{Braun:2014oya}. In particular, the two intersection points 
come together in the fibre over the divisor $\{b_2^2 - 4 c_4 = 0\}$.

The only $I_2$-fibre with components $\tilde A_{II}$ and $\tilde B_{II}$ is over a deformed version of $\calc_{II}$ which will be called $\tilde\calc_{II}$ in the sequel. 
The crucial observation is that over a \emph{generic} base space $\B$, in a sense that will become clear momentarily, the two fibral components $\tilde A_{II}$ and $\tilde B_{II}$ satisfy the two \emph{perturbative} relations
\be \label{pertrelbisec}
[\tilde A_{II} + \tilde B_{II} - \mathfrak f]_{\rm p} = 0, \qquad [\tilde A_{II} - \tilde B_{II}]_{\rm p} = 0,
\ee
where ${\mathfrak f}$ denotes the generic genus-one fibre class.
The second, less obvious relation is inherited from the last relation in (\ref{pertFaFb}) in the model before the Higgsing.  Its geometric origin is a monodromy exchanging $\tilde A_{II}$ and $\tilde B_{II}$ along a path on $\tilde\calc_{II}$ encircling one of the points in the set
\be \label{tilde C2points}
\tilde p =  \{b_2^2 - 4 c_4= 0\} \cap \{c_3 - \frac{1}{2} b_1 b_2 =0 \} \cap \{  - \frac{1}{2} b_0^3 b_2 + b_1^2 c_0 + 2 b_0 b_2 c_0 - b_0 b_1 c_1 + c_1^2 + b_0^2 c_2 - 
 4 c_0 c_2 = 0 \}. 
\ee
This is precisely the intersection of the divisor $\{b_2^2 - 4 c_4 = 0\}$ with $\tilde\calc_{II}$.
Indeed, over a generic base point $U_{\rm bi}$ intersects $\tilde A_{II}$ and $\tilde B_{II}$ in two separate points, one on each fibral curve.
Locally, over a generic base point, the two fibre components can therefore be distinguished according to which of these two separate points they contain.
Over the locus $\tilde p$,
 these two points coincide with each other and with one of the two intersection points of the $I_2$-fibral curves \cite{Mayrhofer:2014haa}. 
 Note that the points in the set $\tilde p=\{\tilde p^\pm_1,\tilde p^\pm_2, \tilde p^\pm_3,\ldots\} $ come in pairs because the polynomial $b_2^2 - 4 c_4$ has an even number of roots loosely speaking given by $b_2 = \pm \sqrt{4 c_4}$.
 One can therefore draw a branch-cut between any pair $\tilde p_i^+$ and $\tilde p_i^-$ such that a path across that branch-cut interchanges 
the two bisection points. Since the only way to distinguish  $\tilde A_{II}$ and $\tilde B_{II}$ locally is via their intersection points with the bisection, this can effectively be thought of as exchanging 
 $\tilde A_{II}$ and $\tilde B_{II}$.
This implies the existence of a perturbative 3-chain $\Gamma_{\rm p}$ fibred over a closed loop on $\tilde\calc_{II}$ with 
\bea \label{3chaintildeAtildeB}
\partial \Gamma_{\rm p} =  \tilde B_{II}-\tilde A_{II} ,
\eea
see figure \ref{bisectionGamma}. The volume of this perturbative chain vanishes in the F-theory limit. 
Combining the two relations (\ref{pertrelbisec}) leads to $[2 B_{II}-\mathfrak f]_{\rm p} =  0$, which shows the presence of torsion modulo the fibre class \cite{Mayrhofer:2014laa}. As we see here, over a generic base $\B$ this torsion is `perturbative' to the extent that there exists a 3-chain with boundary $2 B_{II} -  \mathfrak f$ whose volume vanishes in the F-theory limit.

For later purposes note that the set of points $\tilde p$ in (\ref{tilde C2points}) reduces, for $c_4 \equiv 0$, to the Yukawa points $p$ (see eq. (\ref{tripleintU1model})) in the $U(1)$ model. 
More precisely the pairs of distinct roots of $b_2^2 - 4 c_4=0$ for $c_4 \neq 0$ coalesce for $c_4=0$. Thus as $c_4 = 0$ the branch-cut between the two different roots disappears, and with it the monodromy exchanging the two  curves in the generic fibre over $\calc_{II}$. This is in perfect agreement with the fact that for $c_4=0$ the two fibral curves become distinguishable due to the splitting of the bisection into two sections with different intersection numbers with both fibre components.

 \begin{figure}[t!]
  \centering
    \includegraphics[width=0.55\textwidth]{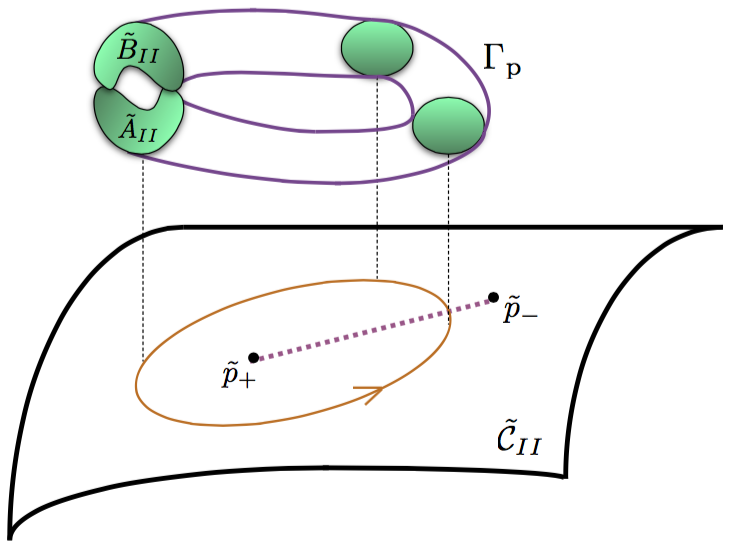}
      \caption{\label{np1}\footnotesize 3-chain $\Gamma_{\rm p}$ connecting $\tilde A_{II}$ and $\tilde B_{II}$.}\label{bisectionGamma}
\end{figure}

As a result of the perturbative 3-chain (\ref{3chaintildeAtildeB}) there exists, in the M-theory effective action, a perturbative bilinear term of the form
\bea \label{bilinear-genusonea}
\tilde\Phi_{\tilde A_{II}} \Phi_{\tilde B_{II}} + c.c.
\eea
in addition to the trivially realised operators $\Phi_{\tilde A_{II}} \tilde \Phi_{\tilde A_{II}}$ and $\Phi_{\tilde B_{II}} \tilde \Phi_{\tilde B_{II}}$.
This term arises from an incoming M2-brane on $\tilde A_{II}$ which is transported along  $\Gamma_{\rm p}$ into an outgoing M2-brane on $\tilde B_{II}$. By starting and ending arbitrarily close to one of the monodromy points  $\tilde p^\pm_i$ the volume of $\Gamma_{\rm p}$ can be taken arbitrarily small already in M-theory so that this bilinear term is expected to be of order one. See also  \cite{Garcia-Etxebarria:2014qua} for a different interpretation of a similar coupling.

Consistently, the coupling (\ref{bilinear-genusonea})   originates from 
(\ref{Yuk-coupling2}) upon giving a VEV to $ \Phi_{A_{I}}$. 
If we introduce the notation $ \Phi_{\tilde{II}}$ for the ${\cal N} = 1$ chiral field localised on $\tilde\calc_{II}$ in the F-theory effective action, then the operator (\ref{bilinear-genusonea}) points to a perturbative mass term of the form $ \Phi_{\tilde{II}} \,  \Phi_{\tilde{II}}$. Note that in the language of the original $U(1)$ model, a KK state of charge $-1$ and $0$ combine in  (\ref{bilinear-genusonea}). This is a result of the fact that  in the bisection model the circle reduction from four to three dimensions involves a \emph{fluxed $S^1$} \cite{Anderson:2014yva,Garcia-Etxebarria:2014qua}, or, equivalently, that the Higgs field $\Phi_{A_{I}}$ has KK charge $+1$ \cite{Mayrhofer:2014haa,Mayrhofer:2014laa}.
The presence of these mass terms must be detected perturbatively by computing the massless perturbative spectrum. 
The massless states lie, by definition, in the kernel of the perturbative mass matrix. 
Finally, note that the above-mentioned fact that the monodromy points $\tilde p$ reduce to the Yukawa points in the limit $c_4 \equiv 0$ is in perfect agreement with this field theoretic relation between  the operators (\ref{bilinear-genusonea})  and (\ref{Yuk-coupling2}) because (\ref{bilinear-genusonea})  relies on the monodromy around $\tilde p$ in the same way as the Yukawa  (\ref{Yuk-coupling2}) depends on the existence of the fibre enhancement over the Yukawa points.

Let us now briefly comment on the second, alternative Higgsing process in which the fibral curve $B_I$ shrinks to zero, followed by a deformation into an $S^3$ \cite{Mayrhofer:2014laa}.
The resulting F-theory compactification is defined in terms of a Weierstrass model (\ref{Weier1}) with  non-generic form $f$ and $g$  \cite{Braun:2014oya,Morrison:2014era}.
At the same time as  $B_I$ shrinks to zero volume so does $B_{II}$. The resulting $I_2$-singularity over $\tilde\calc_{II}$,  the deformed version of $\calc_{II}$, 
can only be resolved in a non-crepant manner.
Apart from performing a non-K\"ahler small resolution, one can first blow-up the codimension-two locus $\tilde\calc_{II}$ on the base into a divisor and then resolve the $I_2$-singularity. 
As argued in \cite{Mayrhofer:2014laa} for compactifications to six dimensions, the resolved non-Calabi-Yau fibration $\hat Y$ exhibits a non-trivial $\mathbb Z_2$-torsional component in $H_2(\hat Y, \mathbb Z)$, which is responsible for the appearance of a $\mathbb Z_2$ gauge symmetry in F-theory. 
At any rate we can think of the non-crepant resolution as providing a fibral curve $\tilde B_{II}$ over $\tilde\calc_{II}$ which is purely torsional, i.e. 
\bea \label{non-pertZ2rel}
2 \, [\tilde B_{II}]  = 0 \in H_2(\hat Y, \mathbb Z).
\eea
M2-branes wrapping $ [\tilde B_{II}] $ give rise to states $\Phi_{\tilde B_{II}}$ of $\mathbb Z_2$ charge $+1$ and, from the perspective of M-theory, KK charge zero.

Due to the intermediate singularity it is much harder to determine if  $\tilde B_{II}$ is torsion also in the perturbative sense of section \ref{sec_Zksymmetry}. However, the F-theory effective actions obtained by compactification on the genus-one fibration with bisection and the Weierstrass model must coincide because both describe the same profile for the axio-dilaton $\tau$ and thus the same brane dynamics \cite{Braun:2014oya,Morrison:2014era}. 
This  leads us to the conclusion that the $\mathbb Z_2$ must indeed be perturbatively realised in the sense described in section \ref{sec_Zksymmetry} - at least over a \emph{generic} 3-dimensional base $\B$. 
This implies the existence of a perturbative coupling $ \Phi_{\tilde B_{II}}  \Phi_{\tilde B_{II}}  + c.c.$ in addition to the trivial perturbative coupling $ \tilde \Phi_{\tilde B_{II}}  \Phi_{\tilde B_{II}}$.
Therefore, the states  $ \Phi_{\tilde B_{II}}$ and $ \tilde \Phi_{\tilde B_{II}}$ are indistinguishable at the perturbative level.
Again a computation of the perturbative massless spectrum is required to determine which states lie in the kernel of the mass matrix. 
It would be interesting to derive the nature of the perturbative 3-chain $\Gamma_{\rm p}$ with $\partial \Gamma_{\rm p} = 2 \,\tilde B_{II}$ directly.

\subsection{Perturbative $U(1)$ symmetry and non-perturbative breakdown}

From our discussion so far we conclude that for a generic base space $\B$ all couplings allowed by the $\mathbb Z_2$ symmetry are realised at the perturbative level. 
Generic means in this case that the triple intersection number (\ref{tripleintU1model}) is positive such that the Yukawa points $p$ exist. 
As stated already this condition is equivalent to the existence of the points $\tilde p$ in (\ref{tilde C2points}). 
The triple intersection number counting the number of such points is given by the topological intersection number $\int_{B} [c_4] \wedge [c_3] \wedge 2 [ c_1]$. 
These divisor classes depend on a choice of a class $\beta$ with $0 \leq \beta \leq 2 \bar K$ in the way reproduced e.g. in table 1 of \cite{Mayrhofer:2014haa}. A choice of $\beta$ in this range defines the bisection fibration.
The triple intersection number counting the number of points $\tilde p$ is then given by $\int_{B_3}  (-2 \beta + 4 \bar K)\wedge  (-\beta + 3 \bar K) \wedge (2 \beta + 2 \bar K)$.
For a fibration on base $B$ and class $\beta$ where this topological invariant is negative or zero, the set $\tilde p$ is empty. This is what we refer to as the non-generic case. 

In absence of points $\tilde p$, the bisection fibration lacks a monodromy action interchanging the two fibral curves $\tilde A_{II}$ and $\tilde B_{II}$ along a perturbative 3-chain $\Gamma_{\rm p}$. Rather, while it is still true that $[\tilde A_{II}] - [\tilde B_{II}]=0 \in H_2(\hat \Y, \mathbb Z)$, this relation does no longer hold at the perturbative level. 
Instead a 3-chain bounded by $\tilde A_{II} - \tilde B_{II}$ must extend into $\hat \Y$ away from $\hat \Y|_{\tilde\calc_{II}}$. Such an object has two legs in the base and one leg in the fibre and acquires non-vanishing volume in the F-theory limit. A  coupling with schematic structure (\ref{bilinear-genusonea}) is now induced by an M2-instanton wrapping this 3-chain. It is non-perturbatively suppressed and cannot arise at the level of F-terms (unless the M2-brane forms a BPS bound state with an M5-instanton as discussed in our companion paper \cite{paper2}). In this non-generic situation,
$\Phi_{\tilde A_{II}}$ and $\Phi_{\tilde B_{II}}$ do not mix via bilinear terms at the perturbative level.
This can also be seen by noting again that the points $\tilde p$ are precisely the locus over which the two intersection points of the bisection $U_{\rm bi}$ with the fibre over ${\tilde\calc_{II}}$ coincide \cite{Mayrhofer:2014haa}. 
In absence of $\tilde p$ the bisection splits into a sum of two sections over the curve ${\tilde\calc_{II}}$, $U_{\rm bi}|_{\tilde \calc_{II}} = \tilde S + \tilde U$. This induces the notion of a perturbative $U(1)$ charge under which the states $\Phi_{\tilde A_{II}}$ and $\Phi_{\tilde B_{II}}$ have opposite charge: The generator of this perturbative `massive' $U(1)$ symmetry is the difference $\tilde S - \tilde U$ of the two sections defined over $\tilde\calc_{II}$. This $U(1)$ symmetry  is non-perturbatively broken by the described M2-instantons. 
By the same reasoning we predict that in this non-generic situation also the Weierstrass model exhibits non-perturbative $\mathbb Z_2$-torsion: the M2-brane with boundary $2 \, \tilde B_{II} $ must now stretch  away from $\tilde\calc_{II}$ into the 4-fold along a 3-chain with non-vanishing volume in the F-theory limit.

\section{Non-perturbative relations, massive $U(1)$s and D1-instantons at weak coupling} \label{sec:Weakcoupling}

In the previous sections we have discussed various examples of perturbative and non-perturbative relations. As already mentioned, in presence of non-perturbative relations the corresponding M2-instantons
should be interpretable as F1-D1 instantons from the IIB perspective. The relation with F1 worldsheet instantons has been discussed in the simple local models of section \ref{sec:localnp}.   We would now like to better understand the non-perturbative relations corresponding to D1-instantons. As we will see
the existence of such non-perturbative relations and of the associated M2-brane instantons depends on the global structure of the internal space,  in contrast to  the examples of section \ref{sec:localnp}. We will furthermore connect the non-perturbative homological relations to  St\"uckelberg massive $U(1)$s  in perturbative IIB theory. In order to prepare the ground, we will start with a detailed discussion of this material. Then, we will uplift it to F-theory and explain its link with the results of  section \ref{sec:general}, presenting also an explicit model to illustrate the general discussion.

\subsection{Axionic gauging and D1-instantons in Type IIB} \label{axgauging}

Consider a Type IIB orientifold compactification on a Calabi-Yau three-fold $\X$ with holomorphic orientifold involution $\sigma: \X\rightarrow \X$. 
The D7-branes organise in pairs (D7${}_a$, D7${}'_a$)  wrapping divisors $(D_a, D'_a)$ with $D'_a=\sigma_*(D_a)$, or appear as orientifold invariant D7$_\alpha$, wrapping divisors $\hat D_\alpha = \sigma_*(\hat D_\alpha)$. For simplicity, let us assume that the are no multiple D-branes.  The field-strengths on the D7-branes are denoted by $F^a$, $F^{a\prime}$ and $\hat F^\alpha$ respectively such that $F^{a\prime}=-\sigma^*(F^a)$ and $\sigma^*(\hat F^\alpha)=-\hat F^\alpha$. In particular $\hat F^\alpha$ is odd on $\hat D_\alpha$.

 It is well-known that, even in absence of bulk fluxes, the R-R gauge potential $C_2$ is gauged under D7 gauge transformations $A^a\rightarrow A^a+\d \lambda^a$,  $A^{a\prime}\rightarrow A^{a\prime}+\d \lambda^{a\prime}$ (with $\lambda^{a\prime}=-\sigma^*(\lambda^a)$) and $\hat A^\alpha\rightarrow 
\hat A^\alpha+\d \hat\lambda^\alpha$. The derivation of this result from a fully 10D perspective is reviewed in appendix \ref{app:gauging}.
 In particular, consider the case of gauge parameters that are constant on the internal divisors.  Hence the orientifold projection implies that $\lambda^{a\prime}=-\lambda^a$ and  $\hat\lambda^a=0$
 and the corresponding gauging is given in \eqref{RRshiftgeom}. 
 This gauging is the microscopic origin of the St\"uckelberg mechanism in the 4D effective theory. 
 The latter sees only the cohomological part of \eqref{RRshiftgeom}, which can be written in terms of Poincar\'e dual cycles as
\be
[C_2]\rightarrow  [C_2]+\frac{1}{2\pi}\sum_a\lambda^a\, \Big([D_a]- [D'_a]\Big) \label{axiongeom}
\ee
This gauging gives rise \cite{Jockers:2004yj,Plauschinn:2008yd} to what was called in \cite{Grimm:2011tb} the geometric St\"uckelberg mechanism  in the 4D effective theory. The corresponding $U(1)_a$ gauge field receives a mass if and only if $D_a$  and $D_a'$ are not homologous. 

The above gaugings imply that D-brane instantons coupling to $C_2$
 can induce operators in the effective action which by themselves are charged under the $U(1)$ symmetry \cite{Blumenhagen:2006xt,Ibanez:2006da,Florea:2006si,Haack:2006cy}.
The relevant instantons  are provided by Euclidean  D1-branes and fluxed D3-branes. We will analyse fluxed D3-instantons and their F/M-theory description in 
the companion paper \cite{paper2}. 
 For now let us focus on the contribution of D1-instantons wrapping a certain two-cycle $\Sigma\subset X$ which is odd under the orientifold involution, $\sigma_*\Sigma=-\Sigma$. 
There are $h^{1,1}_-(\X)$ such homologically inequivalent non-trivial 2-cycles, which may be represented as the difference of two effective  holomorphic cycles,
\be\label{SigmaCC}
\Sigma=\calc-\calc' \qquad  {\rm with}  \quad \calc'=\sigma_*\calc.
\ee
Such instantons could support a non-trivial  flux, in which case they should be rather thought of as D1-F1 bound states. However, the following discussion is not affected by such possible additional ingredients and we therefore focus for simplicity on a pure Euclidean D1-brane instanton, sometimes dubbed E1-brane. 
An instanton of this type can be thought of as a Euclidean D1-brane wrapping $\calc$ together with orientifold image along $\calc'$. Since the holomorphic involution compatible with O7/O3-planes maps the Euclidean ${\rm D1}$-brane to a Euclidean anti-D1 ($\overline{\rm D1}$) brane, we must include a minus sign for $\calc'$. Hence a D1-instanton on $\Sigma$ may be considered a D1-$\aDo$ bound state, which breaks all four supercharges preserved by the ${\cal N}=1$ supersymmetric O7-D7-background and thus contributes to the D-terms of the effective action.
This contribution is weighted by $e^{-S_{\rm D1}}$, with 
\be
S_{\rm D1}=\frac{\pi}{g_s}{\rm vol}(\Sigma)-\pi\ii\int_{\Sigma}C_2
\ee
(including a factor of $\frac12$ due to the orientifold projection), where the volume is measured in string frame. Crucially, note that the combination appearing in $S_{D1}$ is not
an ${\cal N}=1$ chiral superfield of the effective action, in agreement with the fact that this instanton is not half-BPS.

From \eqref{axiongeom} it follows that under the $U(1)_a$ gauge symmetries this factor transforms as
\be
e^{-S_{\rm D1}}\quad \rightarrow \quad e^{\ii q_a\lambda^a}e^{-S_{\rm D1}}
\ee
with quantised charges
\be
q_a= \Sigma\cdot D_a,
\ee
where we have used $\Sigma\cdot D_a=-\Sigma\cdot D'_a$. Therefore, as generally discussed in \cite{Blumenhagen:2006xt,Ibanez:2006da,Florea:2006si,Haack:2006cy}, such instantonic contributions can in principle give rise to interaction terms with a suitable combination of the chiral fields with total charges $q_a$. 

In order to see concretely how such interaction terms arise, consider first a D7-brane in flat space and zoom in the locally flat patch around one intersection point of a D7$_a$-brane and the 
 Euclidean D1-brane. Quantisation of  the open string stretching between the two D-branes shows that there is just one non-exited zero-mode state in the Ramond sector  \cite{Blumenhagen:2006xt,Ibanez:2006da,Florea:2006si}. 
 Suppose that $\Sigma$ intersects $D_a$ in $n_a^+$ points with poisitive intersection number and in $n_a^-$ points with negative intersection number. 
 At each of the $n_a^+$ intersection points, labeled by $i=1,\ldots, n_a^+$, a fermionic zero mode $\eta^{i}_{a}$ with charge $(1,-1)$ under $U(1)_{a}\times U(1)_{\rm D1}$ resides, while at each of the $n_a^-$ intersection points  a fermionic zero mode  $\tilde\eta^{\tilde\imath}_{a}$ with charge $(-1,1)$ under $U(1)_{a}\times U(1)_{\rm D1}$ localises. 
The total $U(1)_a$ charge of these zero modes is given by 
 \be
 q_a = n^+_a - n^-_a = \Sigma \cdot D_a.
 \ee
  This results in a total charge $-q_a$ of the zero-mode path-integral measure, which precisely compensates the charge of the factor $e^{-S_{\rm D1}}$. 
Note that the chiral index of charged modes at the intersection with orientifold-invariant branes vanishes.

 Suppose now that  the D1-brane instanton intersects two mutually intersecting D7-branes wrapping $D_a$ and $D_b$ at some common point contained in the matter curve $\calc_{ab}\equiv D_a\cap D_b$  with {\em opposite} intersection numbers. Then the D1 effective action can contain (gauge invariant) interaction terms  of the schematic form
 \be
 \eta_{a} \, \Phi_{ab} \, \tilde\eta_{b},
 \ee
 where $\Phi_{ab}$ is some matter field localised on $\calc_{ab}$.
 Analogously, terms of the form $ \eta_{a } \Phi_{ab'}\tilde\eta_{b'}$ can also be present when $\Sigma$ intersects $D_a$ and $D_{b'}$ on $\calc_{ab'}\equiv D_a\cap D_{b'}$ and with opposite intersection numbers. 
 
 Of course the presence of such terms appears to depend on the specific form of $\Sigma$ and then it does not seem topological in nature. In particular, a volume-minimising $\Sigma$,
 which extremises the exponentially suppressed contribution $e^{-S_{\rm D1}}$ to the path integral may not intersect two 7-branes at some common point. However, one has to recall that one has to integrate over all D1-configurations in the path-integral, which is of the schematic form
 \be
 \int \cald[\Sigma]\prod\d\eta \, \d\tilde\eta \, e^{-S_{\rm D1}+\ldots} .
 \ee  
 Hence, due to the presence of the charged zero-mode contribution $\prod\d\eta \, \d\tilde\eta$    in the path-integral measure, only configurations $\Sigma$ that not only have the proper intersection numbers discussed above but that also intersect the 7-branes at points located on matter curves can contribute to the path integral: Indeed only in this case can the appearance of the interaction terms  $ \eta_{a}\Phi_{ab}\tilde\eta_{b}$ and $ \eta_{a}\Phi_{ab'}\tilde\eta_{b'}$ for these specific configurations saturate the fermionic measure. The path-integral therefore `localises' on this restricted class of D1-configurations, among which the volume minimising ones will dominate as usual.

Take  for instance two D7-branes wrapping $D_1$ and $D_2$ and their orientifold images and  suppose that $\Sigma$ is as in (\ref{SigmaCC}), with
${\calc} \cdot D_1 =1$, ${\calc}' \cdot D_1 =0$ and ${\calc} \cdot D_2 =0$, ${\calc}' \cdot D_2 =1$.
This implies not only that $q_1=-q_2=+1$, but also  that there are precisely two fermionic zero modes $\eta_{1}$ and $\tilde\eta_{2}$ with opposite charges.
If $D_1$ and $D_2$ intersect along a non-trivial curve $D_1\cap D_2$, the instantonic path-integral localises on those two-cycles which intersect $D_1$ and $D_2$ at some common point of $\calc_{12}=D_1\cap D_2$. This point supports an interaction term of the form $\eta_{1} \, \Phi_{12} \,  \tilde\eta_{2}$ and then the path-integral gives a term proportional to
\be
\int\d\eta_{1} \, \d\tilde\eta_{2} \,e^{-S_{\rm D1}+\eta_{1} \, \Phi_{12} \, \tilde\eta_{2}}= \Phi_{12}\,e^{- S_{\rm D1}} .
\ee
Hence, we see that such an instantonic contribution creates a ``tadpole" for  $\Phi_{12}$ in the effective action, which can induce a decay of the particle $\Phi_{12}$ into nothing.  Of course this effect is damped by the exponential suppression $e^{-\frac{\pi}{g_s} {\rm vol}(\Sigma)}$, which is very large in the weak-coupling limit $g_s\ll 1$. To compute the precise form the operator in the effective action we must integrate also over the non-charged zero-modes of the instanton, which we are suppressing in the above expression. In particular, as stressed already above, since the D1-instanton is not half-BPS, this term cannot arise at the level of the superpotential, but rather as a D-term due to the appearance of uncharged fermionic zero modes beyond the universal $\theta^\alpha, \alpha=1,2$ required for an superpotential term. 
The same argument can be repeated for the case $q_1=-q_2=k$, $k> 1$, and then the D1-instanton can give rise an operator proportional to 
\be\label{opins}
(\Phi_{12})^k\,e^{-S_{\rm D1}}.
\ee

\subsection{D1 versus M2 instantons: local discussion}
\label{sec:D1FT}

In perturbative Type IIB orientifold compactifications with constant axio-dilaton $\tau= C_0 + \frac{\ii}{g_s}$ we have seen how a non-zero chiral index of charged chiral fermionic zero-modes obstructs the contribution to the path-integral of a D1-instanton intersecting just one D7-brane with non-trivial intersection number.
The same conclusion can be reached by considering the backreacted background and thus generalises to general F-theory compactifications, as follows.
The backreaction of a  D7-brane  induces a monodromy $\tau\rightarrow \tau+1$ around the D7-brane.
A global definition of $\tau$ requires  the introduction of branch-cuts associated with the $T$ generator of the non-perturbative Type IIB SL($2,\mathbb{Z}$) duality group
as in \cite{Gaberdiel:1997ud}. A D1-brane which passes through the $T$-cut is transformed into a $(1,1)$-string. Hence a D1-brane which transversely intersects a single D7-brane
is, by itself, not locally consistent. 

On the other hand, consider for instance a local patch $w,v,u$ of the IIB compactification 3-fold
in which a first D7-brane is located at $w=0$ while a second D7-brane is located at $v=0$. Locally around the intersection line $w=v=0$, the axio-dilaton has the form
\be
\tau(w,v,u)=\frac{1}{2\pi\ii}\log (wv)+\text{(regular)}.
\ee
Suppose a Euclidean D1-brane wraps the non-compact two-cycle $\cals$ specified by
$u=0$ and $w=\bar v$. It thus intersects both D7-branes at $u=w=v=0$, with opposite intersection number. If we  introduce a world-volume complex coordinate $\zeta$, the embedding is given by $w=\bar v=\zeta$ and the pull-back of $\tau$
to the D1-world-volume becomes 
\be
\tau_{\rm D1}(\zeta,\bar\zeta)=\frac{1}{2\pi\ii}\log |\zeta|^2+\text{(regular)},
\ee
 which does not have any monodromy. Thus, a D1-brane that meets two D7-branes at their intersection locus, with opposite intersection numbers, appears to be consistent. In fact, a fully consistent configuration is obtained by inserting an F1-string connecting the two D7-branes and localised at their intersection, filling an external semi-infinite word-line and  ending on the D1-instanton. The necessity of including such an F1-string corresponds to the required insertion of the charged operators $\Phi_{12}$ in (\ref{opins}).

This is maybe better understood by using the dual M-theory description, which provides  a nice geometrical realisation of what we have just described in IIB language. Under the duality the D1-instanton gets mapped to an M2-brane wrapping the 2-cycle submanifold $\Gamma$ obtained by fibering the elliptic fibre $B$-cycle over the non-compact 2-cycle $\cals$ in the base. Since by encircling the D7-brane the $A$- and $B$-cycles get mixed by  the monodromy $B\rightarrow B+A$, there is an obstruction to the existence of such a $\Gamma$ in the case in which $\cals$   transversely intersects a single D7-brane, see Fig.~\ref{fig:D1D7}.

\begin{figure}[h!]
  \centering
    \includegraphics[width=0.5\textwidth]{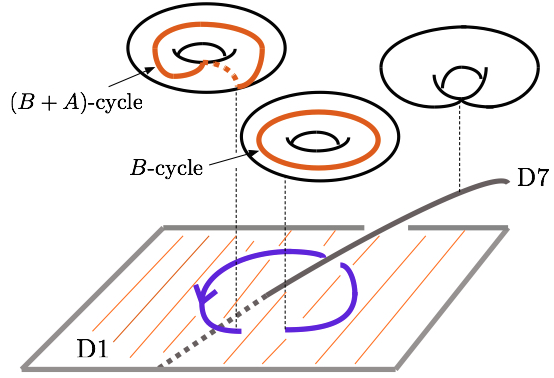}
      \caption{\footnotesize $T$-monodromies in the fibre obstruct the uplift of a Euclidean D1-brane intersecting a D7-brane.}\label{fig:D1D7}
\end{figure}

As above, consider in turn two intersecting D7-branes described  locally  by  D7$_1=\{w=0\}$ and D7$_2=\{v=0\}$. In the dual M-theory picture, such a system can be described by a local Weierstrass model in the form (\ref{ys0}) with $b_2=1-wv$, $b_4=-wv$ and $b_6=-wv$ \cite{Braun:2014nva}.
The base curve $w=v=0$ supports a family of conifold singularities  $y=s=w=v=0$. 
For later convenience let us rescale
$y \rightarrow -\varepsilon y$, $s \rightarrow -\varepsilon s$, $w \rightarrow -\varepsilon w$, $v \rightarrow -\varepsilon v$, so that (\ref{ys0}) becomes
\be \label{r1r2r3ell}
y^2 = s^2 - w \, v (1 - \varepsilon s )^2  + \varepsilon \, s^3 = (s-r_1) \, (s-r_2) \, (s-r_3)
\ee
with
\be
\begin{aligned}
r_1 = -\frac{1}{2} (\varepsilon \, w \, v + \sqrt{4 v w + \varepsilon^2 \, w^2 \, v^2}) 
\, ,\quad 
r_2 = -\frac{1}{2} (\varepsilon \, w \, v - \sqrt{4 v w + \varepsilon^2 \, w^2 \, v^2}) 
\, ,\quad 
r_3 =  \frac{1}{\varepsilon}
\quad .
\end{aligned}
\ee
As usual, the elliptic fibre can be represented by two copies of the complex $s$-plane glued together along two branch-cuts connecting, for instance, $r_1$ with $r_2$ and  $r_3$ with $\infty$. The $A$ and $B$-cycle can then be chosen as in Fig.~\ref{fig:elliptic}. Notice that, indeed, if we encircle say the D7$_1$ locus $w=0$ (for fixed $v\neq 0$), then the branch points $r_1$ and $r_2$ get exchanged. One recovers the monodromy $B\rightarrow B+A$ so that one cannot consistently fibre the $B$-cycle over a non-compact 2-cycle $\cals$ intersecting just D7$_1$, as we have already stressed. 

\begin{figure}[h!]
  \centering
    \includegraphics[width=0.5\textwidth]{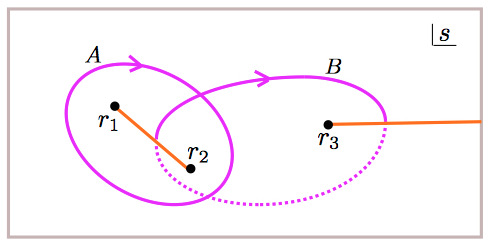}
      \caption{\footnotesize Branch-cuts and 1-cycles in the complex $s$-plane associated with the elliptic fibre (\ref{r1r2r3ell}). }\label{fig:elliptic}
\end{figure}

On the other hand, let us choose $\cals$ locally defined by $u=0$, $w=\bar v=\zeta$ as above so that it intersects both D7$_1$ and D7$_2$ at the same base point $w=v=u=0$ and with opposite intersection number. Then the elliptic fibration restricted to $\cals$ 
becomes
\be
y^2 = s^2 - |\zeta|^2 (1 - \varepsilon s )^2  + \varepsilon \, s^3 
\ee
with branch points 
\be
\begin{aligned}
r_1 = -\frac{1}{2}  |\zeta|(\sqrt{4  + \varepsilon^2 \, |\zeta|^2}+\varepsilon |\zeta| ) 
\, ,\quad 
r_2 =  \frac{1}{2}  |\zeta|(\sqrt{4  + \varepsilon^2 \, |\zeta|^2}-\varepsilon |\zeta| )
\, ,\quad 
r_3 =  \frac{1}{\varepsilon}
\quad .
\end{aligned}
\ee
We clearly see that now there is no non-trivial monodromy obstructing the $B$-cycle fibration  as we move in the $\zeta$-plane over $\cals$. We thus obtain a well-defined $\Gamma$. 

 \begin{figure}[h!]
  \centering
    \includegraphics[width=0.8\textwidth]{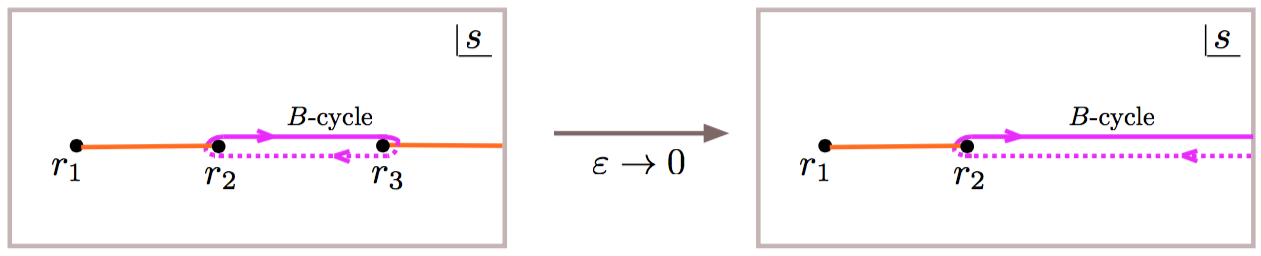}
      \caption{\footnotesize Local limit of elliptic fibre. }\label{fig:elliptic_dec}
\end{figure}

However, in order to understand  $\Gamma$ at the base point $w=v=u=0$, we need to resolve the ambient conifold singularity by a small resolution.  First of all, let us choose the $B$-cycle as in Fig~\ref{fig:elliptic_dec} (in which, for simplicity, we have assumed $\varepsilon$ to be real).  Since a small resolution is a local operation, we can zoom into the singularity by taking the limit $\varepsilon\rightarrow 0$.  In this limit, $r_1=-|\zeta|$, $r_2=|\zeta|$ while the third root $r_3$ disappears. The fibration takes the form of an $A_1$-fibration
\be \label{Weierlocal}
y^2 = s^2  - w \, v  . 
\ee
The $B$-cycle over $\cals$ becomes semi-infinite and can be parametrised as 
\be\label{3chainpara}
s=\sqrt{\alpha^2 + |\zeta|^2}\,,\qquad y=\alpha
\ee 
with $\alpha\in \mathbb{R}$, see Fig.~\ref{fig:elliptic_dec}. We then see that $\Gamma$ passes through the bulk conifold singularity at $\zeta=\alpha=0$. We can now perform a standard small resolution described by the complete intersection
\bea \label{localres1}
(s+y) \,  \lambda_1 = w \, \lambda_2, \qquad (s-y) \, \lambda_2 = v \, \lambda_1
\eea
with homogenous coordinates $[\lambda_1 : \lambda_2]$ parametrising a resolution $\mathbb P^1$. 
As a result, the former point $\alpha= \zeta = 0$ on $\Gamma$ is manifestly resolved into the full $\mathbb P^1$, so that $\mathbb P^1\subset-\del\Gamma$. In other words, at the base point $\zeta=0$ the small resolution opens up a $\mathbb P^1$-boundary on the local M2-brane which is dual to the Type IIB D1-instanton, see Fig.~\ref{fig:D1D7D7}.
 
 \begin{figure}[h!]
  \centering
    \includegraphics[width=0.5\textwidth]{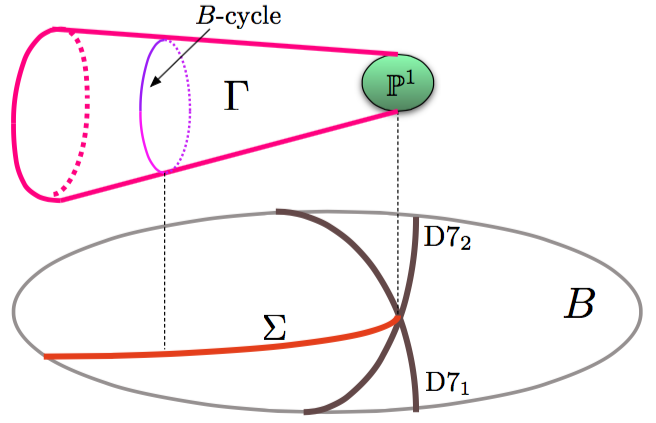}
      \caption{\footnotesize 3-chain $\Gamma$ ending on resolution $\mathbb P^1$. }\label{fig:D1D7D7}
\end{figure}
 
 Of course, such a Euclidean M2-brane instanton, having a boundary, is still inconsistent by itself. This parallels the fate of the D1-brane instanton in Type IIB, which by itself and without open-string insertions does not contrite to the path-integral. But a simple cure to this problem is to introduce a second, incoming physical M2-brane with worldvolume $\gamma^{\rm in}\times \mathbb P^1$, where $\gamma^{\rm in}$ is an incoming external world-line ending at the spacetime point of insertion of the M2-instanton on $\Gamma$.
 The boundaries of both  M2-branes cancel and the two objects glue perfectly together, forming a consistent configuration.  

Once one shrinks the  $\mathbb{P}^1$ to zero size and goes to the dual IIB description, the  time-like M2-brane describes a propagating single particle state of a chiral field $\Phi_{12}$ living at the intersection of the two D7-branes. Then the above semiclassical M2-configuration can be regarded as the M-theory counterpart of the D1-instanton dressed by open-string insertions, which produces the 4D interaction of the form  $\Phi_{12} \, e^{-S_{\rm D1}}$ in a compactified model.

 Let us stress  once again 
that the structure of  $\Gamma$ away from the curve $w=v=0$ depends on the global properties of the full fibration. 
This is evident if the global model admits a weak coupling limit. Here the associated D1-instanton 
must wrap a non-trivial element of $H^-_2(\X)$ on the double cover Calabi-Yau $X$ and then, clearly, its existence depends on the global properties of $X$. In particular, if the D1-instanton does not intersect other matter curves, 
 it must couple to a two-form axion in $H_-^2(\X)$ which is gauged under the relative $U(1)$ gauge field associated with the two D7-branes. This $U(1)$ is then massive. Hence, the corresponding chain $\Gamma$ must  cap off away from the matter curve so to trivialise the resolution $\mathbb P^1$ in the bulk, $\del\Gamma=-\mathbb{P}^1$. As a result, the local small resolution described above cannot be extended to a global  K\"ahler resolution. Indeed, this is the scenario envisaged in \cite{Braun:2014nva} to characterise the presence of a massive $U(1)$,
though our discussion shows that global information is needed to guarantee the closing of $\Gamma$ away from the matter curve. As we will see in the next section, the relation between the M-theory formulation 
and the expected weakly coupled IIB picture can be made precise by going to the stable degeneration limit, 
where one can explicitly see the relation between globally defined 3-chains
and odd classes in $H_2(\X)$.

\subsection{D1 versus M2 instantons: global discussion via the stable degeneration}
\label{sec:D1M2global}

We now discuss in more detail the global structure of the 3-chain $\Gamma$ locally constructed in the previous section.
To better illustrate the general ideas, we will apply them to a model with matter charged only under a massive $U(1)$ and no further gauge groups, which represents the perhaps  most economic F-theory model in which an M2-brane can violate a $U(1)$ selection rule. Such a model  has been introduced and analysed in \cite{Braun:2014nva}. The  elliptic fibration $\Y$ is given by a Weierstrass model in the form (\ref{ys0}) with the choice
\be \label{b2b4b6restr}
b_2 = a_1^2 + w \, a_{21}, \qquad b_4 = w \, \eta, \qquad b_6 = w \, \chi.
\ee
Here $w$ represents the section of a line bundle defining an effective divisor $\{w=0\}$  in the base and $a_{21}$, $\eta$, $\chi$ are generic sections of appropriate line bundles.
The factorisation of the discriminant 
\bea
\Delta:=27 g^2 + 4 f^3 = w \, \Delta'
\eea
indicates an $I_1$ singularity over $w=0$. The intersection $\{w=0\} \cap \{\Delta' = 0\}$ splits into the codimension-two loci
\be\label{C1C2a}
\calc_I = \{w=0\} \cap \{b_2 = 0\}, \qquad \calc_{II} = \{w=0\} \cap \{\chi = 0\}.
\ee
For maximally generic $a_1$, $a_{21}$, $\eta$ and  $\chi$,  $(f,g,\Delta)$ vanish to orders $(1,1,2)$ and $(0,0,2)$ along $\calc_I$ and $\calc_{II}$, respectively. The first indicates a type $II$ cuspidal singularity in the fibre over $\calc_I$, which does not correspond to a singularity of the elliptic fibration $\Y$; no massless matter is localised here. Over $\calc_{II}$, by contrast, an $I_2$ singularity of $\Y$ does indicate the existence of localised massless matter.
Since the massless gauge group of the F-theory model is trivial,  this matter is completely uncharged. Correspondingly, the $I_2$ singularity does not admit a crepant resolution \cite{Braun:2014nva}.

We now consider the weak-coupling of this model. First recall that, in general, the relation of the Weierstrass model with perturbative Type IIB theory is established by performing the well-known Sen limit \cite{sen1,sen2}
\bea \label{Senlimit1}
b_4 \rightarrow \epsilon \, b_4, \qquad b_6 \rightarrow \epsilon^2 \, b_6.
\eea
The resulting form of the discriminant 
\bea
\Delta \simeq \epsilon^2 \, b_2^2 \, (b_4^2 - b_2 \, b_6 ) + {\cal O}(\epsilon^3)
\eea
identifies $\{b_2=0\}$ as the location of the O7-plane and $\{b_4^2 - b_2 \, b_6 =0\}$ as the position of the 7-branes on the double cover Calabi-Yau 3-fold $\X$ of the base $\B$. This $X$ is described as the hypersurface $\xi^2 = b_2$ in a suitable ambient space. The orientifold involution $\sigma$ acts as $\xi \rightarrow - \xi$. 

For the model (\ref{b2b4b6restr}) the position of the 7-branes on $\B$ in the Sen limit factorises as $\{w \, (w \, \eta^2 - b_2 \chi)=0\}$. As argued in \cite{Braun:2014nva}, the uplift of $\{w=0\}$ to the double cover $\X$ splits into a non-homologous brane-image brane pair on
\be\label{D12}
D_1: \{w=0\} \cap \{\xi = a_1\}, \qquad D_1': \{w=0\} \cap \{\xi = - a_1\}
\ee
in the ambient space of $\X$, while the uplift of the brane locus $\{w \, \eta^2 - b_2 \chi=0\}$ describes an invariant 7-brane on the divisor
\be
D_2: \{w \,  \eta^2 - b_2 \chi=0\} \cap \{\xi^2 = b_2\}.
\ee
The generically cuspidal locus $\calc_I$ uplifts to the simultaneous intersection of $D_1$ and $D_1'$ with the O7-plane, while the matter locus $\calc_{II}$ uplifts to the intersection
$D_1 \cap D_2$ and its orientifold image $D_1' \cap D_2$.

To fully capture the essentials of the Sen limit geometrically we describe it as a stable degeneration as introduced in  \cite{Clingher:2012rg,Donagi:2012ts} and very clearly reviewed in
\cite{Braun:2014nva,Braun:2014pva}.
This construction promotes the F-theory Calabi-Yau 4-fold $Y$ to a family $Y_\epsilon$ of 
4-folds parametrised by the complex structure parameter $\epsilon$. The regime $\epsilon \rightarrow 0$ describes the Sen limit as in (\ref{Senlimit1}).
For varying $\epsilon$, $Y_\epsilon$
is described by the Weierstrass model 
\bea
y^2 - (s^3 + b_2 s^2 z^2 + 2 b_4 s \epsilon z^4 + b_6 \epsilon^2 z^6) =0
\eea
 and can be regarded as the fibre of a $(4+1)$-fold ${\cal Y}$. 
The singularity of the family ${\cal Y}$ at $\epsilon = 0$ is resolved by performing a blow-up $(s,y,\epsilon) \rightarrow (s \lambda, y \lambda, t \lambda)$ with Stanley-Reisner ideal $s \, y \, z, y \, s \, t, z\, \lambda$.
After this resolution, the central fibre $Y_0$, corresponding to the Type IIB zero-coupling point, splits into two components $Y_T$ ($t=0$) and $Y_E$ ($\lambda=0$), given respectively by
\begin{subequations}
\begin{align}
Y_T& :  \qquad y^2 - s^2( b_2 z^2 + s \lambda)=0, \\
 Y_E& : \qquad y^2 - (b_2 s^2 + 2 b_4 s t + b_6 t^2) = 0.
\end{align}
\end{subequations}
Both $Y_E$ and $Y_T$ are $\mathbb P^1$-fibrations over the base $\B$.
Note that the cubic term in $s$ is relegated completely to $Y_T$, while $Y_E$ contains only monomials up to quadratic order in $s$.
The resolved central fibre has the form 
\be
Y_0=Y_E\cup_{\X} Y_T .
\ee
$Y_0$ exhibits a normal crossing singularity at $\X=Y_E\cap Y_T$ (the so called ``normal crossing divisor"), which can be regarded as a $\mathbb{Z}_2$ double cover of $\B$ with branch-locus $b_2=0$, defining the O7-plane position. Indeed, $\X$ is given by the locus $Y_T|_{\lambda=0}$  or equivalently $Y_E|_{t=0}$, i.e.
\bea
\X: \qquad y^2 - s^2 b_2 = 0.
\eea
For each point in $\B$ this describes two points $y = \pm \sqrt{b_2} s$ exchanged by the $\mathbb Z_2$-monodromy as one encircles the O7-plane at $b_2=0$. Hence, remarkably, $X$ is precisely  the Calabi-Yau double cover of $\B$ on which the weakly-coupled orientifold IIB theory is defined.

The $\mathbb P^1$-fibration $Y_T$ is nowhere degenerate. By contrast,
the information about the D7-brane system, which in the perturbative limit is localised at the vanishing locus of $\Delta_E = b_4^2 - b_2 \, b_6$, is encoded in $Y_E$, whose fibre locally splits into two $\mathbb P^1$s over $\{\Delta_E = 0\}$,
\bea
Y_E|_{\Delta_E = 0, b_2 \neq 0}: \qquad    (y+ \sqrt{b_2} s + \frac{b_4}{\sqrt{b_2}} t)(y- \sqrt{b_2} s - \frac{b_4}{\sqrt{b_2}} t) = 0.
\eea
The two fibre components are however exchanged by a $\mathbb Z_2$ monodromy around $\{b_2=0\}$ due to the branch-cut of the square-root.

If we specialise this general discussion to the massive $U(1)$ model (\ref{b2b4b6restr}) then we again see from
\bea
Y_E|_{w=0}: \qquad    (y+ a_1\,  s) \, (y - a_1 s) = 0
\eea
the appearance of two rational fibres over $\{w=0\}$. These represent the brane-image brane pair $D_1$ and $D_1'$ on $\X$, which are exchanged as one encircles the orientifold such that $b_2 \rightarrow e^{2 \pi \ii} b_2$ and thus $a_1 \rightarrow - a_1$. 
Of special interest for us is the $I_2$ singularity at the point $\{y=0\} \cap \{s=0\}$ in the fibre of $Y_E$ over the matter curve $\calc_{II}  = \{w=0\} \cap \{ \chi=0\}$. This singular point in the fibre is precisely the intersection point of the two $\mathbb P^1$ fibre components over $\calc_{II}$. 
As discussed in great detail in \cite{Braun:2014nva}, thanks to the absence of the cubic term $s^3$, the fibration $Y_E$ can globally be brought into conifold form
\bea
(y+a_1 s) (y-a_1 s) = w \, ( a_{21} \, s^2 + 2 \eta \, s \, t + \chi \, t^2). 
\eea
This singularity admits  a standard small resolution described as the complete intersection
\bea \label{hatWE}
\hat Y_E:  \qquad  (y+ a_1 s) \,  \lambda_1 = w \, \lambda_2, \qquad (y- a_1 s) \,  \lambda_2 =  ( a_{21} \, s^2 + 2 \eta \, s \, t + \chi \, t^2) \, \lambda_1
\eea
with $[\lambda_1 : \lambda_2]$ homogeneous coordinates of the resolution 2-cycle $\mathbb P^1_E$ replacing the singular conifold point. In this respect the situation for $Y_E$ is very similar to the local model in the limit $\varepsilon \rightarrow 0$ and its small resolution considered in the previous section.

After these preliminaries we are now in a position to examine more closely the structure of chains and cycles on $Y_0$ relevant for our M2-instantons.
The homology of $Y_0$ can be deduced from the homology of $Y_E$ and $Y_T$ with the help of the Mayer-Vietoris exact sequence. We review this powerful machinery in appendix \ref{app-MYCS}, in which  we show that  \be \label{H2Y0}
H_2(Y_0)\simeq  V_2(Y_0) \equiv \frac{H_2(Y_E)\oplus H_2(Y_T)}{  \{\iota_{E*}\Sigma\oplus -\iota_{T*}\Sigma|\Sigma\in H_2(\X)\}}
\ee
and we can think of    $H_3(Y_0)$ as the sum of the two spaces $V_3(Y_0)$ and $T_2(Y_0)$ with
\be \label{H3Y0b}
\begin{aligned}
V_3(Y_0)&\equiv  \frac{H_3(Y_E)\oplus H_3(Y_T)}{  \{\iota_{E*} \Gamma\oplus -\iota_{T*}\Gamma|\Gamma\in H_3(\X)\}}\,,\\
T_2(Y_0)&\equiv \{\Sigma\in H_{2}(\X)| \iota_{E*} \Sigma=0\text{ and }\iota_{T*}\Sigma=0\}\,.
\end{aligned}
\ee
Here $\iota_T: \X \rightarrow Y_T$ and  $\iota_E: \X \rightarrow Y_E$ denote the inclusion maps for $\X$ inside $Y_T$ and $Y_E$, respectively. 
The meaning of (\ref{H2Y0}) is that the  non-trivial  2-cycles in $Y_0$ are  generated by the 2-cycles in  $Y_E$  and   $Y_T$ with pairs of 2-cycles in  $Y_E$  and   $Y_T$ considered equivalent if they are both  homologous to the same 2-cycle $\Sigma \in H_2(\X)$ to avoid overcounting. In particular, consider a non-trivial 2-cycle $y_E$ in $Y_E$ which is homologous to a non-trivial 2-cycle in $\X$, which in turn is trivial in $Y_T$. Such 2-cycle  $y_E$ must be considered trivial in $Y_0$ since it can be trivialised in $Y_T$ by passing through $\X$. Of course, the same conclusion holds for 2-cycles in $Y_T$ which can be trivialised in $Y_E$ by passing through $\X$.
Since $Y_T$ is a nowhere   degenerate $\mathbb{P}^1$ bundle, $H_2(Y_T)$ is generated by the rational fibre class and the push-forward of the 2-cycles in the base by the global section $u_T:\B\rightarrow Y_T$ defined by $z=0$, $y=s=\lambda=1$. Within $H_2(Y_E)$, one identifies as an additional potential contribution the elements of $H_2(Y_E)$ localised at the singularities. We will discuss an explicit example of such contribution in section \ref{sec:cohoSD}. On the other hand, one must check that such elements do not trivialise in $Y_T$. As we will see momentarily, in the massive $U(1)$ model the resolution $\mathbb P^1_E$ within $\hat Y_E$ in (\ref{hatWE}) is an example of such a 2-cycle in $H_2(\hat Y_E)$ trivialised within $Y_0$.

Before turning to this question, let us first discuss (\ref{H3Y0b}).
Consider in particular $T_2(Y_0)$ and notice that the subspace $\{\Sigma\in H_{2}(\X)| \iota_{T*}\Sigma = 0\}$ can be identified with the space of 2-cycles $\Sigma\in H^-_2(\X)$ that are \emph{odd} under the orientifold action $\sigma: X \rightarrow X$. In order to see this, take any $\Sigma \in H^-_2(\X)$ and write it as
\be
\Sigma=\calc-\calc'
\ee
where $\calc,\calc'\in H_2(\X)$ and $\calc'=\sigma_*\calc$. Then, by considering $\X$ as a $\mathbb{Z}_2$ double cover over $\B$ with projection $p: \X \rightarrow \B$, both $\calc$ and $\calc'$ project to the same 2-cycle  $\cals$  in the base
\be
\cals=p_*\calc=p_*\calc'.
\ee
We can now construct a 3-chain $\Gamma_T\in Y_T$  trivializing $\Sigma$ as follows. Take any point   $q\in \cals$. This uplifts to two points $f_\pm=\{y=\pm \sqrt{b_2}\, s \}\in \X$, where $f_+\in \calc$ and $f_-\in \calc'$, which can be regarded as two points in the $\mathbb{P}^1$-fibre of $Y_T$ over $q$. Since the fibre is never degenerate we can construct a 1-chain $\gamma_T$ inside the $\mathbb{P}^1$-fibre over $q$ such that
\be
\del\gamma_T=f_--f_+\,.
\ee
The 3-chain $\Gamma_T$ is then obtained by fibering $\gamma_T$ over any point $q\in \cals$, so that
\be
\del\Gamma_T=\calc'-\calc= -\Sigma\,.
\ee
Consistently, if $\cals$ is transported around the orientifold locus, then $\cals\rightarrow -\cals$ but at the same time $f_+\leftrightarrow f_-$ so that there is no change of orientation on $\Gamma_T$, which  is then globally well defined, see Fig.~\ref{fig:Gamma_T}.
Notice that for this construction to work it is crucial that we start from an odd 2-cycle $\Sigma\in H^-_2(\X)$. Indeed, consider instead an {\em even} 2-cycle $\Sigma\in H^+_2(\X)$. This can be pushed-forward to a non-trivial 2-cycle $p_*\Sigma$ in the base. On the other hand, the projection $p: \X\rightarrow \B$ can be seen as the restriction to $X$ of the projection $\pi_T:Y_T\rightarrow \B$, so that $\pi_{T*}\iota_{T*} \Sigma=p_*\Sigma$. Since $p_*\Sigma$  is non-trivial  in $H_2(\B)$, $\iota_{T*} \Sigma$ cannot be trivial in $H_2(Y_T)$.

 \begin{figure}[t!]
  \centering
    \includegraphics[width=0.55\textwidth]{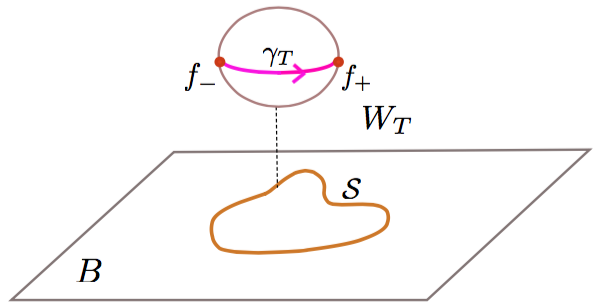}
      \caption{\footnotesize Fibering $\gamma_T$ over ${\cal S}$ yields a 3-chain $\Gamma_T$ in $Y_T$.}\label{fig:Gamma_T}
\end{figure}

We have therefore found that  $T_2(Y_0)\subset H^-_2(\X)$.  On the other hand, in order for $\Sigma\in H^-_2(\X)$ to contribute to $T_2(Y_0)$, and then to $H_3(Y_0)$, one must require that $\Sigma$ trivialises in $Y_E$ as well. This is generically possible only if $\cals$ does not intersect the 7-brane locus $\Delta_E=0$. Indeed, in this case, by restricting $Y_E$ to $\cals$ one gets a nowhere  degenerate $\mathbb{P}^1$-bundle and one can construct a trivialising 3-chain $\Gamma_E$ as done for $\Gamma_T$. 
This is achieved by fibering over $\cals$ a 1-chain $\gamma_E$ in the $\mathbb{P}^1$-fibre of $Y_E$ such-that
\be
\partial \gamma_E = f_+- f_-\,.
\ee
Clearly $\del\Gamma_E=\Sigma$ and then, gluing these two 3-chains together results in a non-trivial 
3-cycle $\Gamma_T+\Gamma_E$ in $Y_0$. An M2-brane wrapping $\Gamma_T+\Gamma_E$ represents the uplift of a D1-instanton wrapping the odd 2-cycle $\Sigma\subset X$, which is not charged under any D7-brane $U(1)$ gauge symmetry.

Instead, if $\cals$ intersects  $\Delta_E=0$ across a single D7-brane, the construction of  the  chain $\Gamma_E$ is obstructed.   The obstruction is due to the impossibility of consistently fibering the 1-chain $\gamma_E$ around the  D7-brane, similarly to what has been discussed in section \ref{sec:D1FT} concerning the obstruction in fibering the $B$-cycle around an $I_1$ locus. Indeed, locally (both in the base and in the fibre) around the singular locus, $Y_E$ and the complete elliptic fibration $\Y$ as well as  $\gamma_E$ and the $B$-cycle can be practically identified.

Finally, the main point of this discussion is what happens when  $\Sigma\in H^-_2(\X)$ intersects a matter curve in $\X$. In this case, for suitable  intersection numbers on $\Sigma$ with this curve, one can construct a 3-chain $\Gamma_E$ on $Y_E$ which passes through the singularity over the uplift of the matter curve. This $\Gamma_E$ appears as a proper trivialising 3-chain of $\Sigma$ inside $Y_E$, but once the singularity is resolved, $\Gamma_E$ can develop a second boundary on a combination of resolution 2-cycles in $\hat Y_E$.
In such a case, $\Gamma_E+\Gamma_T$ constitutes a 3-chain which trivialises the combination of resolution 2-cycles in $Y_0$. As explained above, such combination of resolution 2-cycles does not contribute to $H_2(Y_0)$.

For the massive $U(1)$ model, this scenario precisely applies to the resolution 2-cycle $\mathbb P^1_E$ over the matter curve $\calc_{II}$ in $\hat Y_E$.
Recall that in $Y_E$ we are fibering over $\cals$ the 1-chain $\gamma_E$ connecting the point $f_+$ to the point $f_-$.   Over the matter curve $\calc_{II}$ these two points are given by $y \pm a_1 = s = 0$.
Therefore prior to the resolution of $Y_E$ the 1-chain $\gamma_E$ in the fibre over $\calc_{II}$ passes from the two points $f_\pm$ to the singular point $y=s=0$. When this point is resolved into $\mathbb P^1_E$, the latter turns into a boundary of the 3-chain $\Gamma_E$. This is exactly the same phenomenon described for the local 3-chain $\Gamma$ given in (\ref{3chainpara}), which ends on the resolution $\mathbb P^1$ appearing in (\ref{localres1}). The non-compact 2-cycle in the base considered there is the local version of $\cals$.
What we see in the stable degeneration limit is that $\Gamma_E$ has two boundaries in $\hat Y_E$, one on $X$ and one on $\mathbb P^1_E$: 
\be
\del\Gamma_E=\Sigma-\mathbb P^1_E\,.
\ee
Moreover, while the resolution 2-cycle $\mathbb P^1_E$ is indeed non-trivial in $H_2(\hat Y_E)$ as stressed in \cite{Braun:2014nva}, we find that it is in fact trivialised within $Y_0$ through the 3-chain 
\be
\Gamma\equiv \Gamma_E +\Gamma_T\,,
\ee
 since 
 \be
 \del\Gamma =-\mathbb P^1_E\,.
 \ee

An M2-brane wrapping the 3-chain $\Gamma$ uplifts a ${\rm D}1$ instanton on the odd 2-cycle $\Sigma$ in $\X$.
Crucial for the existence of this chain -- and for the contribution of the ${\rm D}1$ instanton -- is that on $\X$, $\Sigma$ indeed intersects the two 7-branes meeting at the matter curve with opposite intersection numbers. As stressed several times, it is only in this case that the obstruction to the formation of a 3-chain due to the monodromies in the fibre can be overcome. In subsection \ref{subsec:explicitD1} we will explicitly construct the ${\rm D}1$-instanton in a global realisation of this model.

\subsection{The dual picture: (massive) $U(1)$s and axions via stable degeneration}
\label{sec:cohoSD}

What we  have just described has a dual formulation in terms of $U(1)$s  and axions, massive or not. Indeed, by using the cohomological Mayer-Vietoris exact sequence, see appendix \ref{app-MYCS}, one can compute the cohomology groups on $Y_0$ in terms of those of $Y_T$, $Y_E$ and $\X$. The result is 
\be
H^2(Y_0)\simeq V^2(Y_0)\equiv\{\omega_E\oplus\omega_T \in H^{2}(Y_E)\oplus H^{2}(Y_T)| \iota^*_{E}\omega_E=\iota^*_{T}\omega_T  \}
\ee
and $H^3(Y_0)$ can be thought of as the sum of  $V^3(Y_0)$ and $T^2(Y_0)$
with 
\begin{subequations}\label{GRH3}
\begin{align}
V^3(Y_0)&\equiv\{\omega_E\oplus\omega_T\in H^{3}(Y_E)\oplus H^{3}(Y_T)| \iota^*_{E}\omega_E=\iota^*_{T}\omega_T  \}\, ,\label{GRH3a}\\
T^2(Y_0)&\equiv\frac{H^{2}(X)}{\{\omega_X\in H^{2}(X)| \omega_X=\iota^*_{E}\omega_E-\iota^*_{T}\omega_T\}}\, .\label{GRH3b}
\end{align}
\end{subequations}

Regarding $H^2(Y_0)$, let us focus on the classes associated with non-trivial elements $\omega_E\in H^{2}(Y_E)$. They can contribute to  $H^2(Y_0)$ if either $\iota^*_E\omega_E=0$ in $H^{2}(X)$ or if a non-trivial $\iota^*_E\omega_E\in H^{2}(X)$ can be extended to $Y_T$.
Notice that this second possibility requires  $\omega_X\equiv\iota^*_E\omega_E$ to be an orientifold {\em even} class in $\X$. Indeed, any even class $\omega_X\in H^2_+(X)$ can be written as $\omega_X=p^*\omega_B$, where $\omega_B\in H^2(\B)$. Then $\omega_T\equiv\pi_T^*\omega_B$ defines a non-trivial class in $Y_T$ and provides the extension of $\omega_X$ to $Y_T$. Since all elements of $H^{2}(Y_T)$ are of this form, up to the 
the class  $[U_T]$  Poincar\'e dual to the global section $u_T: \B\rightarrow Y_T$, we see that odd 2-cocycles on $\X$ cannot be extended to $Y_T$. This argument also shows that the possible contributions to $H^2(Y_0)$ are  basically given by $[U_T]$ and by the elements of $H^{2}(Y_E)$ which satisfy one of the above conditions.

An instructive example is obtained in a $U(1)$ restricted model \cite{Grimm:2010ez}, which in the parametrisation used in the present paper corresponds to setting $b_6=a_3^2$. In this case $Y_E$ becomes 
\be
(y+a_3 t)(y-a_3 t)=s(b_2s+2b_4 t)\,.
\ee
Then, as discussed in \cite{Braun:2014nva}, one can construct an element  
$[S_+-S_-]\in H^2(Y_E)$
via Poincar\'e duality from the effective divisors
\be\label{WES}
S_\pm=\{y\pm a_3 t=0, s=0\}\,.
\ee  
The pull-back of  $[S_+-S_-]$ to $\X$ is just given by the Poincar\'e dual of the restricted divisor $(S_+-S_-)|_\X$ in $\X$. But 
$S_\pm|_\X$ are obtained by imposing the additional condition $t=0$ on (\ref{WES}). Since $\{y=s=t=0\}$ is not part  $Y_0$, being in the Stanley-Reisner ideal of the ambient geometry, $S_\pm|_\X=0$ in homology and then $[S_+-S_-]$ pulls-back to a trivial class in $H^2(\X)$. By the above discussion, this implies that $[S_+-S_-]$ contributes to $H^2(Y_0)$ and indeed can be identified with a massless $U(1)$ of the complete model. 
Furthermore notice that the resolved $\hat Y_E$ acquires a fibral $\mathbb{P}^1_E$, which contributes to $H_2(Y_E)$, as can be seen from its non-vanishing intersection with  $S_+-S_-$. Such  $\mathbb{P}^1_E$ cannot be regarded as the push-forward of a non-trivial 2-cycle in $\X$, since otherwise it should non-trivially intersect with  $(S_+-S_-)|_\X$,  which is not possible. Hence, this $\mathbb{P}^1_E$ contributes to $H_2(Y_0)$, according to the discussion  following eq.~(\ref{H3Y0b}). 

Consider now the massive $U(1)$ model reviewed in section \ref{sec:D1M2global}. As discussed in \cite{Braun:2014nva}, $Y_E$ can still be written in a conifold-like form, which allows us to identify the divisors 
\be\label{WES2}
\tilde S_\pm=\{y=\pm a_1 s\}\cap \{w=0\}\,.
\ee 
These define a non-trivial element  $[\tilde S_+-\tilde S_-]\in H^2(Y_E)$ \cite{Braun:2014nva}, as in the restricted $U(1)$ model.
However, differently from that case, if we now restrict these divisors to $\X$ by setting $t=0$, one still gets non-trivial divisors $D_1\equiv \tilde S_+|_\X$ and $D'_1\equiv S_-|_\X$ in $\X$. Furthermore, as the notation suggest, $D_1$ and $D'_1$ exactly represent the pair of  D7-branes under the orientifold action characterising this model, see (\ref{D12}), so that by construction  $[D_1-D'_1]$ is a non-trivial element of $H^2_-(\X)$ \cite{Braun:2014nva}. As discussed above, this is exactly the case in which an element of $H^2(Y_E)$ {\em does not} contribute to $H^2(Y_0)$, and then cannot be identified with a massless $U(1)$. 

From  (\ref{GRH3}) one actually extracts more information on the fate of $[D_1-D'_1]$ in the massive $U(1)$ model. Indeed, from 
(\ref{GRH3}) we see that part of $H^3(Y_0)$ can be roughly identified with a subset of $H^2(X)$. More precisely, 
an element  $\omega_X\in H^2(X)$ contributes to $H^3(Y_0)$ if it  cannot be written as  $\omega_X=\iota^*_{E}\omega_E-\iota^*_{T}\omega_T$. By the above discussion, we see that such $\omega_X$ is necessarily {\em odd}, since an even 2-cocycle can be written as  $\iota^*_{T}\omega_T$ for some $\omega_T\in H^2(Y_T)$. This necessary condition has a clear  weakly-coupled IIB interpretation. Indeed,  odd 2-cocycles in $\X$  correspond to  $C_2$ axions in $\X$, which can potentially uplift to $C_3$ axions in F/M-theory. On the other hand,  an odd 2-form $\omega_X\in H^2(X)$ that can be written as $\omega_X=\iota^*_{E}\omega_E$, for some $\omega_E\in H^2(Y_E)$, still does not contribute to $H^3(Y_0)$. An example of such odd form is provided precisely  by the class $[D_1-D'_1]\in H^2(\X)$  just discussed for the massive $U(1)$ model.

This perfectly parallels the IIB discussion of section \ref{axgauging}. The fact that $[D_1-D'_1]$ defines a non-trivial odd class in $X$ implies that the geometric St\"uckelberg is at work. Indeed,  $[D_1-D'_1]$ extends to a non-closed 2-cochain ${\rm w}_2$ in $Y_0$ (which restricts to the 2-cocycle  $[\tilde S_+-\tilde S_-]$ in  $Y_E$) and then  corresponds to a  {\em massive} $U(1)$. Furthermore, we expect the existence of a gauged axion which is `eaten' by this massive $U(1)$. Such axion is precisely given by  the trivial element of $H^3(Y_0)$ associated with $[D_1-D'_1]\in H^2(\X)$ by the Mayer-Vietoris exact sequence  as in (\ref{GRH3b}), i.e.\ it is given by a 3-coboundary $\alpha_3$ in $Y_0$ such that $\alpha_3=\d {\rm w}_2$. 

The weakly-coupled IIB R-R axion can be then identified with the odd 2-form $c_2\equiv [D_1-D'_1]\equiv{\rm w}_2|_\X$ and is naturally paired with the odd 2-cycle $\Sigma$ which, as discussed in section \ref{sec:D1M2global}, is homologous to the resolution 2-cycle $\mathbb{P}^1_E$. The latter in turn is trivialised by the 3-chain $\Gamma$ in $Y_0$. We then have the sequence of identities  
\be
\int_\Sigma c_2=\int_{\mathbb{P}^1_E}{\rm w}_2=-\int_{\Gamma}\alpha_3\,.
\ee
This provides an explicit realisation of the strict interrelation between massive $U(1)$s, axions, brane instantons and charged matter insertions that has been one of the main themes of the present paper.

A final comment: Physical consistency would require the existence of an appropriate extension away from the weak-coupling limit of the above results obtained in the stable degeneration limit. 
In fact, there is a mathematical machinery, the so-called Clemens-Schmid exact sequence (see for instance \cite{Donagi:2012ts} for a brief review), that formalises  the relation between the (co)homology of the central fibre and that of the generic fibre. Indeed, its application to our setting confirms our physical expectations.
However, the proof of the  Clemens-Schmid exact sequence assumes a K\"ahler resolution of the singularities over the matter curve in $Y$, which is not generically available in the present situation. 
Thus, strictly speaking,  the  Clemens-Schmid exact sequence cannot  always be invoked in our setting, even though the above physical argument suggests a more general validity of (parts of) it or at least the existence of an appropriate generalisation thereof.

\subsection{Realisation of  M2/D1-instantons in a concrete model} \label{subsec:explicitD1}

In this section we illustrate the above general discussion by an explicit realisation.
We begin by noting that the model (\ref{b2b4b6restr})  can be engineered as a Tate model (\ref{PTdef}) with a `split $I_1$ singularity' over $\{w=0\}$ induced by specifying
\be\label{Tsplit}
a_2 = a_{21} \, w, \qquad a_3 = a_{31} \, w, \qquad a_4 = a_{41} \, w, \qquad a_6 = a_{61} \, w.
\ee 
The standard relation between the Weierstrass model and the Tate model,
\be
b_2 = \frac14a_1^2 + a_2, \qquad b_4 = \frac14a_1 a_3 + \frac12 a_4, \qquad   b_6 =\frac14 a_3^2 +  a_6,
\ee
yields 
 \be \label{etachi}
\eta = \frac{1}{4} (a_1 a_{31} + 2 a_{41}), \qquad \chi = \frac{1}{4} (a_{31}^2 w + 4 a_{61}).
\ee
Generically, such models do not admit a well-defined Sen limit because of the appearance   of conifold singularites  in the associated Calabi-Yau $3$-fold $\X$ {\cite{Donagi:2009ra,Krause:2012yh,Esole:2012tf}. However, this problem can be avoided for appropriate choices of the base. In this section we focus on a specific example satisfying this condition.
The base $\B$ is given by the toric hypersurface defined in table \ref{tab:scalingB}, which has anti-canonical class $\bar K=P+U$.    
This space has been constructed in \cite{Blumenhagen:2009up}. It is a $\mathbb Z_2$ quotient of a Calabi-Yau 3-fold $\X$ related via a double del Pezzo transition to the quintic Calabi-Yau hypersurface (see table \ref{tab:scalingX}). The $\mathbb Z_2$ involution acts by $w_1 \leftrightarrow w_2$ and $v_1 \leftrightarrow v_2$ on $\X$. Its fix-point locus  is the divisor $\{v_1 w_1 - v_2 w_2 = 0\}$ in class $H+U+V$. To identify $\B$ as the quotient of $\X$ one substitutes $v \rightarrow v_1 \, v_2$, $w \rightarrow w_1 \, w_2,$ $h \rightarrow v_1 \, w_1 +v_2 \, w_2$.

\begin{table}
\centering
\begin{tabular}{c||cccccc}
 & $u_1$ &  $u_2$   & $u_3$  &  $v $ & $h$  & $w$ \\
\hline
\hline
$ P$  & 1 & 1 & 1 & 2 & 1 & 0  \\
$U$   & 0 & 0 & 0 & 1 & 1 & 1 
\end{tabular}
\caption{The base space $\B$ is defined as a generic hypersurface of degree $(5,2)$ with respect to divisor classes $(P,U)$ in the above toric ambient space  \cite{Blumenhagen:2009up}. The intersection numbers on $\B$ are $P^2 U = 2$, $P U^2 = -1$, $U^3= -1$.}\label{tab:scalingB}
\end{table}

One can further specify the split $I_1$ model on $\B$ such that $\X$ represents its Calabi-Yau double cover in the weak-coupling limit. To this end we  pick $a_1$ to be generic and parametrise it as
\be
a_1 = \alpha \, h + w \, p_1(u),
\ee
 where $\alpha$ is a constant and $p_1(u)$ is a generic polynomial of degree $1$ in $(u_1, u_2, u_3)$. Furthermore we take 
\be
a_{21} = -\frac{1}{2} \alpha \, h \, p_1(u) - \alpha^2 \, v - \frac{1}{4} p_1(u)^2 \, w.
\ee
This non-generic form of $a_{21}$ implies that, at weak coupling,  the O7-plane  locus $\{b_2\equiv \frac14 \alpha^2(h^2-4 wv)=0 \}$ in $\B$ uplifts to the $\mathbb{Z}_2$ fix-point locus $\{v_1 w_1 - v_2 w_2 = 0\}$ in $\X$. The remaining coefficients  $a_{n1}$ for $n=3,4,6$ in (\ref{Tsplit}) are taken to be the most generic polynomials of degree $n P + (n-1) X$ on the base $\B$. It is useful to parametrise $a_{61}$ as 
\be
a_{61} =\beta h^4 \, v \, + h^5 \, r_1(u) +  {\cal O}(w) ,
\ee
where $\beta$ is another constant and $r_1(u)$  a generic polynomial of degree 1. With this parametrisation 
 the two loci (\ref{C1C2a}) over which the fibre singularity enhances are given by
\be\label{singcurves}
\calc_I = \{w=0\} \cap \{h=0\}, \qquad  \calc_{II} = \{w=0\} \cap \{h \, r_1(u) + \beta v=0\}.
\ee

In order to better interpret these curves, let us now investigate in more detail the weak-coupling limit of this model \cite{Blumenhagen:2009up} -- see section \ref{sec:D1M2global}. In this limit the D7-branes are located at 
\be\label{Sen-model1}
\Delta_E=w  \, [w \,  \eta^2 - \frac{\alpha^2}{4}(h^2 - 4 v \, w) \,  \chi  ]=0
\ee
with $\eta$ and $\chi$ as in (\ref{etachi}). Then, we recognise the presence of two D7 branes, one at $w=0$ and one at $w \,  \eta^2 - \frac{\alpha^2}{4} (h^2 - 4 v \, w) \,  \chi=0$, on the base $\B$. These uplift in $\X$ to, respectively,  
 a D7-brane and its orientifold image along divisors $D_1$ and $D'_1$,  and an invariant  Whitney-type brane along a divisor $D_2$, with
\bea
 D_1: \{w_1 =0\},  \qquad  D'_{1}:  \{w_2=0\}, \qquad  D_{2}: \{P_{8,7,7}(u, v, w) =0\},
\eea
where $P_{8,7,7}(u, v, w)$ is a certain polynomial of degrees $(8,7,7)$. Then the $U(1)$ gauge group associated with the difference of the gauge groups on $D_1$ and $D'_{1}$ remains as a massive $U(1)$, while their sum is projected out. At $D_1 \cap D_2$ massless matter with $U(1)$ charges $\pm 1$ is localised. No perturbative matter is localised at the intersection of $D_1$ (or $D_1'$) with the O7-plane in class $H + U + V$, nor at the intersection of $D_2$ with the O7-plane.

\begin{table}
\centering
\begin{tabular}{c||ccccccc}
 & $u_1$ &  $u_2$   & $u_3$  &  $v_1 $ & $v_2$  & $w_1$ & $w_2$ \\
\hline
\hline
$ H$  & 1 & 1 & 1 & 1 & 1 & 0 & 0  \\
$U$   & 0 & 0 & 0 & 0 & 1 & 1 & 0  \\
$V$   & 0 & 0 & 0 & 1 & 0 & 0 & 1   \\
\end{tabular}
\caption{The Calabi-Yau 3-fold $\X$ is defined as a generic hypersurface of degree $(5,2,2)$ with respect to divisor classes $(H,U,V)$ in the above toric ambient space  \cite{Blumenhagen:2009up}. The non-zero intersection numbers on $\X$ are $H^2 U = 2$, $H U^2 = -2$, $U^3= 2$, $H U V =1$, $U^2 V = -1$ and those obtained by replacing $U \leftrightarrow V$.}\label{tab:scalingX}
\end{table}

Coming back to the curves (\ref{singcurves}) on $\B$,  $\calc_{II}$ corresponds to the intersection of  $D_1$ and $D_2$  away from the O7-plane on $\X$ and indeed it supports an $I_2$-singularity that indicates the expected massless matter. Over the curve $\calc_I$ the fibre exhibits a split $I_0^*$ singularity (associated with the algebra ${\mathfrak g}_2$) as indicated by the vanishing orders $(1,1,2,2,3;6)$ of  $(a_1, a_2, a_3, a_4, a_6; \Delta)$. The appearance of this singularity in $\Y$    is a consequence of the non-generic choice of $a_{21}$ as well of the form of  $a_{31}$, $a_{41}$, $a_{61}$ forced upon on us on the specific base space $\B$.\footnote{Recall from our previous discussion that over a general base $\B$ with maximally generic choice of these polynomials, the fibre over $\calc_I$ merely exhibits a Type $II$ singularity without rendering $\Y$ singular.} The locus $\calc_I$ uplifts to the simultaneous intersection of the O7-plane with $D_1$, $D_1'$ and $D_2$;  as stressed above, in the weak-coupling limit no massless states arise here. By contrast, in the specific F-theory model non-perturbative matter states are localised on this curve. The ${\mathfrak g}_2$-singularity in the fibre indicates that 
these must be described by multi-pronged $(p,q)$-strings which are invisible in the perturbative limit.

We are now in a position to exemplify a curve $\Sigma$ on $\X$ which uplifts, in the split $I_1$ F-theory model over base $\B$, to a global 3-chain $\Gamma$ of the kind discussed in the previous sections.
By Poincar\'e duality, we can  unambiguously isolate the curves $\calc_H$, $\calc_U$, $\calc_V$ on $\X$ which are canonically dual to the divisors $H$, $U$, $V$ respectively, i.e.\ such that the only non-vanishing intersection numbers are 
\be
H\cdot \calc_H=U\cdot \calc_U=V\cdot \calc_V=1.
\ee
  These can be more concretely identified with the restriction to $\X$ of the curves canonically associated with the GLSM charges in the toric ambient space listed in the three lines of table \ref{tab:scalingX}. Notice that under the orientifold involution $\calc_H\leftrightarrow \calc_H$ and $\calc_U\leftrightarrow \calc_V$.
  We can then identify two 2-cycles
 \be
 \calc=\calc_U-\calc_H,\quad~~~~~~~~~ \calc'=\calc_V-\calc_H
 \ee
related by orientifold involution as $\calc\leftrightarrow \calc'$. Consider the D1-brane wrapped on the odd 2-cycle
\be
\Sigma=\calc-\calc'.
\ee
Notice that
\be
U\cdot\calc=1,\quad~~~~ H\cdot\calc=-1,\quad~~~~~ V\cdot\calc=0.
\ee
Thus, a D1-brane wrapping $\calc$ intersects only $U$ and $H$ with intersection numbers $+1$ and $-1$ respectively, and therefore
\be \label{intnumD1a}
D_1\cdot\calc=1,\quad~~~~    D'_1\cdot\calc=0,  \quad~~~~     D_2\cdot\calc=-1.
\ee
Analogously, we have the orientifold-image intersection numbers  
\be  \label{intnumD1b}
D_1\cdot\calc'=0,\quad~~~~    D'_1\cdot\calc'=1,  \quad~~~~     D_2\cdot\calc'=-1.
\ee
We may then deform $\calc$ into a single connected (non-holomorphic) 3-chain\footnote{We thank Julius Shaneson  for discussions on this point.} that intersects $D_1$ and $D_2$ at one point of their intersection locus  $D_1\cap D_2$, with intersection numbers $+1$ and $-1$ respectively, as in the local model of section \ref{sec:D1FT}. (Of course, analogously, the image $\calc'$ will intersect $D'_1$ and $D_2$ at $D'_1\cap D_2$.) 

Hence, the ${\rm D1}$-brane wrapping $\Sigma$ has exactly the proper intersection numbers to uplift to an M2-brane that trivialises the matter $\mathbb{P}^1$ localised in the fibre over $D_1\cap D_2$,
\be
\Sigma \cdot D_1 = 1 = - \Sigma \cdot D'_1, \qquad \Sigma \cdot D_2= 0.
\ee
 Despite the vanishing of the topological intersection number of $\Sigma$ with the invariant brane $D_2$ in class $8 H + 7 (U + V)$, a single non-chiral  zero mode at the intersection of $\Sigma$ with $D_2$ is guaranteed to exist from the above specific realisation of $\Sigma$. This exemplifies how the explicit construction of $\Sigma$ goes beyond the computation of the chiral index of instanton zero modes.

In this subsection we have mostly worked at weak coupling and we have focused on the matter curve $\calc_{II}$, at which we can apply the general discussion of the previous sections. However, the other matter curve ${\cal C}_{I}$ provides another intrinsically strongly coupled example of the same general idea. Indeed, the ${\mathfrak g}_2$ singularity  over $\calc_I$ is conceptually on the same footing as the $I_2$ singularity over $\calc_{II}$: In both cases massless states localise in codimension-two which are not charged under any massless $U(1)$ gauge symmetry. Therefore no Calabi-Yau resolution exists for these singularities. The ${\mathfrak g}_2$ singularity is thus a truly non-perturbative version of the same effect we have been studying for the $I_2$ singularity and its associated conifold transition.

Finally, let us notice that crucially, in our explicit example,   $\calc_I$ and $\calc_{II}$ do not intersect. In fact, this locus would be given by $\{w=0\} \cap \{h=0\} \cap \{v=0\}$, but the intersection numbers of the associated divisor classes vanish,
\bea \label{vaninnum}
U \, (P+U) \, (2P +U) = 0.
\eea 
This is essential as otherwise at this locus the vanishing orders $(1,2,3,4,6;12)$ of $(a_1, a_2, a_3, a_4, a_6; \Delta)$
would indicate a non-minimal enhancement beyond $E_8$, which would defy a conventional interpretation of the geometry. 
Note that  (\ref{vaninnum})  corresponds to the condition that the locus $\{w = 0\} \cap \{a_1 = 0\} \cap \{a_{21} = 0 \}$ is trivial. In a general split $I_1$ model, this is exactly the locus that, at weak coupling,  corresponds to the conifold singularity on $\X$ {\cite{Donagi:2009ra,Krause:2012yh,Esole:2012tf} and which then prevents the generic   split $I_1$ model
 to admit a well-defined Sen limit, as we already mentioned at the beginning of this subsection.

\section{Conclusions}\label{sec:Conclusions}

In this article we have studied selection rules for charged matter couplings in 4-dimensional F-theory compactifications.
An important question is to determine whether an operator uncharged under all massless and discrete gauge symmetries actually appears perturbatively in the effective action. By this we mean that the coupling is realised without exponential volume suppression.
To this end we have proposed to distinguish between {\em perturbative} and {\em non-perturbative} homological relations of fibral curves on the Calabi-Yau 4-fold.
A certain coupling  of states is known to be allowed by all massless gauge symmetries and discrete gauge symmetries if the states arise from M2-branes wrapping 
fibral curves which sum up to zero in the  integral homology of the elliptic fibration.
If this homological relation holds also in the fibral homology, the coupling is perturbative in the above sense.
Otherwise a coupling of the associated states can only be mediated by an M2-instanton with non-vanishing volume in the F-theory limit. Such instanton wraps the 3-chain bounded by the set of fibral curves in question and extends away from the matter locus in the base. This leads to an exponential suppression of the coupling by the volume of the 3-chain in the F-theory effective action.

If a  coupling can arise only in this non-perturbative, i.e.\ exponentially suppressed sense, one can view this as the result of a geometrically massive $U(1)$ symmetry in the spirit of \cite{Grimm:2011tb}. The mass of the $U(1)$ boson sits at the KK level.
We have argued in this article that the set of independent homological relations which are only realised as non-perturbative homological relations is in 1-1 correspondence with such massive $U(1)$s, generalizing the results of \cite{Braun:2014nva}.
The non-perturbative effects break this $U(1)$ completely or to a discrete subgroup as first described in \cite{BerasaluceGonzalez:2011wy} in weakly coupled string theory.

In fact, our results shed new light on the role of discrete symmetries in  F-theory: The full effective discrete gauge group is geometrically realised via torsional (co)homology \cite{Camara:2011jg,Mayrhofer:2014laa}.
In particular, if Tor$H^3(\Y,\mathbb Z) = \mathbb Z_k$ (modulo torsion in the base), then the effective action contains a discrete gauge group factor $\mathbb Z_k$.
In this case also Tor$H_2(\Y,\mathbb Z) = \mathbb Z_k$ and there exist $k$-torsional fibral curves. Only those linear combinations of fibral curves whose torsion part vanishes in $H_2(\Y,\mathbb Z)$ give rise to allowed couplings. 
If the torsion charge vanishes in the perturbative homology, then the couplings allowed by the discrete $\mathbb Z_k$ symmetry arise already at the perturbative level.
There can nonetheless be situations in which the perturbative couplings do not exhaust the full set of operators compatible with the gauge and the discrete $\mathbb Z_k$ symmetries.
In this case the effective perturbative symmetry group is enhanced. In general, therefore, only the full symmetry group including the part governing non-perturbatively suppressed couplings coincides with what is dictated by the torsional (co)homology.

A model with discrete gauge group $\mathbb Z_k$ can be related to a model with a massless $U(1)$ by Higgsing a field with charge $k$.  This requires that the Higgs field is a localised matter field as opposed to merely an axion participating in the St\"uckelberg mechanism.  If all couplings allowed by the $U(1)$ gauge group are realised perturbatively, then the same is true for the discrete symmetry after Higgsing.
The F-theory models with discrete symmetry studied so far in \cite{Morrison:2014era,Anderson:2014yva,Klevers:2014bqa,Garcia-Etxebarria:2014qua,Mayrhofer:2014haa,Mayrhofer:2014laa,Cvetic:2015moa} are of this type for sufficiently generic base space. For an F-theory compactification with $\mathbb Z_2$ symmetry we have demonstrated in detail how the perturbative homological relations carry over between the Higgsed phase, described as a bisection fibration \cite{Braun:2014oya,Morrison:2014era,Anderson:2014yva,Klevers:2014bqa,Garcia-Etxebarria:2014qua,Mayrhofer:2014haa,Mayrhofer:2014laa}, and the unhiggsed $U(1)$ \cite{Morrison:2012ei} phase. This has lead us to conjecture that in the associated Jacobian model \cite{Braun:2014oya} the $\mathbb Z_2$ torsion \cite{Mayrhofer:2014laa} must arise in the perturbative homology. 
It would be interesting to prove this claim directly in the Weierstrass model and to identify the associated 3-chains.

By contrast, models where the full discrete gauge group is exhausted only at the non-perturbative level therefore either require that also in the unhiggsed $U(1)$ phase not all couplings arise perturbatively - this is possible for non-generic base spaces as we have discussed  - or that no direction in the complex structure moduli space exists in which the discrete gauge group enhances to a massless $U(1)$ gauge group at finite distance in moduli space.
In F-theory both options are less generic because of the rich spectrum of $[p,q]$-strings which can realise both the required Higgs field and account for its interactions.  
In perturbative Type II orientifolds, on the other hand, models with discrete $\mathbb Z_k$ gauge group related to a massless $U(1)$ by a St\"uckelberg mechanism (unlike a Higgs mechanism involving a localised open string Higgs field) are known to arise. From an F-theory perspective these models form a proper subclass.

The 3-chains which lead to a volume-suppression of couplings in F-theory can crucially depend on the global structure of the compactification. For fibrations with a perturbative Sen limit, this happens when the M2-instanton wrapping the 3-chain corresponds to a D1-instanton in the weakly-coupled IIB theory.
We have been able to make this very precise by relating the 3-chain to an orientifold-odd 2-cycle on the Calabi-Yau double cover of the base. 
Along the way we have gained further insights into the F-theory uplift of orientifold-odd forms and cycles by exploiting the Mayer-Vietoris sequence in the stable degeneration limit introduced in \cite{Clingher:2012rg} (see also \cite{Braun:2014pva,Braun:2014nva}). 
A challenge for future studies will be to estimate the volume of the 3-chain also in situations without obvious weak coupling limit in order to quantitatively determine the value of the suppressed coupling.
From a more formal perspective, it will be illuminating to investigate the precise relation between the notion of perturbative homological equivalence introduced in this paper and standard concepts of refined equivalence between cycles in algebraic geometry such as algebraic or rational equivalence.

 M2-brane instantons wapping 3-chains in Calabi-Yau 4-folds are by themselves not half-BPS and can therefore only contribute to the D-terms. The generation of F-terms requires that the M2-instanton form a half-BPS bound state with a suitable  M5-instanton along a vertical divisor of the elliptic fibration. In the companion paper \cite{paper2} we will study such configurations and relate them to M5-instantons with suitable field strength for their chiral 2-form.

\bigskip

\centerline{{\bf Acknowledgements}}

\noindent We are indebted to Andreas Braun, Andres Collinucci, Christoph Mayrhofer, Eran Palti  and Roberto Valandro for many important discussions and comments. 
We further thank Lara Anderson, Martin Bies, Herb Clemens, Antonella Grassi, Hans Jockers, Ling Lin, Dave Morrison, Julius Shaneson and Oskar Till for helpful discussions and correspondence.
This work was partially funded by DFG under Transregio 33 `The Dark Universe' and by
the Padua University Project CPDA144437.

\newpage

\centerline{\LARGE \bf Appendix}
\vspace{0.5cm}
\begin{appendix}

\section{Splitting curves in the $SU(5)$ model}
\label{app:split}

In this appendix we describe how, in the $SU(5)$ model described in section \ref{sec:SU(5)ex}, the split (\ref{P1split}) of  $\mathbb P^1_{3D}$ over $p_2$ arises. Generally, to determine the fibre structure over specific loci of $\B$ one has to analyse the factorisation properties of $\hat P_T$ given in (\ref{PTeq}) restricted to those loci.
In order to perform this explicitly over the curve  $\calc_{\bf 5}$ we apply the same reasoning as in \cite{Lin:2014qga} and first note that 
away from $a_{21} = 0$ the defining polynomial  $\cald$ given in (\ref{C10C5}) can be factorised as
\be
\cald = \frac{1}{a_{21}}   \cald^+     \cald^- \qquad \text{with}\quad 
\cald^\pm   = a_{21} a_{32} - \frac{1}{2} \Big(a_1 a_{43} \pm a_1 \sqrt{a_{43}^2 - 4 a_{21} a_{65}}   \Big). 
\ee
Let us denote by $\calc_{\bf 5}^\pm$ the two loci $\{w=0\} \cap \{\cald^\pm=0\}$. Hence, away from $a_{21} = 0$ we may write $\calc_{\bf 5}\simeq \calc_{\bf 5}^+\cup \calc_{\bf 5}^-$. However
the two components $\calc_{\bf 5}^+$ and $\calc_{\bf 5}^-$ are not disjoint. Apart from having the Yukawa coupling points $a_1=0=a_{32}$ in common,  they 
are glued together along the vanishing locus of the square-root. Thus $\calc_{\bf 5}$ is really irreducible. 
The advantage of the above presentation of $\cald$ is that we can solve $\cald^\pm=0$ for $a_{32}$ and plug this expression into the equation of the hypersurface equation.
This introduces rational expressions in $a_{21}$, which is sufficient for our purposes as we restrict ourselves to the patch where $a_{21} \neq 0$.
One then finds that over generic points in this patch, the rational fibre of $E_3$ splits because  the hypersurface constraint factorises into the polynomials
\be
\begin{aligned}
 \hat P_T|_{\{e_3= 0\} \cap  \{\cald^\pm = 0\} } &= \frac{1}{2 a_{21}} {\hat P}^\pm_{3C} \,   \hat P^\pm_{3D}, \\
 {\hat P}^\pm_{3C}   &=     x + e_4 \frac{a_{43} \pm \sqrt{a_{43}^2 - 4 a_{21} a_{65}}  }{2 a_{21}}   , \\
 \hat P^\pm_{3D}     &=   - 2 a_1 a_{21}  y + 2 a_{21}^2 x e_2 + e_2 e_4 (a_{21} a_{43} \mp a_{21}  \sqrt{a_{43}^2 - 4 a_{21} a_{65}}  )  .
  \end{aligned}
\ee
This means that over $\{\cald^\pm=0\}$ the fibre of $E_3$ factorises into two fibral curves which we denote by $ (\mathbb P^1)^\pm_{3C}$  and $ ( \mathbb P^1)^\pm_{3D}$ and which are given explicitly as the loci
\be
\begin{aligned}
 (\mathbb P^1)^\pm_{3C} &= \{e_3=0\} \cap \{ {\hat P}^\pm_{3C} =0\} \cap \cald^\pm  \cap D_1, \\
 (\mathbb P^1)^\pm_{3D} &= \{e_3=0\} \cap \{ {\hat P}^\pm_{3D} =0\} \cap \cald^\pm  \cap D_1, 
 \end{aligned}
\ee
where $D_1$ is any divisor on the base that intersects $\cald^\pm$ in a single point.
As we pass from $\{\cald^+=0\}$ to $\{\cald^-=0\}$ by crossing the branch-cut, the square-root changes sign. This exchanges $( \mathbb P^1)^+_{3C}$  with $ (\mathbb P^1)^-_{3C}$ and  $ (\mathbb P^1)^+_{3D}$  with $ (\mathbb P^1)^-_{3D}$.
Nonetheless over the full curve $\calc_{\bf 5}$ the fibre of $E_3$ splits into two globally distinct components \cite{Lin:2014qga}  $\mathbb P^1_{3C}$  and $ \mathbb P^1_{3D}$ .
As we approach one of the Yukawa points $p_2$ along a path $\gamma_\pm$ starting on ${\cal C}_{\bf 5}^\pm$, we observe the factorisation 
\be
 \hat P_{3D}^\pm |_{p_2}  =   2 a^2_{21} \,  \hat P_{32} \hat P_{3C}^{\pm}|_{p_2} \qquad\text{with}\quad  \hat P_{32} =  e_2\,.
\ee

Indeed, at $p_2$ the fibre contains both curves $\mathbb P^1_{3\pm} \equiv (\mathbb P^1)^\pm_{3C}$, given explicitly as
\bea
\mathbb P^1_{3\pm} = \{e_3=0\} \cap \{\hat P_{3C}^{\pm}=0\} \cap \{a_1=0\} \cap \{a_{32}=0\}.
\eea 
These two distinct fibral curves are reached either via a splitting of $\mathbb P^1_{3D}$ along $\gamma_\pm$, or  by continuing $\mathbb P^1_{3C}$ towards $p_2$ along $\gamma_\mp$.
This implies that $\mathbb P^1_{3+}$ and $\mathbb P^1_{3-}$  at $p_2$ are connected via a 3-chain given by fibering $\mathbb P^1_{3C}$ over a path $\gamma$ leaving the Yukawa point via ${\cal C}_{\bf 5}^+$, crossing to ${\cal C}_{\bf 5}^-$ and approaching the Yukawa point again. This is a manifestation of the fact that the ${\bf 5}$-curve self-intersects \cite{Hayashi:2009bt} at $p_2$. For the computation of the weight vectors associated with the individual fibral curves we refer to the analysis in  \cite{Krause:2011xj}.

\section{Axionic gauging}\label{app:gauging}

We work with `twisted' R-R forms, defined by $C= e^{B_2}\wedge C^{\rm standard}$, where $C=\sum_{k} C_k$ and  $C^{\rm standard}=\sum_{k} C^{\rm standard}_k$, with $C^{\rm standard}_k$ denoting the more commonly used RR potentials.  Away from localised sourced, the associated `twisted'  field-strengths are locally defined by $G_{k+1}=\d C_k$. By using `twisted' RR forms and units $2\pi\sqrt{\alpha'}=1$, the D-brane WZ-coupling is given by $2\pi\int C\wedge e^{\frac{F}{2\pi}}$. Furthermore,  in the D7/O7  backgrounds described in section \ref{axgauging}, the R-R Bianchi identity for $G_3$
takes the form
\be\label{IIBBI}
\d G_3= -\frac{1}{2\pi}\sum_a\Big[\delta_{2}(D_a)\wedge F^a+\delta_{2}(D'_a)\wedge F^{a\prime}\Big] -\frac{1}{2\pi}\sum_\alpha\delta_{2}(\hat D_\alpha)\wedge \hat F^\alpha,
\ee  
where we have omitted curvature induced terms, which are not relevant for  the following discussion. For the sake of generality we allow for a non-vanishing supersymmetric expectation  value of the world-volume field-strengths on the internal divisors. This means that, for instance, we can substitute $F^a$ with $\d A^a+ F^a$ in the above formulas, where $F^a$ now represents the self-dual non-vanishing flux, such that $[\frac{F^a}{2\pi}]$ defines a non-trivial element  of $H^2(D_a,\mathbb{Z})$, while $A^a$ represents the dynamical gauge field. Correspondingly, we can split $G_k$ into $G^0_k+\tilde G_k$, where $G^0_k$ solves (\ref{IIBBI}) with the background D7-brane fluxes, while $\tilde G_k$ ($\d \tilde G_k=0$) represent the dynamical fluctuations around the background configuration.

Then  \eqref{IIBBI} reduces to 
\be\label{IIBBI2}
\begin{aligned}
&\d \Big\{\tilde G_3+\frac{1}{2\pi}\sum_a\Big[A^a\wedge \delta_{2}(D_a)+A^{a\prime}\wedge \delta_{2}(D'_a)\Big] +\frac{1}{2\pi}\sum_\alpha\hat A^\alpha\wedge \delta_{2}(\hat D_\alpha)\Big\} =0.
\end{aligned}
\ee  
This can be  solved by introducing the dynamical R-R-potential $\tilde C_2$,
\be\label{IIBBIsol}
\begin{aligned}
\d \tilde C_2&=\tilde G_3+\frac{1}{2\pi}\sum_a\Big[A^a\wedge \delta_{2}(D_a)+A^{a\prime}\wedge \delta_{2}(D'_a)\Big] +\frac{1}{2\pi}\sum_\alpha\hat A^\alpha\wedge \delta_{2}(\hat D_\alpha).
\end{aligned}
\ee  
Since $G_3$ is gauge invariant, $C_2=C_2^{(0)}+\tilde C_2$ must transform as 
\be\label{RRshift}
\begin{aligned}
C_2&\rightarrow  C_2+\frac{1}{2\pi}\sum^a\Big[\lambda^a\, \delta_{2}(D_a)+\lambda^{a\prime}\, \delta_{2}(D'_a)\Big] +\frac{1}{2\pi}\sum_\alpha\hat\lambda^\alpha\, \delta_{2}(\hat D_\alpha)
\end{aligned}
\ee  
under the $U(1)$ gauge transformations $A^a\rightarrow A^a+\d \lambda^a$,  $A^{a\prime}\rightarrow A^{a\prime}+\d \lambda^{a\prime}$ (with $\lambda^{a\prime}=-\sigma^*(\lambda^a)$) and $\hat A^\alpha\rightarrow \hat A^\alpha+\d \hat\lambda^\alpha$.

In particular, suppose that the $U(1)$ parameters depend just on the external coordinates (hence defining 4D gauge transformations) or,  in other words, that they are constant on the internal divisors. Hence the orientifold projection implies that $\lambda^{a\prime}=-\lambda^a$ and  $\hat\lambda^\alpha=0$, so that
\be
C_2\rightarrow  C_2+\frac{1}{2\pi}\sum_a\lambda^a\, \Big[\delta_{2}(D_a)- \delta_{2}(D'_a)\Big] .  \label{RRshiftgeom}
\ee
This implies that under the 4D $U(1)_a$ gauge transformations the R-R field $C_2$   is shifted by cohomologically non-trivial terms and that (\ref{RRshift}) provides the ten-dimensional origin of this effect.

For completeness, we remark  that one can repeat the same steps  for the R-R potential $C_4$, starting from its Bianchi identity
\be
\d G_5= \frac{1}{8\pi^2}\sum_a\Big[\delta_{2}(D_a)\wedge F^a\wedge F^a+\delta_{2}(D'_a)\wedge F^{a\prime}\wedge F^{a\prime}\Big]+ \frac{1}{8\pi^2}\sum_\alpha\delta_{2}(\hat D_\alpha)\wedge \hat F^\alpha\wedge \hat F^\alpha
\ee
and arriving at the microscopic gauging 
\be\label{10DC4shift}
C_4\rightarrow  C_4- \frac{1}{4\pi^2}\sum_a\Big[\lambda^a\, F^a\wedge \delta_{2}(D_a)+\lambda^{a\prime}, F^{a\prime}\wedge\delta_{2}(D'_a)\Big]+ \frac{1}{4\pi^2}\sum_\alpha \hat\lambda^\alpha\, \hat F^\alpha\, \wedge\delta_{2}(\hat D_\alpha).
\ee
In the case of purely 3-dimensional gauge transformations, this reduces to
\be
C_4\rightarrow  C_4- \frac{1}{4\pi^2}\sum_a\lambda^a\Big[F^a\wedge \delta_{2}(D_a)- F^{a\prime}\wedge\delta_{2}(D'_a)\Big].
\ee
Hence (\ref{10DC4shift}) provides the ten-dimensional origin of the well-known 4-dimensional St\"uckelberg gauging induced by D7-brane world-fluxes.


\section{Mayer-Vietoris exact sequence} \label{app-MYCS}

In this appendix we compute the homology and cohomology groups of the central fibre $Y_0=Y_E\cup_X Y_T$ appearing in section \ref{sec:Weakcoupling} in the context of the stable degeneration limit of an F-theory compactification.
The Mayer-Vietoris exact  sequence (see for instance \cite{MR1867354})  allows one to express these (co)homologies in terms of those of $Y_E$, $Y_T$ and the normal-crossing divisor $X$. 

\subsection{Homology of the central fibre}
\label{sec:hom}

Let us start from the homology of $Y_0$. The Mayer-Vietoris exact sequence  reads
\be\label{holMV}
\ldots \xrightarrow{\del_*} H_k(X) \xrightarrow{(\iota_{E*},-\iota_{T*})}  H_k(Y_E)\oplus H_k(Y_T) \xrightarrow{j_{E*}+j_{T*}} H_k(Y_0)\xrightarrow{\del_*} H_{k-1}(X)\xrightarrow{(\iota_{E*},-\iota_{T*})} \ldots
\ee
Here $\iota_E: X\rightarrow Y_E$, $\iota_T: X\rightarrow Y_T$, $j_E: Y_E\rightarrow Y_0$ and $j_T: Y_T\rightarrow Y_0$ are the inclusion maps. On the other hand, $\del_*$ is non-trivial on $k$-cycles 
$y\subset Y_0$ which can be written as the union $y=y_E\cup_x y_T  $ of two $k$-chains $y_E\subset Y_E$ and $y_T\subset Y_T$ with common boundary on $X$, that is, $\del y_E =-\del y_T=x$, where $x\subset X$ is a $(k-1)$-cycle. Then, on such a cycle $y=y_E\cup_x y_T  $,
\be
\del_* y=x .
\ee
The long exact sequence (\ref{holMV}) gives the short exact sequence
\be\label{splithom}
0\longrightarrow V_k(Y_0) \longrightarrow H_k(Y_0)\longrightarrow T_{k-1}(Y_0) \longrightarrow 0 ,
\ee
where
\be\label{kerimhom}
\begin{aligned}
V_k(Y_0)&\equiv{\rm coker}(\iota_{E*},-\iota_{T*})_k=\frac{H_k(Y_E)\oplus H_k(Y_T)}{ \{\iota_{E*}x_k\oplus -\iota_{T*}x_k|x_k\in H_k(X)\}}\,,\\
T_{k-1}(Y_0)&\equiv {\rm ker}(\iota_{E*},-\iota_{T*})_{k-1}=\{x_{k-1}\in H_{k-1}(X)| \iota_{E*} x_{k-1}=0\text{ and }\iota_{T*}x_{k-1}=0   \}\,.
\end{aligned}
\ee
In other words, the cycles in $H_k(Y_0)$ are of two types. The first ones belong to $V_k(Y_0)$ and are the $k$-cycles in $Y_E$ and/or $Y_T$, modded out by $H_k(X)$, in order not to overcount them. The second ones belongs to $T_{k-1}(Y_0)$ and are of the form  $y=y_E\cup_x y_T $  described above, so that they are indeed associated  with $(k-1)$-cycles $x\subset X$  which are trivial in both $Y_E$ and $Y_T$.

Let us consider more explicitly  $H_2(Y_0)$ and $H_3(Y_0)$. Assuming $H_1(X)=0$ (which for instance holds when $X$ is a Calabi-Yau), we have $T_1(Y_0)={\rm ker}(\iota_{E*},-\iota_{T*})_{1}=0$, so that
\be
H_2(Y_0)= V_2(Y_0)=\frac{H_2(Y_E)\oplus H_2(Y_T)}{  \{\iota_{E*}x_2\oplus -\iota_{T*}x_2|x_2\in H_2(X)\}}.
\ee
On the other hand,  
\be
H_3(Y_0)\simeq V_3(Y_0)\oplus T_{2}(Y_0)\,,
\ee
where
\be
\begin{aligned}
V_3(Y_0)&\equiv   \frac{H_3(Y_E)\oplus H_3(Y_T)}{  \{\iota_{E*}x_3\oplus -\iota_{T*}x_3|x_3\in H_3(X)\}}          \,,\\
T_{2}(Y_0)&\equiv     \{x_{2}\in H_{2}(X)| \iota_{E*} x_{2}=0\text{ and }\iota_{T*}x_{2}=0\}  \,.
\end{aligned}
\ee


\subsection{Cohomology of the central fibre}
\label{sec:coho}

Let us pass to the dual (integer) cohomology. The dual Mayer-Vietoris sequence is 
\be\label{coholMV}
\begin{aligned}
\ldots  \xrightarrow{}  H^{k-1}(Y_E)\oplus H^{k-1}(Y_T)&\xrightarrow{\iota^*_{E}-\iota^*_{T}}  H^{k-1}(X)\xrightarrow{\del^*}H^{k}(Y_0)\\
&\xrightarrow{(j_E^*,j^*_T)} H^{k}(Y_E)\oplus H^{k}(Y_T) \xrightarrow{\iota^*_{E}-\iota^*_{T}} H^{k}(X)\ldots
\end{aligned}
\ee
This induces the short exact sequence 
\be\label{splitcohom}
0\longrightarrow T^{k-1}(Y_0) \longrightarrow H^k(Y_0)\longrightarrow V^k(Y_0) \longrightarrow 0,
\ee
where
\be
\begin{aligned}
V^k(Y_0)&\equiv {\rm ker}(\iota^*_{E}-\iota^*_{T})_{k}=\{\omega_E\oplus\omega_T\in H^{k}(Y_E)\oplus H^{k}(Y_T)| \iota^*_{E}\omega_E=\iota^*_{T}\omega_T  \}\,,\\
T^{k-1}(Y_0)&\equiv{\rm coker}(\iota^*_{E}-\iota^*_{T})_{k-1}=\frac{H^{k-1}(X)}{\{\omega_X\in H^{k-1}(X)| \omega_X=\iota^*_{E}\omega_E-\iota^*_{T}\omega_T\}}\,.
\end{aligned}
\ee

We will focus on  $H^2(Y_0)$ and $H^3(Y_0)$. Assuming that $H^1(X)=0$ (as it happens if $X$ is a Calabi-Yau), $T^{1}(Y_0)=0$ and  then 
\be
H^2(Y_0)= V^2(Y_0)=\{\omega_E\oplus\omega_T \in H^{2}(Y_E)\oplus H^{2}(Y_T)| \iota^*_{E}\omega_E=\iota^*_{T}\omega_T  \}\,.
\ee
On the other hand 
 \be
 H^3(Y_0)\simeq V^3(Y_0)\oplus T^{2}(Y_0),
 \ee
where 
\be
\begin{aligned}
V^3(Y_0)&=\{\omega_E\oplus\omega_T\in H^{3}(Y_E)\oplus H^{3}(Y_T)| \iota^*_{E}\omega_E=\iota^*_{T}\omega_T  \}\,,\\
T^{2}(Y_0)&=\frac{H^{2}(X)}{\{\omega_X\in H^{2}(X)| \omega_X=\iota^*_{E}\omega_E-\iota^*_{T}\omega_T\}}\,.
\end{aligned}
\ee


\end{appendix}

\bigskip
\bibliography{references}  
\bibliographystyle{custom1}

\end{document}